\documentclass[letter, twocolumn, amsfonts, amssymb, amsmath, reprint, showkeys, nofootinbib, twoside]{revtex4-2}
\usepackage[english]{babel}
\usepackage[sort&compress]{natbib}
\usepackage[utf8]{inputenc}
\usepackage[colorinlistoftodos, color=green!40, prependcaption]{todonotes}
\usepackage{amsthm}
\usepackage{mathtools}
\usepackage{physics}
\usepackage{xcolor}
\usepackage{graphicx}
\usepackage{adjustbox}
\usepackage{placeins}
\usepackage[T1]{fontenc}
\usepackage{lipsum}
\usepackage{csquotes}
\usepackage{booktabs}
\usepackage[hypertexnames=false]{hyperref}
\hypersetup{colorlinks=true,citecolor=blue,linkcolor=blue,urlcolor=black}

\renewcommand{\selectlanguage}[1]{}

\newcommand{\figp}[1]{(\textbf{#1})}
\def\gmc{G_\text{mc}}
\def\gse{G_\text{se}}
\def\gsemelt{\gse^\text{melt}}
\def\gsegrow{\gse^\text{grow}}
\def\gsecrit{\gse^\text{crit}}
\def\tmelt{t_\text{melt}}
\def\tcycle{t_\text{cycle}}

\begin{document}

\title{A minimal scenario for the origin of non-equilibrium order} 
\author{Riccardo Ravasio$^{*,1}$, Kabir Husain$^{*,1,2}$, Constantine G. Evans$^{3}$,  Rob Phillips$^4$, Marco Ribezzi$^5$, Jack W. Szostak$^6$, Arvind Murugan$^1$}
\affiliation{
${}^1$ Department of Physics, University of Chicago, Chicago, IL 60637,
${}^2$ Department of Physics and Astronomy, University College London, United Kingdom,
${}^3$ Hamilton Institute, Maynooth University, Maynooth, Ireland,
${}^4$ Department of Physics, Caltech, Pasadena, CA 91125,
${}^5$ Laboratoire de Biochimie, Chimie Biologie et Innovation, ESPCI Paris, Université PSL, Paris, France,
${}^6$ Department of Chemistry, University of Chicago, Chicago, IL 60637,
${}^*$ These authors contributed equally.}

\begin{abstract}
    Life uses non-equilibrium mechanisms to create ordered structures not attainable at equilibrium; the resulting order is assumed to provide functional benefits that outweigh costs of time and energy needed by these mechanisms.
    Here, we show that models of DNA replication and self-assembly, when expanded to include known stalling effects, can evolve error correcting mechanisms like kinetic proofreading and dynamic instability through selection for fast replication alone. We abstract these results into a general framework that predicts a counterintuitive ``order through speed'' effect if the distribution of replication times is wide enough.  We test our results against recent mutational screens of proofreading polymerases. Our work suggests the intriguing possibility that non-equilibrium order can evolve even before that order is directly functional, with consequences for the evolution of mutation rates, viral capsid assembly, and the origin of life.
\end{abstract}

\maketitle

\hspace{1in}

Life is often described as a battle against the second law of thermodynamics: ever since Schr{\"o}dinger's \textit{What is life?} \cite{schrodinger_what_1944}, the ability to maintain a state of unnaturally high order through dissipative mechanisms has been seen as a fundamental characteristic of living matter. Non-equilibrium driving is required to maintain order --- i.e., lower entropy or variation than expected at thermal equilibrium --- in processes ranging from sensing \cite{tu_nonequilibrium_2008,ten_wolde_fundamental_2016}, computation \cite{ouldridge_thermodynamics_2017,mehta_energetic_2012,raichle_appraising_2002} and spatial organization \cite{tan_odd_2022,bacanu_inferring_2023} to the self-replication of heritable information \cite{Neumann1966-qz,ninio_kinetic_1975,hopfield_kinetic_1974,Qian2006-ag,murugan_discriminatory_2014,Ehrenberg1980-qz} itself. 

\begin{figure*}
    \centering
    \includegraphics[scale=1]{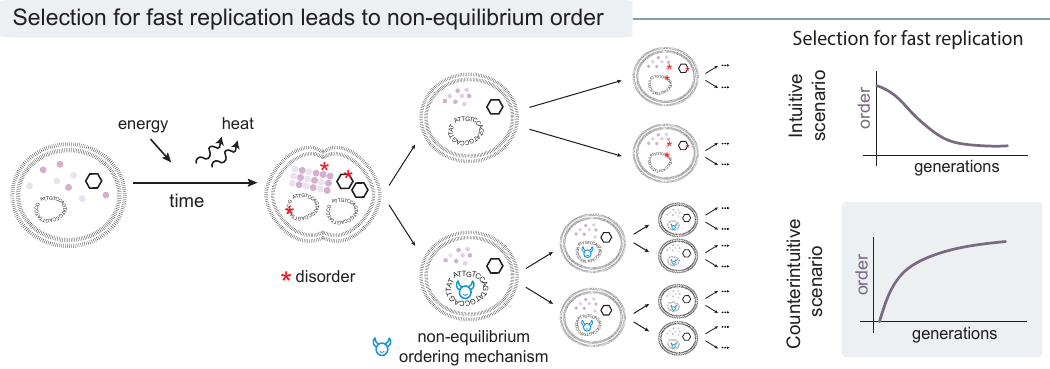}
        \caption{\textbf{An alternative scenario for the origin of non-equilibrium order.} The basic traits of a self-replicator, such as a cell undergoing repeated divisions as shown, are the time to replicate and fidelity in replication, i.e. lack of disorder; both traits are impacted by dissipative fidelity-enhancing mechanisms. Here, we outline a broad set of conditions where selection for fast replication alone, with no selection on fidelity or order, will result in highly dissipative order-maintaining mechanisms. In our scenario, complex non-equilibrium mechanisms that increase order can arise as an inevitable consequence of exponential self-replication itself, even if there is no fitness benefit to that order. }
    \label{fig:speedorderschematic}
\end{figure*}

A distinct question from the functioning of non-equilibrium mechanisms is their spontaneous origin; what conditions might drive equilibrium matter to spontaneously start exploiting energy sources in the environment to achieve higher order? The conditions for the emergence of such coupling are not known, even though several ideas have been proposed. One point of view argues that all physical systems have a spontaneous tendency to evolve dissipative structures \cite{prigogine_self-organisation_1985,gingrich_dissipation_2016,horowitz_minimum_2017,horowitz_spontaneous_2017}. A more common perspective is that these mechanisms originated through a Darwinian process where the fitness benefit of order was sufficient to overcome the associated costs of energy and time \cite{shoval_evolutionary_2012,murugan_speed_2012,lan_energyspeedaccuracy_2012}; see Fig. \ref{fig:speedorderschematic}. 

Here, we suggest another possibility, shown in Fig. \ref{fig:speedorderschematic}:  systems undergoing rapid exponential self-replication can spontaneously evolve non-equilibrium order-enhancing mechanisms if the distribution of replication times is wide enough. We find that this condition on replication times is met naturally in key processes of the central dogma and in molecular self-assembly where errors cause large variations in replication times. In our scenario, order (or the lack of errors) does not need to provide any direct Darwinian fitness benefit; selection can be entirely for faster replication, a selection implicit in any population of self-replicating agents and explicit in some protocells \cite{Chen2004-oh}.

\section*{Templated replication}
We first explore these ideas in templated replication, a process exploited by all known life forms to transmit heritable information across generations. The fidelity of this replication process is a key physical constraint on the origin of life \cite{joyce_protocells_2018, tjhung_rna_2020} and subsequent evolution \cite{good_dynamics_2017}.  

At equilibrium, we might expect a fidelity constrained by the finite differences between base pair free energies; e.g., $\Delta G = G(AT) - G(AG) \approx 1-3$ kcal/mol $\approx 1.688-5\, k_b T$ would indicate an error rate of $\mu \sim 10^{-1}-10^{-2}$ \cite{loeb_fidelity_1982}. However, several DNA polymerases contain an exonuclease domain that \textit{proofreads} \cite{Hopfield1974-wa,Ninio1975-ky}: 
the nascent DNA strand is occasionally passed from the polymerase to the exonuclease domain where bases are excised in 3’ to 5’ direction. 
Using the exonuclease frequently leads to higher accuracy (lower $\mu$) since wrong nucleotides are preferentially discarded; but since right nucleotides are also discarded on occasion, such error correction also increases the time to copy a strand. A simple mathematical model, following numerous published models \cite{Hopfield1974-wa,Ninio1975-ky,munsky_specificity_2009,sartori_kinetic_2013,murugan_discriminatory_2014, Banerjee2017-fc, Banerjee2017-dg}, quantifies this intuitive `higher fidelity, lower speed' trade-off 
~---~see \textit{Materials and Methods}. 

\begin{figure*}
    \centering
    \includegraphics[scale=1]{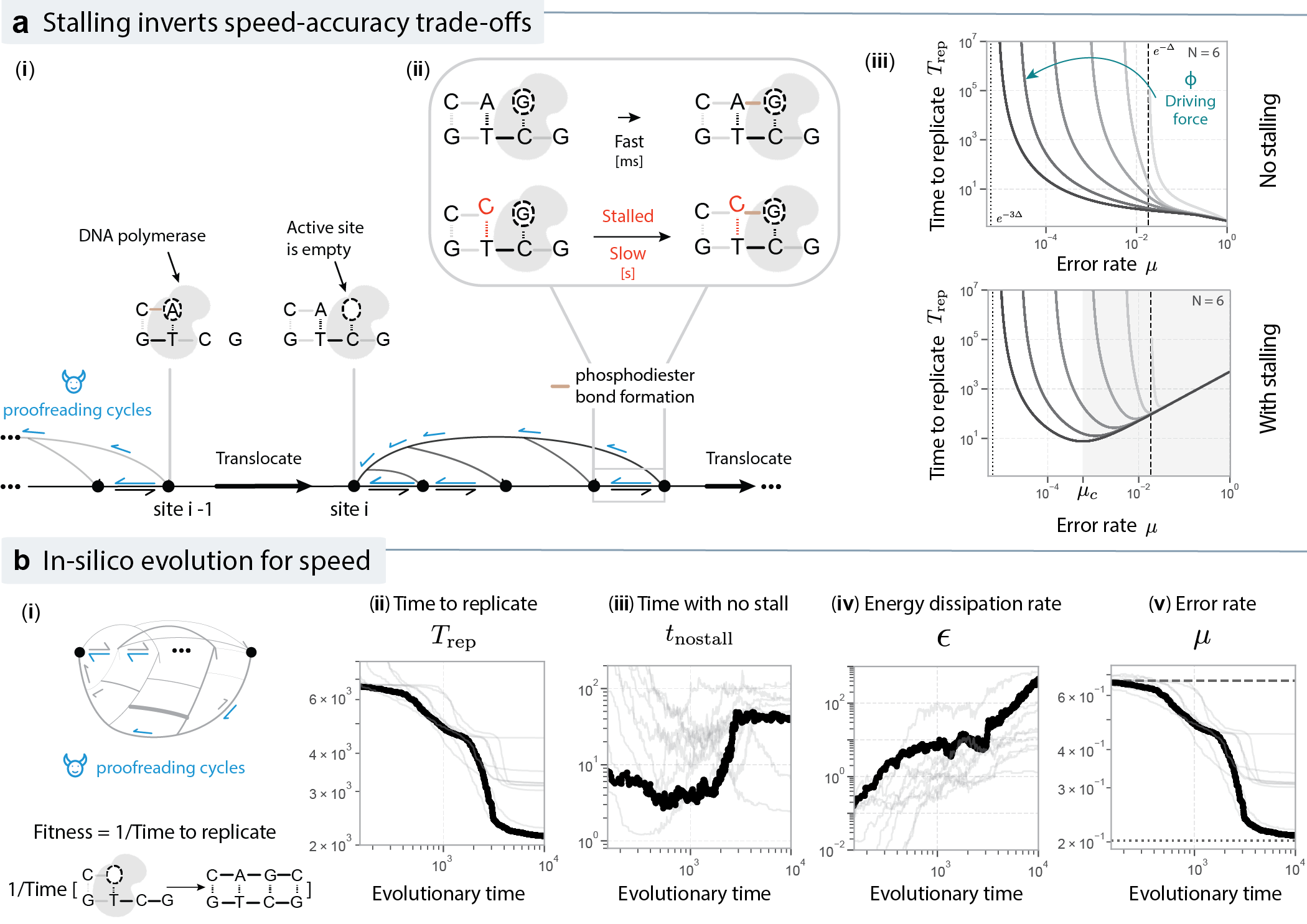}
    \caption{\textbf{Fast replication selects for kinetic proofreading in the presence of stalling.}  \textbf{(a)} We extend canonical models of kinetic proofreading for polymerases (i) by including the stalling effect (ii): incorporation of the correct nucleotide (here, G) at site $i$ is dramatically slowed (up to 1000x \cite{perrino_differential_1989,mendelman_base_1990,huang_extension_1992}) if site $i-1$ has an incorrect base (here, red C) due to misaligned 3' of C and 5' end of G. (iii)   Average time to copy a strand as a function of error rate in proofreading models without and with stalling. Curves correspond to different non-equilibrium driving potentials $\phi$ in the proofreading network. 
    In models with stalling, higher accuracy is linked to higher speed (gray region). \textbf{(b)} (i) \textit{In silico evolution} of a random network model of polymerases that incorporates stalling. The fitness function is set by the time to copy a long strand, with no consideration of accuracy.  (ii,iii,iv,v) Evolution of strand replication time $T_{\rm{rep}}$, copying time  $t_{\rm{nostall}}$ without accounting for stalling, energy $\epsilon$ dissipated by network  and resulting error rate $\mu$ during \emph{in silico} evolution. Typical trajectory is highlighted. 
    }\label{fig:proofreading_unified}
\end{figure*}

We now introduce a key experimentally observed fact ~---~ stalling \cite{perrino_differential_1989,mendelman_base_1990,huang_extension_1992,johnson_structures_2004,ichida_high_2005,baranovskiy_structural_2022,midha_synergy_2023, goppel_kinetic_2021,matsubara_avoidance_2023}~---~ that is not often considered in proofreading models, but completely inverts this intuitive trade-off picture. The incorporation of a wrong nucleotide significantly slows the catalysis of the phosphodiester bond for the \emph{next} base~---~even if correct.  This biophysical effect of geometric nature \cite{johnson_structures_2004}, due to incorrect seating of the prior incorrect base on the template, e.g. misaligned 3' end, 
is intrinsic to the nature of templated replication and is independent of proofreading activity: stalling has been observed even in non-enzymatic RNA replication \cite{rajamani_effect_2010} relevant to the RNA world~---~see \textit{Supplementary Information}. 
The time to replicate a strand of length $L$ is,
\begin{equation}
\label{eq:time_1}
T_{\rm{rep}} \approx  L t_{\rm{nostall}}(\mu) + L \mu  \tau_{\rm{stall}}
\end{equation}
where $t_{\rm{nostall}}(\mu)$ is the time to copy one base, assuming no stalling occurs; this time, shown in Fig.\ref{fig:proofreading_unified}a-iii, increases as the error rate $\mu$ decreases since additional proofreading activity might excise correctly matched bases as well \cite{johnson_exonuclease_2001, singh_excessive_2020,ninio_kinetic_1975,hopfield_kinetic_1974}; $\mu L$ is the number of mutations, each of which causes a stall of time $\tau_{\rm{stall}}$. 

Augmenting the proofreading model with stalling reveals a striking new regime (Fig. \ref{fig:proofreading_unified}a-iii): for large enough error rates $\mu\geq\mu_c$, the trade-off between time to copy a strand and error is reversed: faster polymerases are more accurate at replicating the strand. 
As derived in the \textit{Supplementary Information} using an analytically tractable model, the width $\mu_c$ of this counterintuitive regime increases with the stalling time $\tau_{stall}$ and with increasing non-equilibrium drive $\phi$: 
$
\label{eq:muc}
\mu_c \sim \frac{\tau_{0}}{\tau_{\rm{stall}}} f(\phi)\, , 
$
where $\tau_0$ is the average incorporation time of a base in the absence of stalling and $f$ is a decreasing function of $\phi$. These results suggest that (i) selection for fast replication alone can lead to the evolution of mechanisms that result in higher fidelity; (ii) such higher fidelity will be preferentially achieved through highly dissipative (higher $\phi$) error-correcting mechanisms. 

To test these predictions, we carried out \emph{in silico} evolution of a DNA polymerase, with fitness agnostic to the copying fidelity $\mu$. Instead, fitness is entirely set by the speed of replicating a strand of length $L$, $F=1/T_{\rm{rep}}$; this time $T_{\rm{rep}}$ includes both proofreading and stalling time as in Eq. \ref{eq:time_1}. 

The polymerase was modeled by a generic chemical network composed of $N$ enzyme-substrate states, as sketched in Fig. \ref{fig:proofreading_unified}b. 
We initialize the network kinetic parameters $k_{ij}$ randomly, but satisfying detailed balance. However, $k_{ij}$ can evolve to break detailed balance, representing energy sources such as NTP hydrolysis: {$k_{ij}=k
_{ij}^{\rm{eq}}e^{\beta \phi}$, where $\phi$ is the driving force and $1/\beta=k_B T$}~---~see \textit{Materials and Methods} and Extended Data Fig. \ref{fig:SI_proofreadingth}. 

The results of the evolution for speed $F=1/T_{\rm{rep}}$ of a network of size $N=7$ are summarized in Fig. \ref{fig:proofreading_unified}b. Over cycles of mutation and selection, the replication time $T_{\rm{rep}}$ falls as expected but surprisingly, so does the error rate $\mu$ ~---~despite no selection for fidelity. Further, the time to copy ignoring stalling $t_{\rm{nostall}}$, \textit{increases} and energy dissipation increase; see Fig.~\ref{fig:proofreading_unified}b.

Thus, a random network, when selected for fast replication, evolved a proofreading mechanism\cite{ } that costs energy and is \emph{slower} (i.e.,  higher $t_{\rm{nostall}}$); but the reduction in errors and stalling ultimately pays off in \emph{faster} replication (i.e., lower $T_{\rm{rep}}$).

\subsection*{Experimental evidence}
An extensive body of theoretical~---~and some experimental~---~work on proofreading reports the intuitive speed-accuracy trade-off. 
How might our work, motivated by experimental observations of stalling \cite{perrino_differential_1989,mendelman_base_1990,huang_extension_1992,johnson_structures_2004,doty_optimizing_2018,baranovskiy_structural_2022}, be consistent with such prior work? Prior studies consider only the impact of a few mutations or one biophysical perturbation, e.g. changing Mg$^{2+}$ levels \cite{johansson_rate_2008}. Such tests cannot probe the nature of any potential tradeoff between speed and fidelity.  More precisely, trade-offs manifest as a Pareto front defined by an inequality between traits such as speed and accuracy. 
A single or few perturbations to a well-adapted polymerase (e.g., the wildtype) will likely reveal variants further away from the Pareto front. Thus, comparing a pair of variants, in isolation, will support the intuitive trade-off, when the real Pareto front -- as revealed by the collection of all mutants -- corresponds to the counterintuitive trade-off, see Extended Data Fig. \ref{fig:SI_reftradeoff}. Thus, testing the ideas here presented necessarily requires comparing many variants created through extensive mutagenesis.

\begin{figure}[h]
    \centering
    \includegraphics[scale=1]{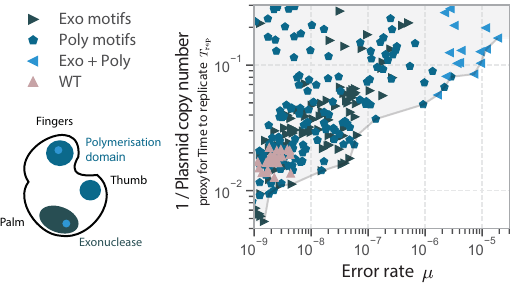}
    \caption{\textbf{Speed-accuracy data for hundreds of proofreading polymerase variants shows a counterintuitive  trade-off.} Error rate vs speed data from the highest throughput mutagenesis of a proofreading DNA polymerase to date \cite{ravikumar_scalable_2018}. Data points of different shape indicate mutations in the exonuclease, polymerase domain or both. DNA plasmid copy number maintained at steady state by a dedicated DNA polymerase is a measure of its total activity that combines speed and processivity; see \textit{Materials and Methods}.}
    \label{fig:data}
\end{figure}

The largest such library of a DNA polymerase mutants was built recently \cite{ravikumar_scalable_2018} in the course of creating the OrthoRep platform \cite{ravikumar_orthogonal_2014}. We plotted the data generated by \cite{ravikumar_scalable_2018} in this process for 213 (single and multiple mutant) variants of the pGKL1 DNA polymerase \cite{jung_yeast_1987}. This viral-origin  Family B DNA polymerase has exonuclease activity and is homologous to the commercially available $\Phi$-29 DNA polymerase. The data in Fig. \ref{fig:proofreading_unified}c shows that the counterintuitive trade-off holds over four orders of magnitude variation in the error rate and two orders of magnitude in the measure of a proxy for speed used in \cite{ravikumar_scalable_2018}. See Extended Data Fig. \ref{fig:SI_chang} for more details and experimental support for the ideas presented here.

\vfill
\section*{Self-assembly}
Nucleic acid replication is not the only instance of our proposed paradigm. We now consider a cell whose replication is limited by the time to assemble a multicomponent structure; the structure is built from $N$ molecular species whose binding interactions $J_{ij}$ include some non-specific interactions. Extant life must build ordered structures like ribosomes~---~with every component in the right place despite non-specific interaction~---~in order to replicate. 

\begin{figure*}
    \centering
    \includegraphics[scale=1]{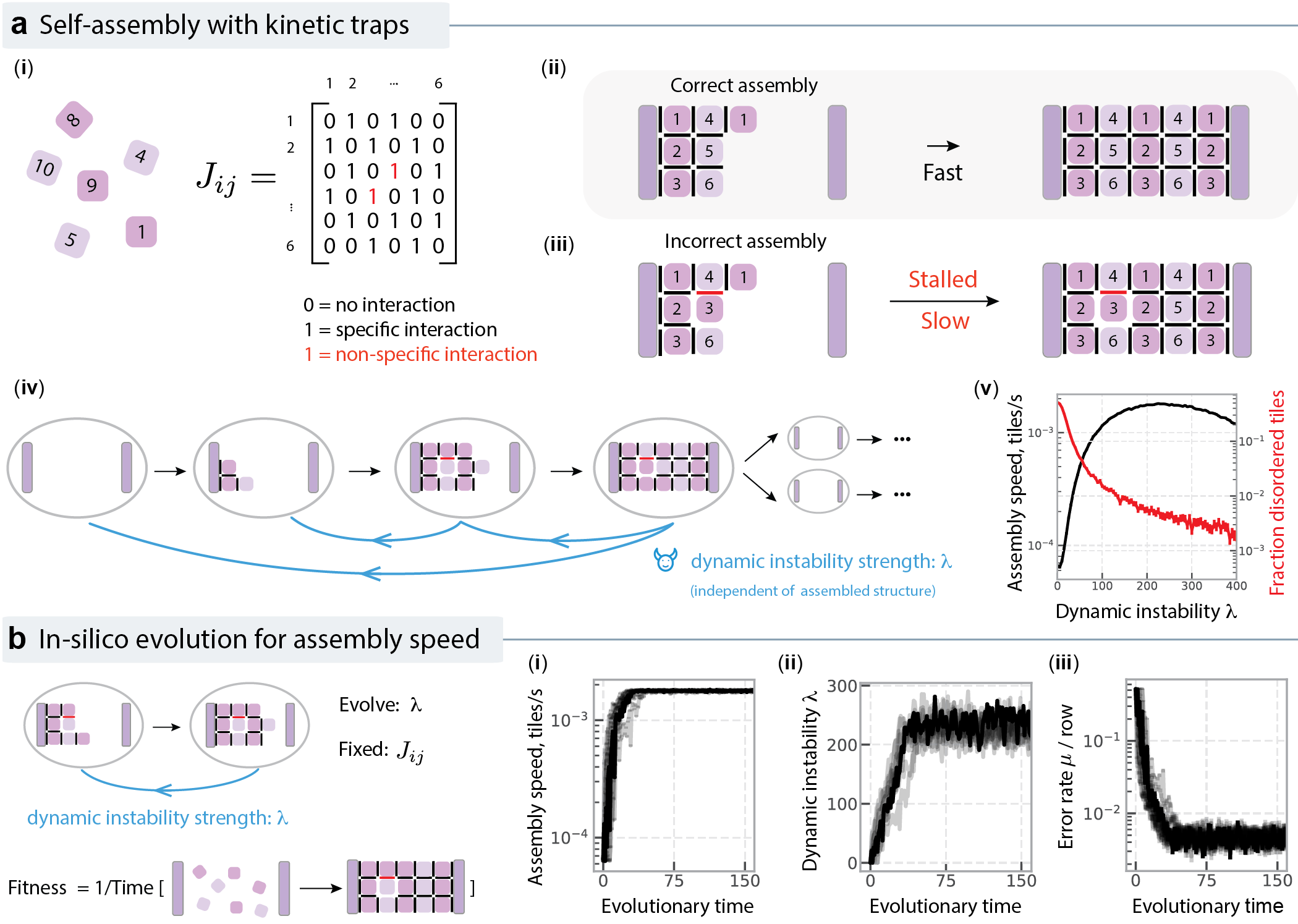}
\caption{\textbf{Fast replication selects for error-correcting dynamic instability in self-assembly.} (\textbf{a}) We consider a cell that divides when it assembles any structure of linear size $L$ using $N$ molecular species; replication is contingent only on structure size and \emph{not} contingent on ordering of components.  The molecules have specific interactions (black entries of $J_{ij}$) that enable assembly of a (ii) regular lattice. (iii) But misincorporations due to random non-specific interactions (red entries of $J_{ij}$) can lead to frustrated partial assemblies that are slow to grow further.  (iv) In our model, cells can potentially evolve a non-equilibrium mechanism that induces frequent irreversible disassembly (dynamic instability~\cite{mitchison_dynamic_1984}). (\textbf{b}) \textit{In silico} evolution with fitness set by the time to assemble \emph{any} structure of length $L$, independent of component ordering. $J_{ij}$ is held fixed but time in conditions of dynamic instability per hour $\lambda$ is allowed to evolve. Despite selection for assembly speed alone and not on order, (ii) dynamic instability strength $\lambda$ increases and consequently (iii) structural errors, defined as variation in component ordering in assembled structures at division, decreases.}
   \label{fig:selfassembly}
\end{figure*}

In contrast, we will assume that our cell replication is \emph{not} contingent on the composition or ordering of components in the assembled structure. Our cells divide upon building \emph{any} structure of linear dimensions at least $L \times W$: 
structural order is not directly functional in our model. This assumption is a theoretical device to highlight the most extreme consequences of our  counterintuitive speed vs structural order relationship.

We carried out \emph{in silico} evolution with a speed-of-assembly fitness function \cite{nguyen_design_2016,bisker_nonequilibrium_2018, marsland_active_2018}. The interaction matrix $J_{ij}$ is fixed but mutations can introduce microtubules-inspired dynamic instability \cite{mitchison_dynamic_1984}; i.e. a layer of non-equilibrium dynamics that allows structures to be partially or entirely disassembled in an irreversible way \cite{hadjivasiliou_selection_2023}~---~see \textit{Materials and Methods} for details. 

Selection for fast self-assembly showed two surprises (Fig. \ref{fig:selfassembly}b). First, the frequency of dynamic instability-\emph{disassembly} $\lambda$ increased (Fig. \ref{fig:selfassembly}b-ii) despite selecting for fast assembly.
Second, despite \emph{not} selecting for ordered assembly, structural variation in assembled structures at the time of cell division \emph{decreased} (Fig. \ref{fig:selfassembly}b-iii); i.e. assembled structures tended to be ordered in a specific way. 

These surprising results can be explained by viewing dynamic instability as a mechanism that speeds up assembly by disassembling misformed structures. As shown in prior work, non-specific interactions can lead to misformed structures \cite{winfree_proofreading_2004,Jhaveri2023-zh} that dramatically slow further assembly. In the absence of dynamic instability, such misformed leading edges might take a long time to spontaneously melt away. Dynamic instability~---~of the right amount~---~might effectively speed up assembly \cite{holy_dynamic_1994} by disassembling these kinetic traps while also fixing errors.

\begin{figure*}
    \centering
\includegraphics[scale=1]{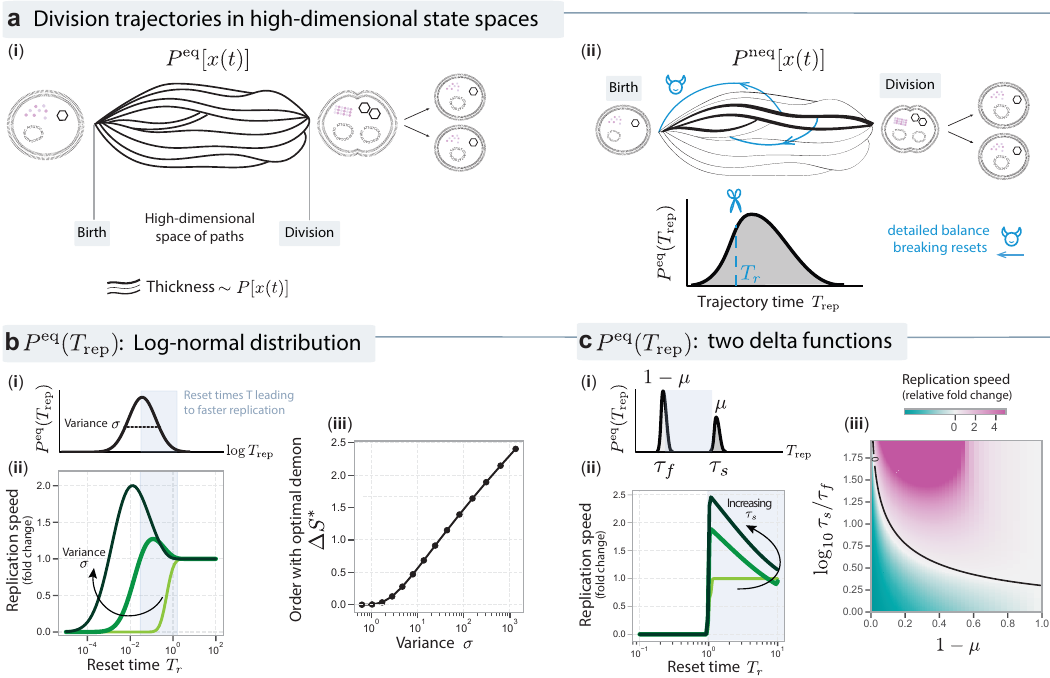}
    \caption{ \textbf{Fast replication selects for Maxwell demons that reduce the entropy of birth-to-division paths in high-dimensional state spaces}   
    (\textbf{a}) A cell divides when it navigates from a Birth to a Division state along any one of many trajectories $x(t)$ through cell state space.  (ii) If the distribution $P^{eq}(T_{\rm{rep}})$ of completion times $T_{\rm{rep}}$ along these trajectories satisfies Eq. \ref{eqn:favorableDemon}, fast replication selects for a Maxwell demon that randomly and irreversibly resets cells back to Birth on a timescale $T_r$. The demon changes the distribution of paths used to reach Division from $P^{eq}[x(t)]$ in (i) to a narrower distribution $P^{neq}[x(t)]$ shown in (ii) and thus increases order. \textbf{(b)} (i,ii) Log-normal distributions $P^{eq}(T_{\rm{rep}})$ with a sufficiently large variance $\sigma$ show increased replication speed when reset at a specific timescale (blue region). (iii) The extra order $\Delta S = D_{KL}(\mathcal{P}^{neq}[x(t)] \vert \vert \mathcal{P}^{eq}[x(t)])$ due to the resetting demon increases with variance $\sigma$ of replication times. Here we compute the order achieved with the optimal demon, $\Delta S^*$, see Eq.~\ref{eq:resetoptimal} in \textit{Materials and Methods} for how the optimal demon is found. \textbf{(c)} (i) $P^{eq}(T_{\rm{rep}})$ with two delta functions corresponding to fast ($\tau_f$) and slow paths ($\tau_s$); slow paths are taken with probability $\mu$. 
(ii,iii) Resetting demons are favored for sufficiently large $\tau_s/\tau_f$ and sufficiently small $\mu$. The solid line in (iii) is the boundary defined by Eq. \ref{eqn:favorableDemon}. }
\label{fig:ResetDemon}
\end{figure*}

\section*{Navigating high-dimensional disordered spaces}

We abstract our results on kinetic proofreading and self-assembly to a more general framework. Consider a cell that divides upon completing $N$ actions out of a library of $M$ possible actions, in any order and possibly with repetition. As shown in Fig. \ref{fig:ResetDemon}a-i, such a cell can be seen as navigating one of many possible trajectories $x(t)$ in a high-dimensional configuration space to go from a Birth state to a mature Division state that can then divide into two cells. These actions could represent the order in which different nucleotides are added during templated replication or molecular components are added to, e.g. a growing viral capsid structure \cite{perlmutter_mechanisms_2015} or different metabolic and synthesis processes that must be completed during a cell cycle \cite{phillips_molecular_2020}.

Equilibrium thermodynamics generally sets a limit on how strictly one can enforce an order of these actions \cite{phillips_molecular_2020}, resulting in a relatively wide probability distribution $\mathcal{P}^{eq}[x(t)]$ over trajectories $x(t)$.  Extant cells exploit non-equilibrium mechanisms to sample from a narrower distribution $\mathcal{P}^{neq}[x(t)]$; e.g. typically, only a specific subset of all possible $M$ actions is taken and, in a precise order, to avoid deleterious consequences. This non-equilibrium order can be quantified by the reduction in the entropy of trajectories $\Delta S = D_{KL}(\mathcal{P}^{neq}[x(t)] \vert \vert \mathcal{P}^{eq}[x(t)])$. 

Can mechanisms enforcing such non-equilibrium order potentially arise from the need to replicate fast alone? Inspired by the \textit{undoing} action of proofreading and dynamic instability \cite{murugan_speed_2012}, we consider Maxwell demons (backward-facing blue arrows in Fig. \ref{fig:ResetDemon}a-ii) that randomly and irreversibly reset \cite{evans_stochastic_2020,evans_diffusion_2011,gordon_local_2011} the system back to the Birth state on a timescale $T_r$, independent of what state the cell might be at. In contrast with classic Maxwell demons that act based on being in the right or wrong state \cite{phillips_molecular_2020}, our mechanism acts solely on time.

To determine when resetting demons will be selected for by the need for fast self-replication alone\cite{evans_stochastic_2020,pal2022-on,evans_diffusion_2011,fuchs_stochastic_2016,reuveni_role_2014}~---~without selecting for order, we first define $P^{eq}(T_{\rm{rep}})$ to be the distribution of replication times induced by trajectories sampled from the ensemble $P^{eq}[x(T_{\rm{rep}})]$. 
The fitness $f$ of a lineage of cells with replication time distribution $P^{eq}(T_{\rm{rep}})$ can be shown to be given by $2 \tilde{P}(-f) = 1$ where $\tilde{P}(w) = \int_0^\infty e^{w t} P^{eq}(t) dt$ is the moment generating function for $P^{eq}(T_{\rm{rep}})$~---~see \textit{Materials and Methods} and \cite{jafarpour_bridging_2018}. 
We can also use similar means to calculate the modified lineage fitness $f(T_r)$ in the presence of a demon that resets on a timescale $T_r$, see \textit{Materials and Methods}. Setting $f(T_r) > f(\infty)$, we find an approximate condition for when speed of replication alone will select a resetting demon of timescale $T_r$ if
\begin{equation}
    E(T_{\rm{rep}} | \mbox{lost}) > \frac{T_r}{1-P_{lost}(T_r)} + E(T_{\rm{rep}} | \mbox{not lost})
    \label{eqn:favorableDemon}
\end{equation}
where $E(T_{\rm{rep}} | \mbox{lost})= \int_{T_r}^{\infty} t P^{eq}(t)dt/P_{lost}(T_r)$, $P_{lost}(T_r)=\int_{T_r}^{\infty}P^{eq}(t) dt$ and $E(T_{\rm{rep}} |\mbox{not lost})= \int^{T_r}_{0} t P^{eq}(t)dt/(1-P_{lost}(T_r))$. Here, ``lost'' represents slow trajectories $x(t)$ with long completion times $T_{\rm{rep}} > T_r$. Intuitively, Eq. \ref{eqn:favorableDemon} amounts to two conflicting requirements on $P^{eq}(T_{\rm{rep}})$ that must be satisfied for some $T_r$: (i) the average time cost of slow trajectories $E(T_{\rm{rep}} |\mbox{lost})$ must be large; (ii) the probability of taking these slow trajectories $P_{lost}(T_r)$ must be small.

While resets are often seen as a way to speed up search \cite{evans_stochastic_2020}, the key point in our context is that resetting demons create non-equilibrium order by also modifying $P^{eq}[x(t)]$ to $P^{neq}[x(t)]$. To see this, note that in the presence of a demon, the Division state can be reached only by a small subset of all possible trajectories $x(t)$, namely the fast ones. 
Such a reduction in entropy of trajectories is a form of non-equilibrium order. For example, if different trajectories $x(t)$ represent $N$ actions being performed in many different orders, the entropy reduction corresponds to ensuring that those $N$ actions are carried out in a very specific temporal order \cite{phillips_molecular_2020}. 
Crucially, in many other cases, the trajectories through state space are themselves not observable unless the system successfully reach the Division state; only these successful trajectories that are not reset leave behind an observable fossil record in the form of a molecular structure, e.g. a completed DNA transcript or self-assembled structure not dissolved by dynamic instability. In this context, the reduced entropy of successful trajectories will be directly reflected in, say, lower error rates in newly synthesized DNA strands or higher structural order of macromolecular structures.

The reduction in trajectory entropy $\Delta S$ due to  resetting \cite{Eliazar2023-kl} can be bounded in many cases of interest by
\begin{equation}
\Delta S  \leq - \log  \int_0^{T_r} P^{eq}(T_{\rm{rep}}) dT_{\rm{rep}}
\label{eq:order}
\end{equation}
See \textit{Materials and Methods} for details. 
As an example, see Fig. \ref{fig:ResetDemon}b for results for a log-normal distribution $P^{eq}(T_{\rm{rep}})$; resets of a specific timescale $T_r$  \emph{reduce} replication time for sufficiently high variance $\sigma$. In this regime, resets restrict the set of trajectories used to reach the Division state, creating non-equilibrium order $\Delta S$ that increases with the variance $\sigma$; see Fig. \ref{fig:ResetDemon}b-iii.

Real systems can show a distribution of replication times $P^{eq}(T_{\rm{rep}})$ for numerous mechanistic reasons, ranging from kinetic traps in self-assembly, stalling in DNA replication or other frustrated states along some trajectories $x(t)$ but not others \cite{jafarpour_bridging_2018, lin_effects_2017, lahini_nonmonotonic_2017}. We studied several families of replication time distributions $P^{eq}(T_{\rm{rep}})$; we find that some distributions never show order-through-speed while others do so in specific regimes. See Fig.~\ref{fig:ResetDemon}c and Extended Data Fig. \ref{fig:SI_resets}.

\section*{Preferential evolution of dissipative order}
\label{sec:dissipative}

Biological systems often have a choice of evolving one of many order-maintaining mechanisms that differ in their dissipation cost. 

We find that selection for fast replication alone will preferentially select for order through \emph{more} dissipative mechanisms over less dissipative mechanisms when such a choice is available, even when both mechanisms can achieve the same order. 

For templated replicationn, we compared kinetic proofreading to an alternate error correcting mechanism, i.e., near-equilibrium polymerization\cite{sartori_kinetic_2013, bennett_dissipation-error_1979}. 
We find that the more dissipative mechanism, kinetic proofreading, shows higher fitness, i.e. replicates faster, when compared at the same error rate $\mu$. 
We can intuitively understand this effect by computing $\frac{\Delta \mu}{\Delta  t_{\rm{nostall}}}$, i.e., the reduction $\Delta \mu$ in error rates obtained for a given proofreading-associated slow down $\Delta t_{\rm{nostall}}$ of copying time in the absence of stalling. As shown in Extended Data Fig. \ref{fig:Dissipation}, this ratio increases with dissipation.

A similar conclusion holds for self-assembly; see \textit{Supplementary Information} for a comparison of a minimally dissipative error correction through near-melting point assembly and dynamic instability.

The above results suggest that the mere fact of exponential proliferation leads to a selection coefficient $s_{\epsilon}$ to dissipate more if the distribution of replication times is wide enough. We can estimate $s_{\epsilon}$ by adding a dissipation cost $F \to F -\lambda \epsilon$ to the fitness and finding the critical $\lambda_c$ that prevents the spontaneous evolution of non-equilibrium order; see Extended Data Fig. \ref{fig:SI-DNA-dissipation} for examples.

\section*{Discussion}

A basic question in the transition from matter to life \cite{Neumann1966-qz} is to understand the conditions that drive matter to non-equilibrium states with higher structural order \cite{lynch_frailty_2007}. 

Our work outlines one minimal scenario: self-replication can intrinsically amplify dissipative mechanisms that enhance order if the distribution of replication times has large enough variance.  Once such order emerges, it can be functionalized~---~e.g. accurately copied RNA can begin to code for functional ribozymes, thus adding further selection pressure to maintain that order \cite{tjhung_rna_2020,laurent_emergence_2024}, as in the conventional picture. Consequently, the route to error-correcting ribozymes \cite{tjhung_rna_2020} in the origin-of-life context might be easier than expected, building on prior stalling-fidelity work \cite{rajamani_effect_2010,goppel_kinetic_2021}. 
More broadly, our work adds to a line of work on the origin of complexity through ratchets that does not postulate any direct adaptive benefits of complexity \cite{gingrich_dissipation_2016,lenski_evolutionary_2003,jenkins_novo_2010,starr_pervasive_2018,pillai_origin_2020}.

Our work has implications for speed and accuracy of enzymes of the central dogma \cite{bustamante_revisiting_2011} such as the ribosome and DNA and RNA polymerases, to the extent they stall upon misincorporations \cite{Joazeiro2019-wq, Mallory2019-yt, midha_synergy_2023} and informs efforts to engineer enzymes of higher specificity without loss of speed \cite{tawfik_accuracy-rate_2014,flamholz_revisiting_2019}. Our results can explain mutation rates too low to be accounted for by mutational load alone \cite{lynch_evolution_2010}.

In self-assembly, our results suggest that disassembly pathways can provide time-efficient error correction when combined with misincorporation-induced pauses seen in   natural systems such ribosomal assembly checkpoints \cite{sanghai_co-transcriptional_2023} and synthetic systems \cite{winfree_proofreading_2004, doty_optimizing_2018,lenz_geometrical_2017}.  Our results suggest, in a twist, that classic annealing protocols 
\cite{Muller2007-cg} for reducing defects in crystal growth can also increase net growth rate under some conditions. 

Finally, resets have been shown to be a broadly relevant strategy for speeding up search in a broad range of contexts \cite{goppel_kinetic_2021,gordon_local_2011,evans_stochastic_2020,reuveni_role_2014,pal2022-on,britain_progressive_2022}. Our work points out that in addition to saving time, reset mechanisms effectively reduce the entropy of paths used to reach a destination state. Such `canalization' into a few paths can be seen as a non-equilibrium version of Waddington's homeorhesis \cite{waddington_canalization_1942
}. 
The reduction in trajectory entropy can show up as higher observable order in, e.g. assembled structures or copied polymers. As a consequence, complex systems can achieve stereotyped reproducible behaviors, despite living in high-dimensional disordered state spaces, through simple non-equilibrium mechanisms that also provide speed benefits. 

\section*{Acknowledgments.}
We thank John Sutherland, Irene Chen, David Huse, Michael Rust,  Erik Winfree,  Chang Liu and members of the Murugan group and the CZI theory group for discussions. AM and JWS received support from the Sloan (G-2022-19518) and Moore (11479) foundations, Matter-to-Life program. This work was supported by the NSF Center for Living Systems (grant no. 2317138). AM acknowledges support from NSF PHY-2310781 and NIGMS of the NIH under award number R35GM151211. JWS acknowledges support from NSF grant (2104708). JWS is an Investigator of the Howard Hughes Medical Institute. RP acknowledges the NIH MIRA 1R35 GM118043-01. CGE acknowledges support from European Research Council grant no. 772766, Science Foundation Ireland grant nos. 18/ERCS/5746 and 20/FFP-P/8843, and the Evans Foundation for Molecular Medicine.  RR and KH acknowledge support from the Yen fellowship.

\renewcommand{\thefigure}{E\arabic{figure}}
\renewcommand{\thetable}{E\arabic{table}}
\renewcommand{\theequation}{E\arabic{equation}}
\setcounter{figure}{0}
\setcounter{table}{0}
\setcounter{equation}{0}

\subsection*{Materials and Methods}

 \subsection*{Kinetic proofreading and stalling}
 \subsubsection*{Models of proofreading}
Following prior work, we model enzyme-substrate transformations relevant for kinetic proofreading using a Markov chain. See Extended Fig. \ref{fig:SI_proofreadingth}. The initial state $E$ ($i=0$) corresponds to the enzyme without any ligand loaded; the final state $EX$ ($i=N-1$)~---~with $X$ either the right or wrong nucleotide loaded~---~corresponds to steps in the polymerization process, e.g. the catalysis of the phosphodiester bond between the current nucleotide and the previous one. From state $i=N-1$, the enzyme irreversibly enters the product formation state, $P_X$, where the enzyme $E$ is released, and product $X$ is formed after having undergone the proofreading process at previous steps. We assume that product formation is characterized by a slow rate compared to other rates in the network. 

The probability of being in state $i$, $p_{i}$, as function of time obeys the Master equation with $N$ states
\begin{align}
    \frac{dp_{i}}{dt} = - \sum_{j\neq i} k_{ji} p_i + \sum_{j\neq i} k_{ij} p_j
\label{eq:markov}
\end{align}
where $k_{ij}$ is the rate of transition from $j \to i$ and $\sum_{i=0}^{N-1}p_i=1$. Consistent with prior work, we model the kinetics of $R$ and $W$ as differing only in their binding energies, with $EW$ being lower than the ones for $ER$ by a factor of $\Delta$. This translates into higher off-rates for some $W$ reactions than the corresponding rates for $R$ reactions by a factor $e^{\Delta}$.

After processing a nucleotide at site $s$ along the chain, the polymerase proceeds irreversibly to the next site $s+1$. The processing of a new ligand at that site is again modeled by a Master equation of the type above. In the limit of slow product formation from the complexes $EX$ (state $i=N-1$) the error rate, i.e. number of mutations in the growing strand, is approximated by the ratio of steady-state probabilities of Eq.~\ref{eq:markov}, 
$\mu \approx \frac{p_{N-1}^W/p_0^W}{p_{N-1}^R/p_0^R}\, .$

The average time to copy one base~---~in the absence of stalling, see below~---~is determined by computing the mean first passage time $t_{\rm{nostall}}$ from the unbound state $E$ ($i=0$) to the absorbing state defined by the product state $P_R$ ($i=N)$. This is computed explicitly to define $t_{\rm{nostall}}$ in the in-silico evolution, Fig. \ref{fig:proofreading_unified}b. In the limit of slow product formation, the first passage time is approximated by the inverse of the flux to the same state $P_R$, $t_{\rm{nostall}} \propto 1/p_{N-1}^R$, i.e. from state $ER$. We use this approximation when computing the times in Fig. \ref{fig:proofreading_unified}a. 

The rate of entropy production $ \epsilon$ is computed as defined in \cite{schnakenberg_network_1976}. 

See \textit{Supplementary Text} for explicit equations.

\renewcommand{\thepage}{E\arabic{page}}
\setcounter{page}{1} 

\subsubsection*{Model of stalling}

\begin{figure*}[h!!!]
    \centering
    \includegraphics[scale=0.8]{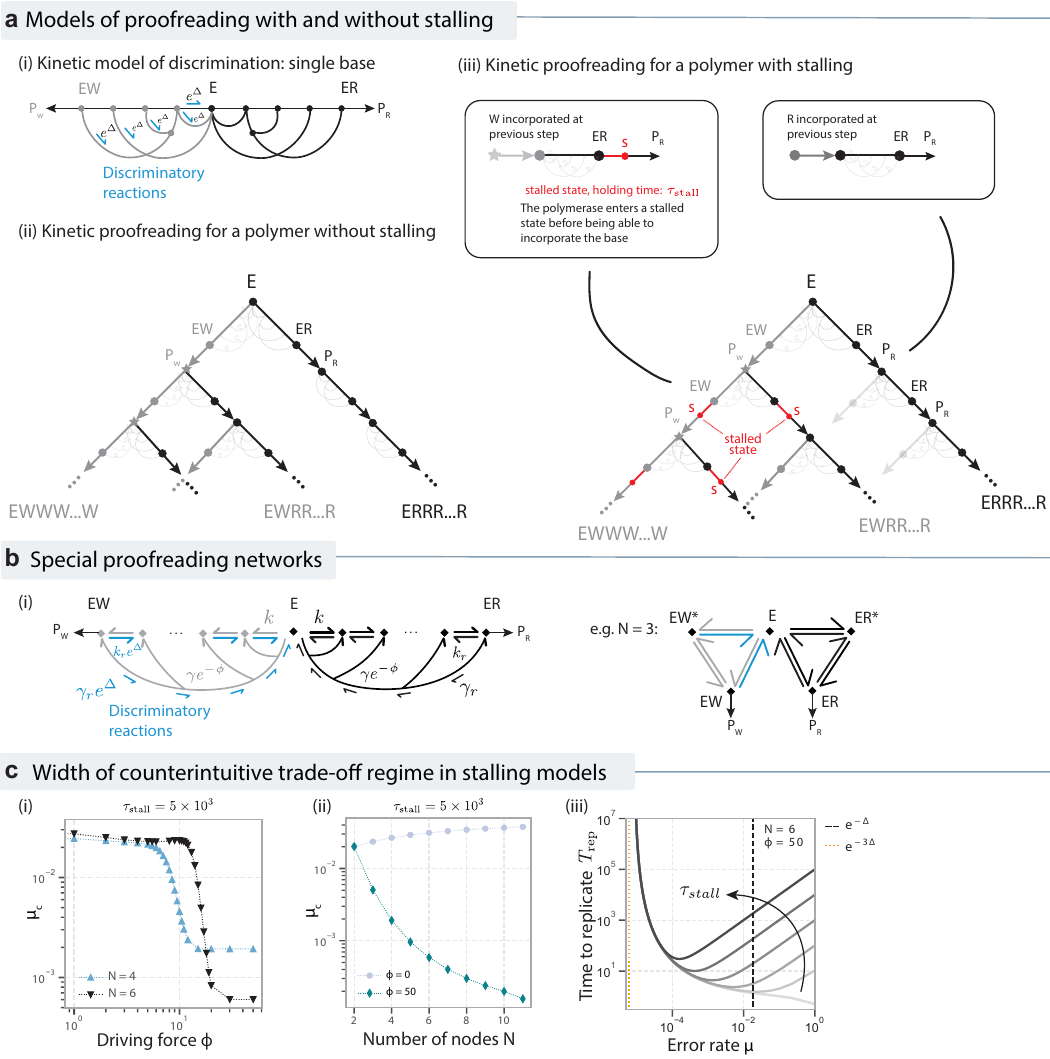}
    \caption{\textbf{Extending simple proofreading models with the stalling effect.} \footnotesize(\textbf{a}) (i) A kinetic network model of proofreading at the single base level; $E$ is the enzyme (e.g., DNA polymerase loaded with a DNA template),  $P_R, P_W$ represent product formation with the right ($R$) and wrong ligands (e.g. nucleotides) respectively. (ii) Any single base model can be chained together as shown to get a model of polymers as shown; each path corresponds to one particular outcome of replication, e.g. all right nucleotide incorporated in $ERRR\cdots R$. (iii) We introduce stalling in the polymer model shown in (ii) by adding an additional state $s$ with holding time $ \tau_{\rm{stall}}$ if the \emph{prior} incorporated base in the tree is wrong (`W'). (\textbf{b}) Specific examples of the single base kinetic proofreading model used to produce the curves of Fig. \ref{fig:proofreading_unified}a in the main paper. A driving force $\phi$ is coupled to the reactions $\gamma$ to control the contribution to the chemical potential of the non-equilibrium discriminatory reactions. For simplicity, the network is shown for $N=3$, representing the typical kinetic proofreading network \cite{hopfield_kinetic_1974, ninio_kinetic_1975}. (\textbf{c}) With stalling, a counterintuitive tradeoff is seen for mutation rates $\mu > \mu_c$. The crossover error rate $\mu_c$ is shown as function of the driving force $\phi$ (i) and the number of nodes $N$ with a value of stalling time $\tau_{\rm{stall}}=5\times 10^3$. (iii) The time to replicate Eq. \ref{eq:time_1} is shown as function of error rate for increasing values of the stalling time $\tau_{\rm{stall}}$ for a network of size $N=6$ and driving force $\phi=50$; the size of the counterintuitive tradeoff region $\mu > \mu_c$ increases with $\tau_{stall}$. See \textit{Supplementary Information} for more details of the model.}\label{fig:SI_proofreadingth}
\end{figure*}

Experiments \cite{perrino_differential_1989,mendelman_base_1990,huang_extension_1992,johnson_structures_2004,ichida_high_2005,baranovskiy_structural_2022,midha_synergy_2023} show that each error slows down the addition of the next base by a factor of $10^1-10^6$, depending on the system and the type of mismatch. We capture this effect in our proofreading model by introducing, upon addition of a mismatch, a time delay before proceeding to the next base~---~see \textit{Supplementary Text} for a discussion of alternative models of stalling and proofreading. We represent this time delay as an extra state in the Markov system Eq.~\ref{eq:markov} after the addition of a mismatch, with characteristic holding time $\tau_{\rm{stall}}$; see Extended Fig. \ref{fig:SI_proofreadingth}.

With stalling, the average time to copy a strand of length $L$ is
\begin{equation}
\langle T \rangle \equiv \sum_{m=0}^L P(m) T(m) \approx L t_{\rm{nostall}}(\mu) + L \mu  \tau_{\rm{stall}},     
\end{equation}
where the approximation holds for low error rates $\mu$, $P(m) = {L\choose m}\mu^m(1-\mu)^{L-m}$ is the probability that $m$ errors occur in the strand, and $T(m)$ is the time to form a product with $m$ errors, see \textit{Supplementary Text} for details.

\subsubsection*{In silico evolution}
We consider a fully connected Markov network with $N$ possible states described by a Master equation as Eq. \ref{eq:markov} above. The kinetic rates depend on the energy of state $i$, $V_i$, and the energy barrier between state $i$ and $j$, $B_{ij}$, thus the model is defined by parameterizing the energy landscape $k_{ji} = e^{V_i - B_{ji}-\phi_{ij}}, k_{ij} = e^{V_j - B_{ij}}$ with $B_{ij} = B_{ji}$. While the network is initialized at equilibrium $\phi=0$, i.e. preserving detailed balance, the parameters $\{V, B, \phi\}$ are allowed to change according to a Metropolis rule optimizing a specified fitness function (defined below) and can thus evolve to go out of equilibrium. The fitness function is chosen to be the speed of replicating a strand of length $L$, i.e. Eq. \ref{eq:time_1}.

For the results in Fig. \ref{fig:proofreading_unified}b, we run the in silico evolution for $N=7$ for typically $10^4$ Monte Carlo steps with an evolution temperature $T_e=10^{-2}$, discrimination factor $\Delta=4$ and stalling time $\tau_{\rm{stall}}=10^3$. Ten replicates are run. See \textit{Supplementary Text} for more details.

\subsection*{Speed-accuracy data for hundreds of DNA polymerases}
 
The wild-type DNA polymerase (TP-DNAP1) in Fig. \ref{fig:data} is a homolog of the DNA polymerase of $\phi29$ and replicates a linear, multi-copy DNA plasmid (pGKL1 p1 for brevity) in the cytosol of yeast, orthogonal to the nuclear replication machinery. p1 is originally part of the killer toxin-antitoxin system of \textit{Kluyveromyces lactis}; in its native context, the DNA polymerase is encoded on the plasmid itself.

The OrthoRep platform \cite{ravikumar_scalable_2018} modifies this system by (i) moving TP-DNAP1 (or mutant variants thereof) to a CEN/ARS plasmid localized in the nucleus, and (ii) introducing a recombinant p1 with an auxotrophic marker on it, among other genes. The replication of recombinant p1 is therefore dependent on the activity of the CEN/ARS-encoded DNA polymerase; however, we caution the reader that some of the data in \cite{ravikumar_scalable_2018} may also include a contribution from the wildtype DNA polymerase encoded on unmodified p1.

We take the copy number of the p1 plasmid to be a proxy for the overall activity of the DNA polymerase. The assay to measure plasmid copy number in \cite{ravikumar_scalable_2018} involves a calibration curve relating qPCR measurements to fluorescence levels of a marker encoded in the plasmid. The error rate of the polymerase is measured with a Luria-Delbr{\"u}ck assay as the rate of reversion of a stop codon in an auxotrophic marker gene (leu2) encoded on the p1 plasmid \cite{ravikumar_scalable_2018}.

\begin{figure*}[h!!!]
    \centering
    \includegraphics[scale=0.8]{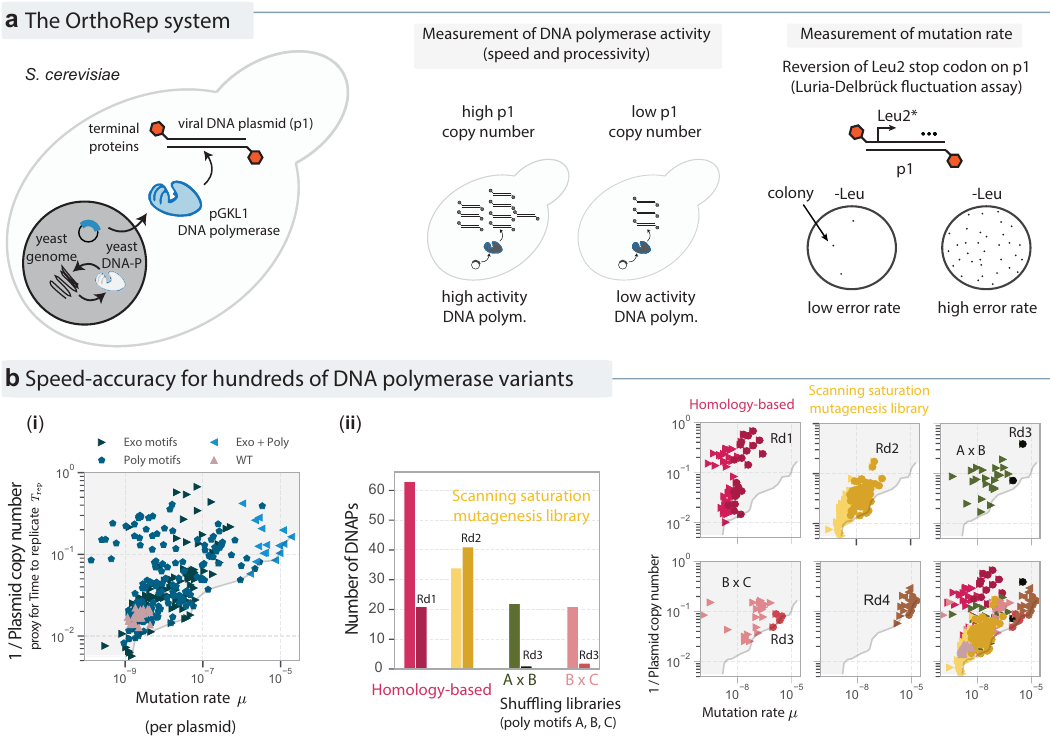}
    \caption{\textbf{The largest dataset to date on speed and accuracy of DNA polymerase variants is provided by the OrthoRep platform.} \footnotesize(\textbf{a}) Sketch of the OrthoRep system \cite{ravikumar_orthogonal_2014,ravikumar_scalable_2018,rix_scalable_2020} in \textit{S. cerevisiae}, where a  cytosolic, multicopy, linear DNA plasmid (pGKL1, p1 for brevity) of viral origin is shown. Plasmid p1 can only be copied by a dedicated DNA polymerase (pGKL1 DNAp, in blue) and not by the host yeast machinery in the nucleus. The orthogonality of the replication system of p1 and the yeast genome allows for measurement of activity and accuracy of the pGKL1 polymerase, independent of yeast DNA replication machinery. Activity is measured by determining the copy number of p1 at steady state.  Accuracy is measured through a Luria-Delbr{\"u}ck fluctuation assay, based on the reversion of a stop codon in Leu2 gene placed on p1 in a Leu-auxotroph strain. (\textbf{b}) (i) Data plotted from Table S2 of \cite{ravikumar_scalable_2018} with all the DNA polymerases for which p1 copy number and mutation rates have been characterized. Mutations in  different domains is shown using color and shape reported in the legend. (ii) Subdivision of the same data as (i), but according to different libraries constructed in \cite{ravikumar_scalable_2018}: based on mutations in known homologs such as $\phi$29, scanning saturation mutagenesis, shuffling, with the definition of the different rounds used for the discovery of highly-error-prone DNA polymerase variants with acceptable activity. }
    \label{fig:SI_chang}
\end{figure*}

All-in-all, p1 replication activity and substitution mutation rate have been measured for 213 unique mutant DNA polymerases, as reported in Table S2 from \cite{ravikumar_scalable_2018}. We plotted and repurposed this data to investigate the relationship between speed and accuracy in Fig. \ref{fig:data}. See Extended Fig. \ref{fig:SI_chang} and \textit{Supplementary Information} for more details.

 \subsection*{Self-assembly}

 \subsubsection*{Tile sets and errors}

The self-assembly system is defined by the tile set. Fig. \ref{fig:selfassembly} shows a simplified tile set, for the simulations we consider 12 tiles organizing in two columns $1-6$ and $7-12$, which then repeat until a given size is reached. Non-specific interactions allow it to form erroneous rows as disordered rows of tiles $1-6$. One such configuration can grow further only if monomers attach by single bonds, i.e. tiles are partially-connected. Such a growth process is slow and effectively represents stalling. 

\subsubsection*{Kinetic simulations}

\begin{figure*}
    \centering
    \includegraphics[width=\textwidth]{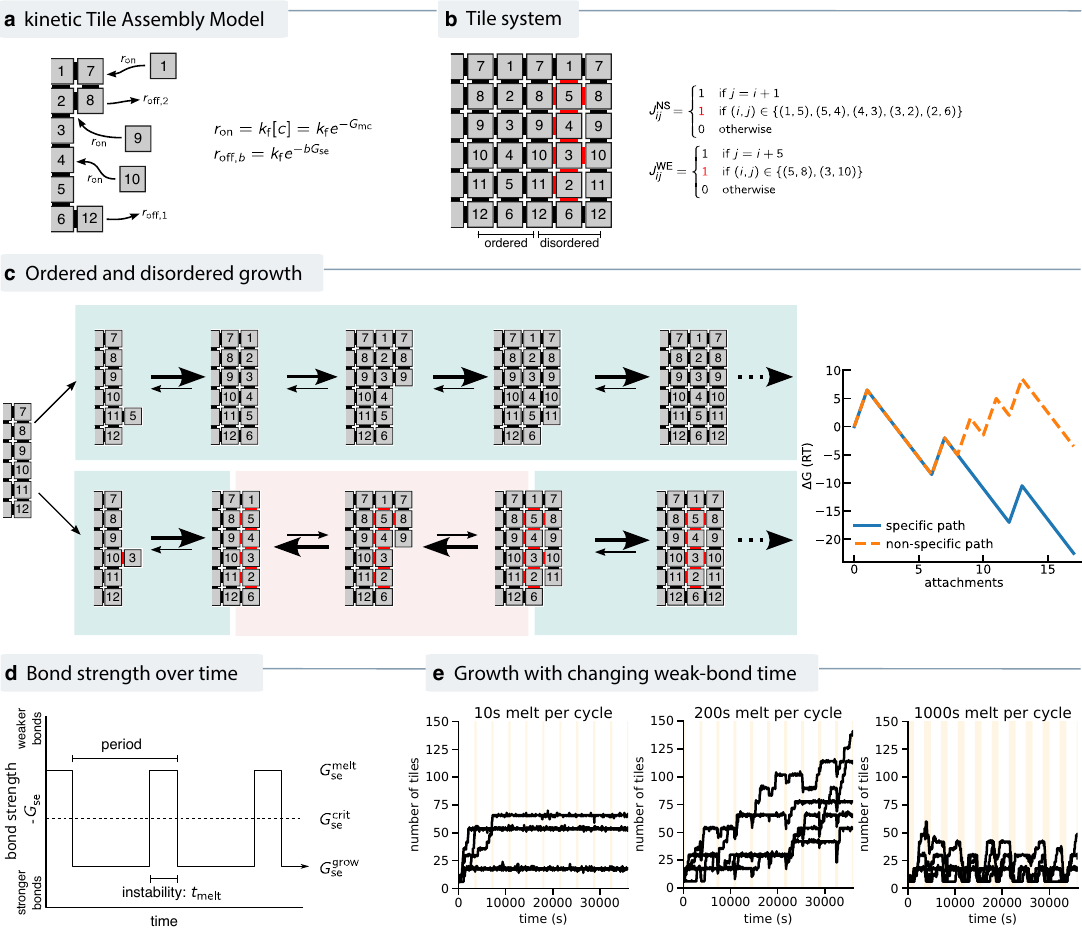}
    \caption{\textbf{Kinetic model of self-assembly and dynamic instability} \footnotesize\figp{a} We use a kinetic Monte Carlo simulation of self-assembly, the kinetic Tile Assembly Model (kTAM) \cite{winfree_simulations_1998,g_evans_physical_2017}, in which tiles attach to locations based on the tile concentration $[c]$ (or equivalently, free energy $\gmc$), and detach based on the total strength of interactions $b$ times a (sign-reversed) free energy $\gse$ (in units of $k_B R$). \figp{b} A ribbon tile system with non-specific interactions exhibiting stalling and errors.  Specific interactions (black) result in fully-connected alternating rows of tiles 1--6 and 7--12.  But non-specific interactions (red) allow a disordered row of tiles 1--6, from which growth cannot continue without incorporating partially-connected tiles. \figp{c} Example pathways for ordered (top) and disordered (bottom) growth.  Both initially grow by primarily favorable, forward-biased steps (green background), but while the ordered pathway can continue favorably, several strongly reverse-biased steps are required for the disordered pathway to continue. \figp{d} We model dynamic instability as periodic pulses of high temperature (equivalently, lower bond strength $\gse$).  The temperature cycles between a high temperature $\gsemelt$ at which the ribbon the melts and a low temperature $\gsegrow$ at which the ribbon can grow, as shown. ($\gsecrit$ separates these melting and growth regimes.) The strength of dynamic instability $\lambda$ is set by the amount of time $\tmelt$ at high temperature relative to the cycle period $\tcycle$.
    \figp{e} Size of five simulated assemblies over time, with $\tcycle = 3600\;\text{s}$, $\gmc=16$, $\gsegrow=9.5$, $\gsemelt=8.4$ for three different $\tmelt = 10s,200s, 1000s$. With a short $\tmelt$, ribbon growth stalls through disordered interactions.  With a longer $\tmelt$, melting cycles detach the growth front, allowing stalled ribbons to grow when bonds become stronger.  With yet longer $\tmelt$, the too much of the ribbon melts each cycle, and thus ribbons do not grow.}
    \label{fig:ext-sa-example}
\end{figure*}

We use a kinetic Monte Carlo simulation, the kinetic Tile Assembly Model (kTAM), in which monomers are modeled as fixed orientation square tiles with four distinct edges.  Tiles attach to empty locations on a square lattice at a rate $r_\text{on} = k_f e^{-\gmc} = k_f c$, where $c$ is the tile monomer concentration, and $\gmc \equiv -\log (c/u_0)$.  Each attached tile detaches at a rate $r_\text{off} = k_f e^{-b \gse}$, where $\gse$ is the unitless, sign-reversed free energy of a strength-one interaction, and $b$ is the total interaction strength between the tile and adjacent tiles in the assembly. See Extended Fig. \ref{fig:ext-sa-example}.

We perform simulations using the Gillespie algorithm, as implemented for the kTAM in rgrow \cite{Gillespie1977-mv,rgrow}.  We assume constant free monomer concentrations, equivalent to the assumption of either small assembly concentrations or a chemostat.

Simulations are run with a stopping condition of a target size of 480 tiles, i.e., 80 rows (or reaching a $10^8$ s time cut-off).  Assembly rate is then defined as the mean of the inverse of the time for simulations to reach the target size, divided by the target size, resulting in a rate in tiles per second. Error rate is calculated considering an error as two tiles in adjacent locations with no bond between them, counting these instances in the first 74 rows of each assembly (to avoid temporary disorder at a growth front), and calculating an error rate per row.

\subsubsection*{Dynamic instability}

\begin{figure*}
    \centering
    \includegraphics[width=\textwidth]{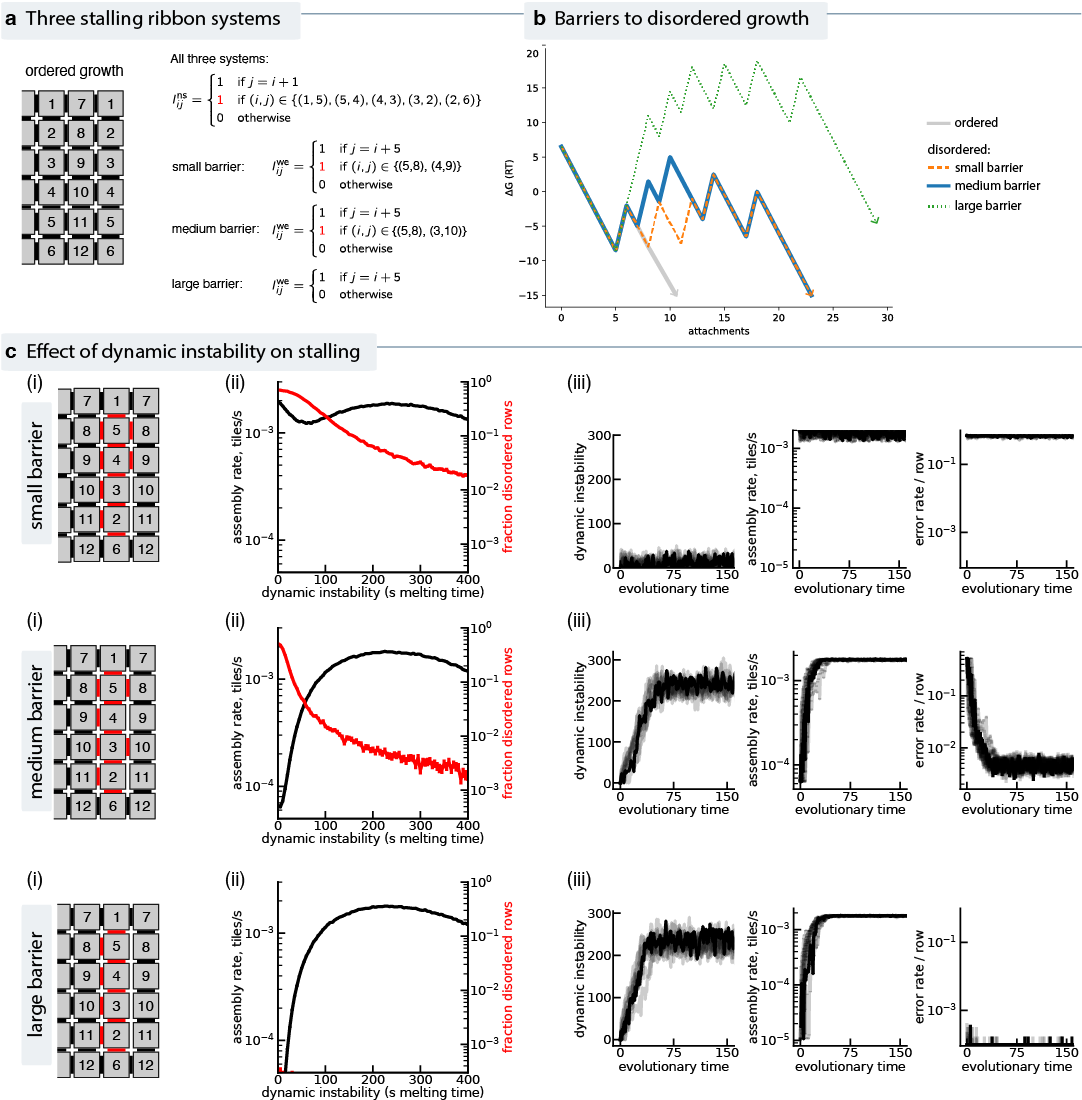}
    \caption{\textbf{Tile sets with too large or too small a stalling barrier do not evolve non-equilibrium order due to a need for speed. } \footnotesize\figp{a} We consider three distinct self-assembling systems that differ in the size of the energy barrier to growth past disorder. All three systems have same set of specific interactions between 12 tiles, and grow as in Fig. \ref{fig:ext-sa-example}, but the three systems differ in their non-specific interactions: four bonds, with three adjacent, in the \textbf{small} barrier system, four bonds spread across the row for the \textbf{medium} barrier system, and two bonds in the \textbf{large} barrier system.  \figp{b} The different non-specific interactions result in different free energy barriers to disordered growth.  \figp{c} (i) Structures and bonds for continued growth after a disordered row in each system.  (ii), as dynamic instability increases (i.e., melting time $t_{melt}$), the small and medium barriers have less disordered growth, and the medium and large barriers have significantly faster growth. (iii), both the medium and large barriers can have melting time $t_{melt}$ evolved to optimize speed, but only in the medium barrier system does this also decrease disordered growth.}
    \label{fig:ext-sa-goldilocks}
\end{figure*}

We model an error correcting mechanism through a dynamic instability by turning down $\gse$, i.e., increasing temperature, on a precise, periodic schedule: for each period of time $t_\text{cycle} = 3600 s$, $\gse$ is set to $8.4$, low enough to melt the growth front of structures, for a time $t_\text{melt}$, then is set to $9.5$, sufficient for structures to grow, for the the remaining cycle time.  The strength of dynamic instability is defined by $\lambda = t_\text{melt}$.  Evolution of $\lambda$ was performed by a hill-climbing algorithm, starting from $\lambda = 0$, and accepting changes to $\lambda$ if the assembly rate, on 768 simulations, increased.  See Extended Fig. \ref{fig:ext-sa-goldilocks}.

\subsection*{Navigating high-dimensional spaces}

We study an abstract model of replication in which  a cell starts at a \textit{Birth} state and must reach a final \textit{Division} state before dividing. These states are connected by a large number of distinct trajectories through the high-dimensional state space of a cell. 

A population of such cells eventually grows exponentially with growth-rate $f$. Denoting by $P(t)$ the distribution of replication times (i.e. the time taken to transit from \textit{Birth} to \textit{Division}), $f$ is given by the solution of the Euler-Lotka-Powell equation,
\begin{equation}\label{eq:eulerlotka}
\int_0^{\infty} dt \, e^{-f \, t} P(t) = \frac{1}{2},
\end{equation}
see, e.g. \cite{jafarpour_bridging_2018} for a derivation and generalisation. The integral on the left hand side is compactly written as $\tilde{P}(-f)$, where $\tilde{P}$ denotes the moment-generating function of $P$: $\tilde{P}(x) \equiv \int dt \, e^{x t} P(t)$.

\subsubsection*{When do resets reduce the time to replicate?}

We now consider a demon that `resets' the cell back to the \textit{Birth} state if it has not divided after time $T_r$. In the absence of the demon, let use denote the distribution of replication times by $P^{eq}(t)$. The action of the demon modifies the distribution of replication times to $P^{neq}(t)$. In the \textit{Supplementary Information}, we derive its moment generating function, finding 
\begin{equation}\label{eq:resetPnoneq}
\tilde{P}^{neq}(w)  = \frac{a_w}{1 - (1-a) e^{w T_r}},
\end{equation}
where $a_w(T_r) = \int_0^{T_r} e^{w t} P^{eq}(t) dt$. Notice that $a$ has the natural interpretation as the probability that the cell reaches the \textit{Division} state without being reset; in the main text, we refer to $1-a$ as $p_{\rm{lost}}$. For brevity, we will suppress the dependence of $a_w$ and $p_{\rm{lost}}$ on $T_r$.

Combining Eq.~\ref{eq:resetPnoneq} with Eq.~\ref{eq:eulerlotka}, the growth rate $f$ of the population in the presence of resets satisfies
\begin{equation} \label{eq:fitnesswithresets}
{2 \int_0^{T_r}dt \, e^{- ft} P^{eq}(t) = 1 - (1-a) e^{- f T_r}}\, . 
\end{equation}
We seek the optimal reset time that maximises the growth rate by differentiating Eq.~\ref{eq:fitnesswithresets} w.r.t. $T_r$. This gives a relationship between the optimal reset time, $T_r^{\ast}$, and the maximum growth rate $f^{\ast}$,
\begin{equation}\label{eq:resetoptimal}
f^{\ast} = \frac{P^{eq}(T_r^{\ast})}{p_{\rm{lost}}(T_r^{\ast})}.    
\end{equation}
Operationally, we numerically solve Eqs.~\ref{eq:fitnesswithresets} and \ref{eq:resetoptimal} simultaneously to obtain $f^{\ast}$ and $T_r^{\ast}$ for specified distributions $P^{eq}(t)$; see the section on \textit{Example distributions} for details.

To gain qualitative insight into when resets are favoured by natural selection, we compute the mean time to replicate, $\tau_{\rm{rep}}$, in the presence of resetting. Differentiating Eq.~\ref{eq:resetPnoneq} w.r.t $w$ and setting $w$ to $0$, we obtain
\begin{equation}
\tau_{\rm{rep}} = \frac{1}{1 - p_{\rm{lost}}} \left( 
\int_0^{T_r} dt \, t P^{eq}(t) + p_{\rm{lost}} \, T_r  \right)
\end{equation}
where we remind the reader that $p_{\rm{lost}}$ depends on $T_r$.

A demon that resets at time $T_r$ is beneficial only if it decreases the average time to replicate; i.e. $\tau_{\rm{rep}}(T_r) < \tau_{\rm{rep}}(T_r = \infty)$. This leads to Eq. \ref{eqn:favorableDemon}~---~details are reported in the \textit{Supplementary Information}.

\subsubsection*{Resets increase order, reduce entropy}
We define a path as the sequence of states, $\boldsymbol{x}\in\{ABDEFG, ACYTGS, \dots \}$, taken by the cell from \textit{Birth} to \textit{Division}. This is a stochastic quantity, and we denote by $P^{eq}(x)$ ($P^{neq}(x)$) its distribution without (with) resets. Note that, in the presence of resets, the path does not include the `futile cycling' that occurs between resets -- only the final, successful path taken to \textit{Division}. 

To quantify the reduction in entropy of paths, we take the Kullback-Leibler (KL) divergence between $P^{eq}(x)$ and $P^{neq}(x)$
\begin{equation} \label{eq:kldefn}
\Delta S \equiv D_{\rm{KL}} \left( P^{neq}(x) \vert\vert P^{eq}(x) \right).
\end{equation}
To bound $\Delta S$, consider the joint distributions of paths $x$ and replication times $t$ in the absence of resets: $P^{eq}(x,t)$. In the presence of the resetting demon, no pair $(x,t)$ with $t > T_{r}$ is possible; consequently, we can write the joint distribution in the presence of resets as $P^{neq}(x,t) \propto P^{eq}(x,t) \, \left[ 1 - \Theta(t-T_{r}) \right]$. 

Normalising, and then integrating out $t$, we obtain an explicit expression for $P^{neq}(x,t)$ that we use in Eq.~\ref{eq:kldefn} to obtain
\begin{eqnarray}
    \Delta S =&& - \log q_0(T_r) \nonumber \\ && +  \sum_x P^{neq}(x) \, \ln \int_0^{T_{\rm{r}}} dt \, P^{eq}(t \vert x).
\end{eqnarray}
As the second term is either negative or zero, we obtain the advertised bound Eq. \ref{eq:order}; see \textit{Supplementary Information} for details.

\subsubsection*{Example distributions}

The fold change replication speed reported in Fig. \ref{fig:ResetDemon}b-c is defined as $f(T_r)/f_{eq}$, where $f_{eq}$ is the growth-rate without the resetting demon (Eq.~\ref{eq:eulerlotka}). $f(T_r)$ is computed numerically from Eq. ~\ref{eq:fitnesswithresets}, with $P^{eq}(t)$ taken to be a log-normal distribution (Fig. \ref{fig:ResetDemon}b), or a `double-delta' distribuition, $P^{eq}(t) = (1-\mu) \delta(t - \tau_f) + \mu \delta(t - \tau_s)$ with $\tau_f$ and $\tau_s$ a fast and slow time-scale, respectively (Fig. \ref{fig:ResetDemon}c). 

The order with optimal demon is computed numerically by using Eq.~\ref{eq:order} with $T_r=T_r^{\ast}$, the optimal reset time computed by simultaneously solving Eqs.~\ref{eq:fitnesswithresets} and \ref{eq:resetoptimal}.

The relative fold-change in replication speed is defined as $f(T_r)/f_{eq}-1$ and computed as described above. Eq. \ref{eqn:favorableDemon} reduces to $\frac{\tau_s}{\tau_f} > 1 + \frac{1}{1-\mu}$ for the double delta and approximates the boundary between resets being beneficial or useless to fitness. It is computed numerically and shown in Fig. \ref{fig:ResetDemon}(c)iii: it qualitatively matches the boundary region of the relative fold-change in replication speed.

\begin{figure*}[h!!!]
    \centering
    \includegraphics[scale=0.9]{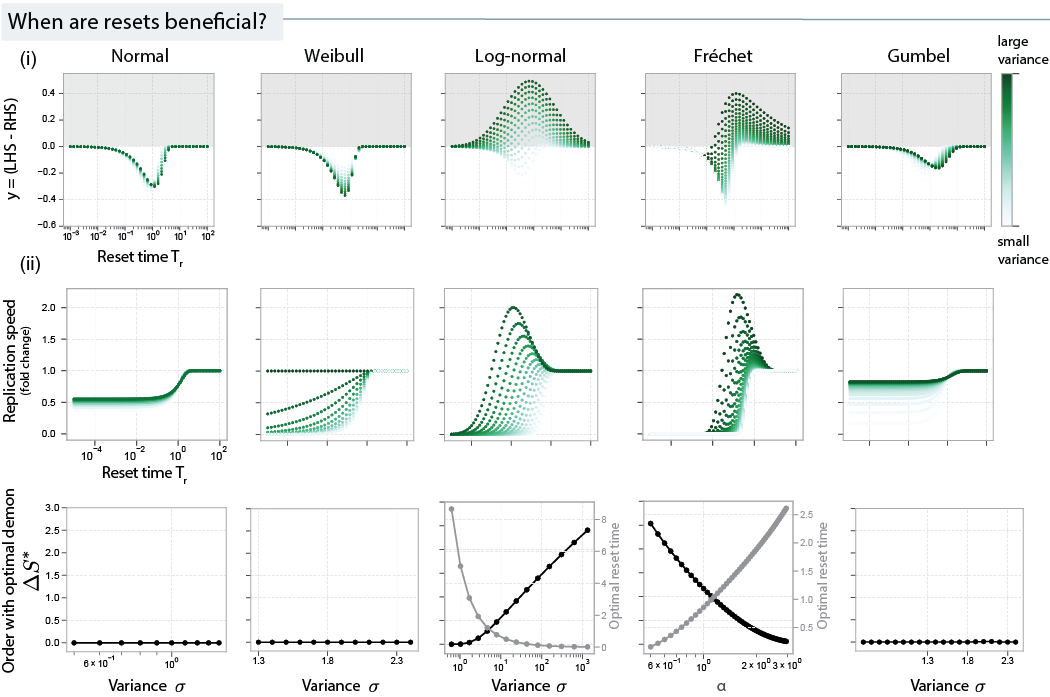} \caption{\textbf{Resets are beneficial only for wide distributions of replication times.}\footnotesize (i) The residual $y=(\rm{LHS - RHS})$ of the inequality in Eq. \ref{eqn:favorableDemon} is plotted as function of reset time $T_r$ and for different values of variance~---~see colorbar: when $y>0$, resets are beneficial, i.e. increase replication speed. We find that $y>0$ for some choices of $T_r$ for the Log-normal and Fréchet distribution. (ii) The fold-change in replication speed (Eq.~\ref{eq:fitnesswithresets}) is shown for different choices of variance parameters and for different probability distributions of replication times: Normal, Weibull, Log-normal, Fr\'echet and Gumbel. The order $\Delta S^*$ with the optimal demon is computed as in  Eq.~\ref{eq:order} imposing Eq.~\ref{eq:resetoptimal}, which sets the constraint for an optimal demon $(f^*,T_r^*)$. See the \textit{Materials and Methods} and \textit{Supplementary Information} for the calculation of how order $\Delta S$ is defined from the distribution of times, leading to Eq.~\ref{eq:order}, which shows that the order of trajectories is bound by the order of times. In the case where resets provide a fitness benefit, for Log-normal and Fr\'echet, the optimal reset time is shown in the plots (grey, right y-axis). We fix the mean of all distributions to unity, but for the plots with the Fr\'echet distribution in (ii), see the \textit{Supplementary Information} for more details.  } 
    \label{fig:SI_resets}
\end{figure*}

Results for other distributions are shown and discussed in the \textit{Supplementary Information} and in Extended Fig. \ref{fig:SI_resets}.
\subsection*{Dissipation}

\subsubsection*{Spontaneous pressure to dissipate}

A given change in error rate $\delta \mu$ results in a change in fitness
\begin{equation}\label{eq:dfdmu}
\frac{\Delta F}{\Delta \mu} = -\frac{1}{T_{\rm{rep}^2}} \left( L \frac{\Delta t_{\rm{nostall}}(\mu) }{\Delta \mu}(\phi) + L \tau_{\rm{stall}} \right) 
\end{equation} 

As seen in Extended Fig. \ref{fig:Dissipation}c, the slope $\frac{\Delta t_{\rm{nostall}}(\mu) }{\Delta \mu}(\phi)$ is negative and increases as the non-equilibrium drive $\phi$ in the proofreading network is increased. Hence, the fitness increase, $\Delta F$, due to a given reduction, $\Delta \mu$, is higher if that reduction was achieved by a more dissipative proofreading mechanism, rather than less dissipative mechanism. Intuitively, dissipative mechanisms achieve the same reduction $\Delta \mu$ using a smaller increase in the time cost $\Delta t_{\rm{nostall}}(\mu)$ error correction. These results suggest that fast self-replication can produce a spontaneous selection pressure $s_\epsilon$ on living organisms to dissipate more energy rather than less.

\subsubsection*{Fitness cost of dissipation}

\begin{figure*}[h]
    \centering
    \includegraphics[scale=0.8]{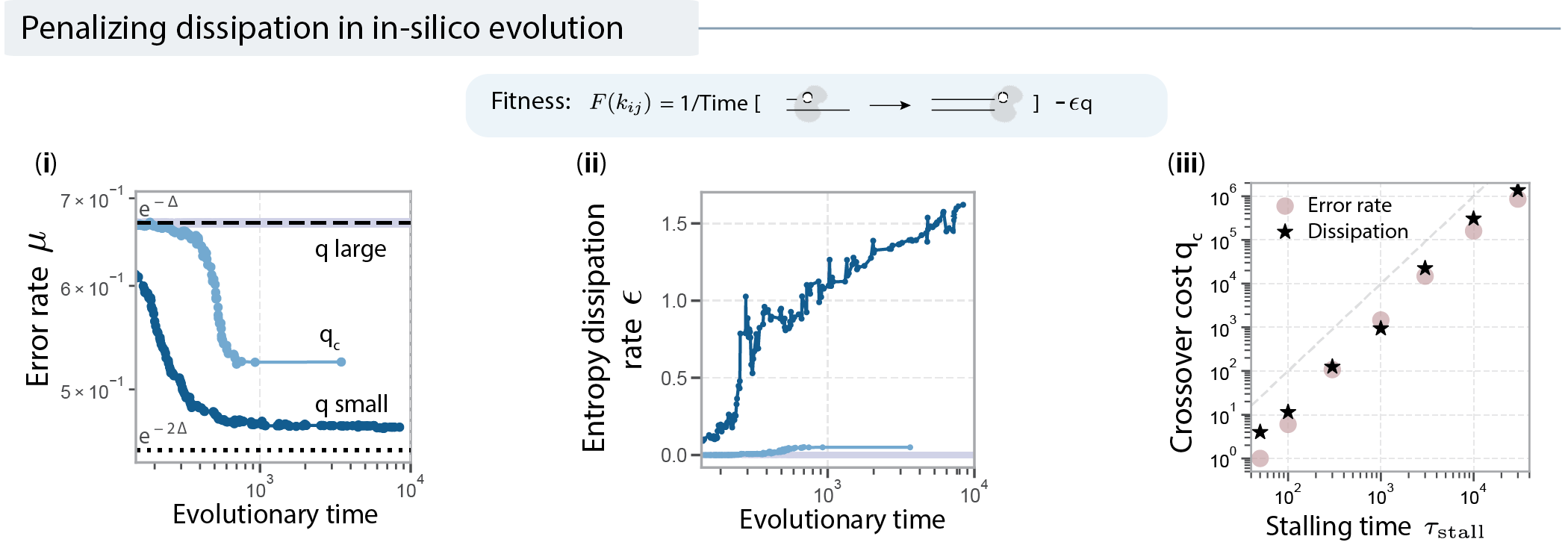}
    \caption{\textbf{In silico evolution with fitness penalty for dissipation} \footnotesize To study the effect of a fitness penalty for dissipation we introduce a cost proportional to dissipation,  $q\epsilon$, in the fitness function, Eq.~\ref{eq:time_1}, introducing a pressure between time minimization and dissipation. The in-silico evolution is performed for a proofreading network of size $N=3$ and $\tau_{\rm{stall}}=10^3$ as in Fig.~\ref{fig:proofreading_unified}, but with the new fitness function including the fitness cost $q\epsilon$, in addition to minimizing time. The effect of such a fitness cost of dissipation is explored for (i) error rate $\mu$, (ii) entropy dissipation rate $\epsilon$, which are shown as function of the evolutionary time. Beyond some critical cost of dissipation $q_c$, in silico evolution no longer evolves proofreading mechanisms, i.e. low $\mu$ or high $\epsilon$. 
    (iii) The crossover cost $q_c$ is identified numerically as the value of $q$ for which error rate no longer decreases or at which dissipation no longer increases. We measure $q_c$ as function of the stalling time, $\tau_{\rm{stall}}$, and show that it increases with stalling time; the dashed line is $q_c\sim\tau_{\rm{stall}}^2$. This crossover cost $q_c$ can be interpreted as the spontaneous fitness pressure to dissipate due to fast self-replication. See \textit{Supplementary Information} for further discussion and details.}
    \label{fig:SI-DNA-dissipation}
\end{figure*}

We can quantify this spontaneous selection pressure to dissipate by introducing a fitness penalty for dissipation as shown in Extended Fig. \ref{fig:SI-DNA-dissipation}. We assume that the fitness function includes a cost proportional to the dissipation rate $\epsilon$, $F(q)=1/T_{\rm{rep}}-q\epsilon\ $. We carried out \emph{in silico} evolution of proofreading networks as earlier, but with this modified fitness function for a network of size $N=3$. The crossover value $q_c$ can be estimated by screening across different values of $q$, monitoring changes in error rate and entropy dissipation rate, which give two independently measured estimates of $q_c$ reported in panel (iii) of Extended Fig. \ref{fig:SI-DNA-dissipation}. Note that $\frac{dF(q)}{d\epsilon} = \frac{dT_{\rm{rep}^{-1}}}{d\epsilon} - q$; at $q_c$, $\frac{dF(q)}{d\epsilon} = 0$, which gives $q_c \sim \frac{dT_{\rm{rep}^{-1}}}{d\epsilon}$, hence $q_c$ is a measure of the spontaneous pressure to dissipate

\subsubsection*{Comparing reversible and irreversible models of proofreading}

\begin{figure*}[h]
    \centering
    \includegraphics[scale=0.8]{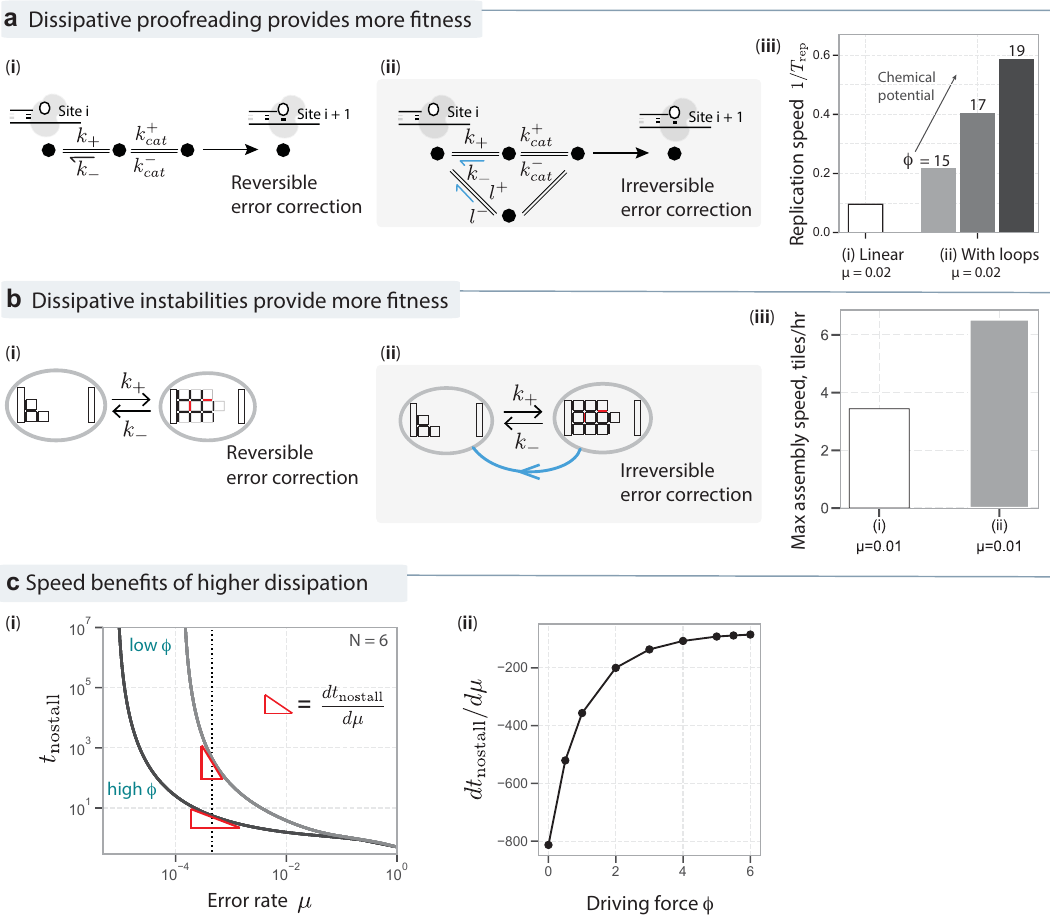}
    \caption{\textbf{Fast replication preferentially selects for more dissipative order-maintaining mechanisms.} \footnotesize(\textbf{a}) We compare polymerases (i) that remove errors by microscopically reversing polymerization pathways and (ii) with exonuclease-based error-correction, i.e. use alternative irreversible pathways. Reversible error correction is achieved in silico with a network of size $N=2$ (linear), while for irreversible error correction we choose a network of $N=6$ (with loops). We fix the stalling time to be $\tau_{\rm{stall}}=5$. Upon selecting for speed, both networks evolve towards lower error rate. However, when compared at the same error rate $\mu = 0.02$, more dissipative mechanisms have higher replication speed. 
     (\textbf{b}) We compare reversible and irreversible error correction in self-assembly: (i) through near-equilibrium self-assembly that exploits the microscopic reversal of assembly pathways, and (ii) through dynamic instability that uses distinct irreversible disassembly pathways. 
    Both mechanisms reduce defects while speeding up assembly; but the assembly speed of dynamic instability-based dissipative mechanisms is higher when compared at the same defect rate $\mu = 0.01$. See the \textit{Supplementary Information} for more details. (\textbf{c}) The preference for dissipative error correction in proofreading can be understood from how $dt_{\rm{nostall}}/d\mu$ depends on the driving force, $\phi$. (i) Plots of the time to replicate with no stalling $t_{\rm{nostall}}$ as a function of error rate $\mu$ shown for an $N=6$ proofreading network driven out of equilibrium to two different extents $\phi$. Both curves show that reducing error by an amount $\Delta \mu$ costs time $\Delta t_{\rm{nostall}}$, but the time cost is lower for highly driven $\phi$ networks.  (ii) The derivative $dt_{\rm{nostall}}/d\mu$ (see Eq. \ref{eq:dfdmu}) is plotted as a  function of the driving force $\phi$ for the same network of size $N=6$. Thus more dissipative mechanisms achieve the same error correction at a smaller cost in time. See \textit{Supplementary Information} for more details.} 
    \label{fig:Dissipation}
\end{figure*}

We compare two distinct models of error correction in both templated replication and self-assembly: an irreversible model, introducing new correcting pathways that are not simply the microscopic reversible of polymerization (or assembly) pathways and a reversible one, where errors are removed by the microscopic reversal of the polymerization (or assembly) pathway itself. We compare these two models at a value of error rate that is achievable by both of them.  The speed to replicate used in Extended Fig.  \ref{fig:Dissipation}(a.iii) is computed as in Eq. \ref{eq:time_1} with $\tau_{\rm{stall}}=5$ and is compared between the equilibrium network ($N=2$) and three values of the driving force $\phi$ in the loop network ($N=6$), $\phi=15,17$ and $\phi=19$, at the same error rate $\mu = 0.02$.  Extended Fig.  \ref{fig:Dissipation}(a.iii) shows that the irreversible error correction mechanism, i.e. kinetic proofreading models with loops, replicate templates faster~---~and would thus be evolutionarily prefered~---~than reversible mechanisms that achieve the same error rate. Note that this effect cannot be explained by the usual picture of dissipation being needed to reduce error rates, i.e. because of the second law, since both networks are being compared when they achieve the same error rate. See the \textit{Supplementary Information} for further discussion and details.

\begin{figure*}[h!!!]
    \centering
    \includegraphics[scale=1]{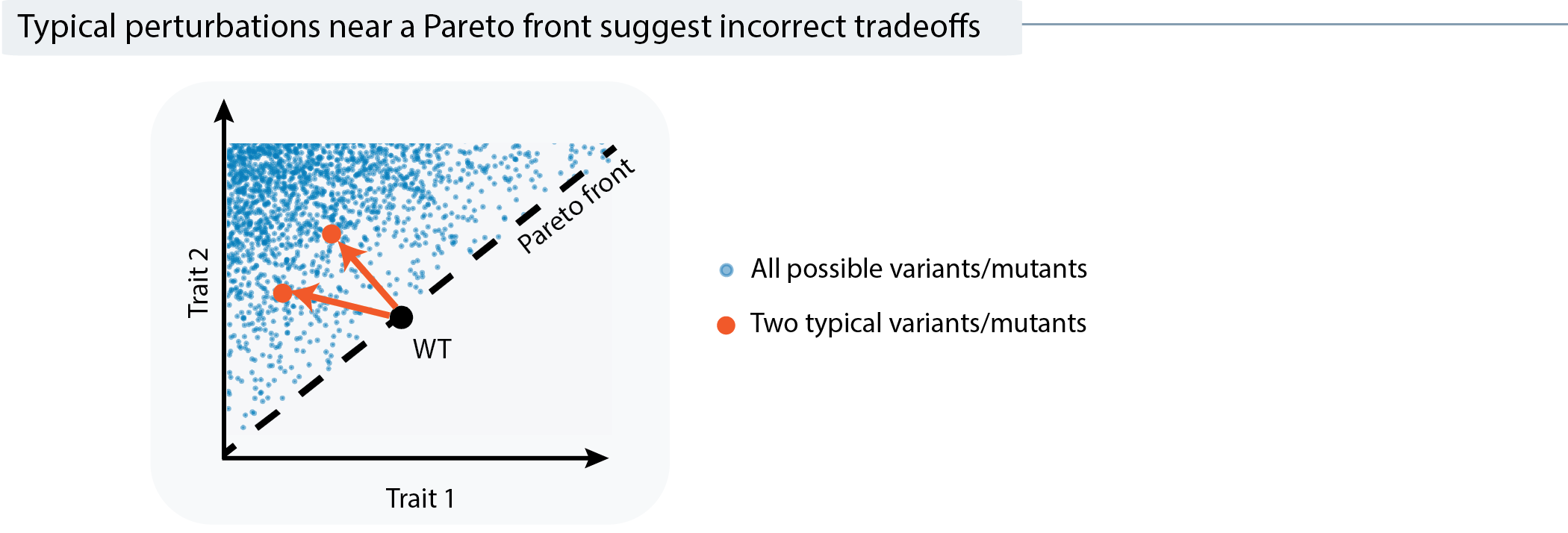}
    \caption{\textbf{Limited data can lead to misleading trade-offs.} \footnotesize A potential hazard in inferring trade-offs from a limited set of perturbations or variants. The sketch shows a scatter plot of two traits over all variants (blue) in a mutational neighborhood of a wild-type (WT) molecule or organism. Together, these variants show a Pareto front, defining a trade-off between the two traits. Generically, the density of variants is expected to be lower in the vicinity of the Pareto front (since each trait is at its maximum with the other trait held fixed). If the wild-type (WT) is near the Pareto front, red points show the perturbation of the WT by two typical mutations, given the distribution of variants in the mutational neighborhood. Red arrows suggest an incorrect trade-off converse to the real trade-off defined by the Pareto front. On the other hand, making all (or many) possible mutations (blue points) will reveal rare perturbations that move the WT along the Pareto front and thus reveal the true trade-off.}
    \label{fig:SI_reftradeoff}
\end{figure*}

\begin{table*}[h]
\footnotesize
\caption{A minimal scenario for the origin of non-equilibrium order.}
\begin{tabular}{lccr}
    {\textbf{Model}} & {\textbf{Key ingredient}} & {\textbf{Non-eq. mechanism}} & {\textbf{Potential relevance}} \\ \midrule
    Replication (Fig. \ref{fig:proofreading_unified})  & Stalling & Kinetic proofreading & DNA replication; transcription; translation; \\
    & & & ribozymes at the origin of life \\
    Self-assembly (Fig. \ref{fig:selfassembly}) & Geometric frustration & Dynamic instability & Viral capsids; microtubules;  
\\
 & & & chaperoned protein folding;\\
 & & &  ribosomal assembly\\
    High-dim. trajectories (Fig. \ref{fig:ResetDemon})  & Wide replication times & Random reset & Cellular replication; cell cycle coordination;
\\
    & & & check points \\
\end{tabular}
    \label{tab:summary}
\end{table*}

\bibliographystyle{naturemag}
\bibliography{paperpile,ravasio,cge_main}

\end{document}


\title{Supplementary Information: A minimal scenario for the origin of non-equilibrium order}
\date{}
\author{R. Ravasio$^*$, K. Husain$^*$, C. G. Evans, R. Phillips, M. Ribezzi, J. W. Szostak, A. Murugan}
\maketitle

\hypersetup{linkcolor=black}
\tableofcontents
\chapter{Kinetic proofreading in templated replication}

This section details theoretical work and experimental support on the speed-accuracy relationship in kinetic proofreading. 

In Sec. \ref{sec:proof_model}, we detail the simplest model of proofreading and stalling that predicts a counterintuitive ``faster is more accurate'' trade-off. We also detail our \emph{in silico} evolution work using random networks with this family of models. 

In Sec. \ref{sec:altstall}, we review experimental work on stalling effects in diverse molecular systems and alternative models that combine proofreading and stalling. 

In Sec. \ref{sec:phi29-changliu}, we discuss our analysis of experimental data on speed and accuracy from the largest mutagenesis library \cite{ravikumar_scalable_2018} of a DNA polymerases studied to date.

In Sec. \ref{sec:priorspeedaccuracy}, we detail prior experimental and theoretical work on speed-accuracy trade-offs, and place our results in the context of these studies. 

\section{A simple model of proofreading with stalling}
\label{sec:proof_model}

Kinetic proofreading, a model of discrimination in enzymatic reactions was introduced in \cite{hopfield_kinetic_1974, ninio_kinetic_1975}. While these models are celebrated for introducing a key idea~---~error-correction through non-equilibrium discard pathways~---~that has been confirmed in numerous systems, these models ignore several other molecular details \cite{Johnson2008-nm} that are important in reality, such as on-rate-based discrimination, induced fit and nucleotide selectivity \cite{loeb_fidelity_1982, banerjee_elucidating_2017, mallory_we_2020, midha_insights_2024}. However, as we show here, the widely studied simplest models \cite{hopfield_kinetic_1974,ninio_kinetic_1975,murugan_speed_2012,lan_energyspeedaccuracy_2012,murugan_discriminatory_2014,ouldridge_robustness_2014,francois_phenotypic_2013,francois_case_2016,estrada_information_2016,cui_identifying_2018,galstyan_allostery_2019,poulton_nonequilibrium_2019,pineros_kinetic_2020}, when combined with stalling, are sufficient to explain the empirical counterintuitive trade-off shown in Fig 3. 

In this section, we focus on this simple model used in Fig 2; later, we discuss key alternatives to our model and the relationship of results here to prior work in Sections \ref{sec:altstall} and \ref{sec:priorspeedaccuracy}. 

We emphasize that it is easy to construct proofreading models, including stalling effects, that predict the intuitive speed-accuracy relationship; such work has been carried out many times \cite{hopfield_kinetic_1974,ninio_kinetic_1975,murugan_speed_2012,bennett_dissipation-error_1979,lan_energyspeedaccuracy_2012,murugan_discriminatory_2014,ouldridge_robustness_2014,francois_phenotypic_2013,francois_case_2016,estrada_information_2016,cui_identifying_2018,galstyan_allostery_2019,poulton_nonequilibrium_2019,pineros_kinetic_2020, banerjee_elucidating_2017, mallory_we_2020, midha_insights_2024}. Our goal here is to demonstrate that alternative simple models can also predict the counterintuitive speed-accuracy relationship observed empirically when hundreds of variants are compared in experiments (Fig 3).

\subsection{Proofreading model at the single base level}
Following prior work \cite{hopfield_kinetic_1974,ninio_kinetic_1975,murugan_speed_2012,lan_energyspeedaccuracy_2012,murugan_discriminatory_2014,ouldridge_robustness_2014,francois_phenotypic_2013,francois_case_2016,estrada_information_2016,cui_identifying_2018,galstyan_allostery_2019,poulton_nonequilibrium_2019,pineros_kinetic_2020}, we model enzyme-substrate transformations relevant for kinetic proofreading using a Markov chain. The probability of being in state $i$, $p_{i}$, as function of time obeys the Master equation with $N$ states
\begin{align}
    \frac{dp_{i}}{dt} = - \sum_{j\neq i} k_{ji} p_i + \sum_{j\neq i} k_{ij} p_j
\label{eq:markov}
\end{align}
where $k_{ij}$ is the rate of transition from $j \to i$ and $\sum_ip_i=1$. It is convenient to write this equation in matrix form, $\partial_t p_i = \sum_j K_{ij} p_j$ with $K_{ij} = k_{ij}$ if $i\neq j$, $K_{ij} = -\sum_{l\neq i}k_{li}$ if $i=j$. The steady state $p_i$ is then in the null-space of $K$, i.e., $\partial_t p_i = \sum_j K_{ij} p_j = 0$. 

\begin{figure*}[h!!!]
    \centering
    \includegraphics[scale=1]{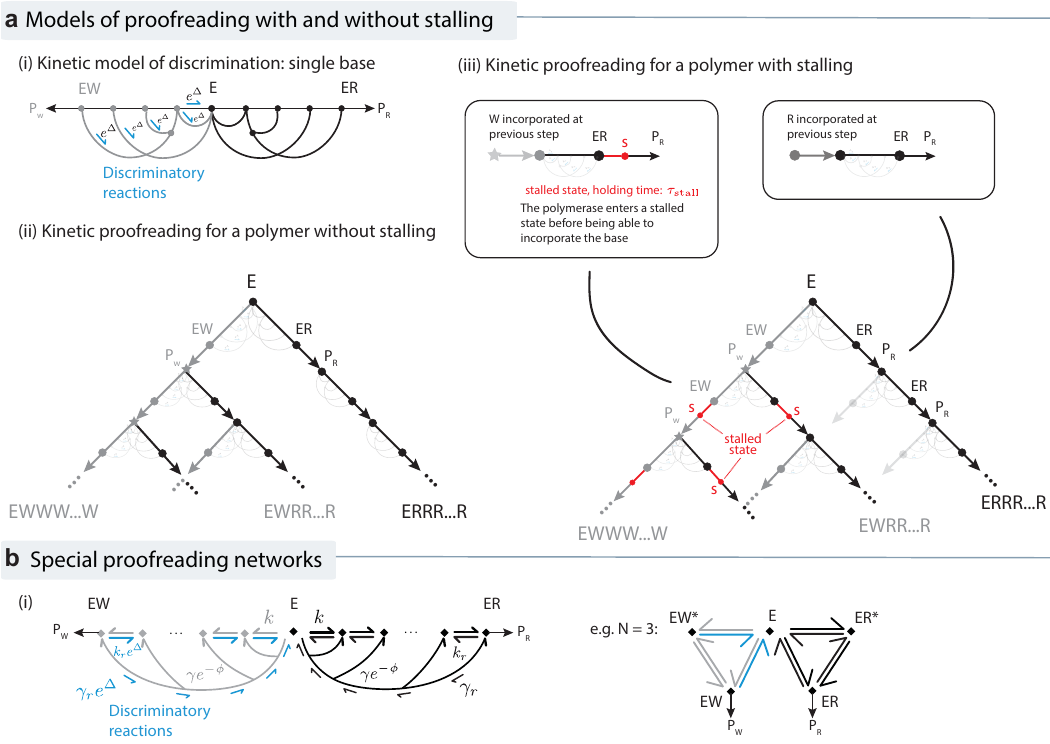}
    \caption{\textbf{Extending simple proofreading models with the stalling effect.} (\textbf{a}) A general network constituting the model of kinetic discrimination at single base level is shown in (i). The extension for a polymer is shown in (ii) where each path corresponds to one particular outcome of replication, e.g. all right nucleotide incorporated in $ERRR\cdots R$. Stalling is introduced in (iii) by adding an additional state $s$ with holding time $ \tau_{\rm{stall}}$. (\textbf{b}) A particular instance of the single base kinetic proofreading model is defined in (i), which is used to produce the curves of Fig. 2a in the main paper. A driving force $\phi$ is coupled to the reactions $\gamma$ to control the contribution to the chemical potential of the non-equilibrium discriminatory reactions defining proofreading. For simplicity, the network is shown for $N=3$, representing the typical kinetic proofreading network \cite{hopfield_kinetic_1974, ninio_kinetic_1975}.}
    \label{fig:SI_proofreadingth}
\end{figure*}

Consistent with prior work, we model proofreading by considering a ``doubled network'' as shown in Fig. \ref{fig:SI_proofreadingth}a(i); here, the left branch models reactions of the enzyme $E$ with the wrong substrate $W$, while the right branch models reactions with the right substrate $R$. Following Hopfield and Ninio's definition \cite{hopfield_kinetic_1974, ninio_kinetic_1975}, the kinetics of $R$ and $W$ differ in the binding energies of $EW$ being lower than the ones for $ER$ by a factor of $\Delta$. This difference translates into higher off-rates for some $W$ reactions than the corresponding rates for $R$ reactions by a factor $e^{\Delta}$, as pictured in Fig. \ref{fig:SI_proofreadingth}a(i). These discriminatory reactions are chosen such that each independent pathway from $E$ to wrong product formation carries one $e^{\Delta}$ factor in its off-rate: if $n$ independent loops are present in the network, there will be $n+1$ independent pathways bringing the discriminatory factor $e^{\Delta}$. This prescription sets the minimal achievable error rate, as discussed below.

\subsubsection{Error or mutation rate rate (per base)}
The error rate is defined as the ratio between the rate of wrong product formation $r(P_W)$ and the rate of correct product formation $r(P_R)$
\begin{equation}
    \mu = \frac{r(P_W)}{r(P_R)}
\end{equation}
In the limit of slow product formation, we can take this to be the ratio of steady-state probabilities. Let $p_i$ be the $i$-th component of the null-space eigenvector of $K^R$ or $K^W$ where $R,W$ indicate the kinetics for the incorporation of the wrong ($W$) or right ($R$) base. The error is the ratio of wrong product formation $r(P_W)$ to right production formation $r(P_R)$ and in the case of slow formation rate $f$ from $ER$ or $EW$ reads \cite{hopfield_kinetic_1974}
\begin{equation}
    \mu \approx \frac{f p_{N-1}^W/p_0^W }{f p_{N-1}^R/p_0^R } = \frac{p_{N-1}^W/p_0^W}{p_{N-1}^R/p_0^R}
\end{equation}
where $f$ is the rate from $ER$ or $EW$ to $P_{R}$ or $P_W$, $0$ is the initial state of the network of size $N$ and $N-1$ is the final state. Note that $p_0^R$ and $p_0^W$ differ, given the different kinetic rates for $R$ and $W$. Indeed, $p^W$ depends on $\Delta$, while $p^R$ does not~---~see Sec. \ref{sec:specificnets} for the prescription used.

\subsubsection{Time for forming product}

In this work, we consider multiple definitions of time to form a product. Definition 1 is used in the in-silico evolution (Fig. 2b, main text), while Definition 2 is an approximation of Definition 1 and is used in the model of kinetic proofreading (Fig. 2a, main text).

\begin{enumerate}

\item \textit{Definition 1: Mean first passage time.} The time to add a base is the mean first passage time of trajectories from state $E$ (initial state, $i=0$) in Fig. \ref{fig:SI_proofreadingth}a(i) to state $P_X$ with $X\in \{R, W\}$ (final state, $i=N)$  
\begin{equation}
    t_{\rm{nostall}}\equiv \tau_{E \to P_X}  = \tau_{MFPT}  = -\sum_{i\neq N}[ \tilde{K}^{+} \vec{\tilde{p}}_0 ]_i\,
    \label{eq:mfpt}
\end{equation}
where $\tilde{K}^{+}$ is the Morse-Penrose inverse of $\tilde{K}$, a modified Markov chain $\tilde{K}$ where $n$ is an absorbing state: $\tilde{K}_{ij} = K_{ij}$ for $j \neq n$ and $\tilde{K}_{in} = 0$ for all $i$. The derivation is discussed below.

Following \cite{munsky_specificity_2009, polizzi_mean_2016}, we determine the mean first passage distribution to state $n$ as follows $\tau_{E \to n} \equiv \int_0^{\infty} dt \, t \, F_{E \to n}(t)$ where $F_{E \to n}(t)$ is the first passage time (FPT) distribution from state $E$ to $n$. We consider a modified Markov chain $\tilde{K}$ where $n$ is an absorbing state: $\tilde{K}_{ij} = K_{ij}$ for $j \neq n$ and $\tilde{K}_{in} = 0$ for all $i$. In our specific case, we choose the product formation state $P_X$ to be the absorbing state. Markov chains $\tilde{K}$ and $K$ have the same first passage distribution $F_{E \to n}(t)$. We can compute the latter by the rate of probability accumulation in state $n$, assuming that at long times state $n$ will be occupied with probability $1$: $F_{E \to n}(t) = \f{\p \tilde{p}_n}{\p t} = \f{\p}{\p t}(1-\sum_{i\neq n}\tilde{p}_i)$
where $\partial_t \tilde{p} = \tilde{K} \tilde{p}$ and initial condition $\tilde{p}(t=0) = (1,0,0,\ldots)\equiv \tilde{p}_0$. 
The mean FPT is $\tau_{E \to n} \equiv \int_0^{\infty} dt \, t \, F_{E \to n}(t) = -\sum_{i\neq n}\int_0^{\infty} dt \, t \, \f{\p\tilde{p}_i}{\p t}$. We expand the solution $\vec{\tilde{p}}$ in the eigenvalues $\lambda$ and singular vectors $\hat{v}_{\lambda}$, $\hat{u}_{\lambda}$ of the rate matrix $\tilde{K}$, $\vec{\tilde{p}}(t) = \l \hat{v}_{\lambda_a} \sum_a e^{\lambda_a t}\, \hat{u}^t_{\lambda_a} \r  \vec{\tilde{p}}_0 $. Inserting this expansion into the integral for $\tau_{E \to n}$, we get
\begin{align*}
\int_0^{\infty} dt \, t \, \f{\p \vec{\tilde{p}}}{\p t} = -\sum_{\lambda_a } \lambda_a \l \hat{v}_{\lambda_a} \hat{u}^t_{\lambda_a} \r \, \vec{\tilde{p}}_0  \int_0^{\infty} dt \, t \, e^{\lambda_a t} \nn \\
= -\sum_{\lambda_a \neq 0} \f{1}{\lambda_a} \l \hat{v}_{\lambda_a} \hat{u}^t_{\lambda_a} \r \, \vec{\tilde{p}}_0 
= - \tilde{K}^{+} \vec{\tilde{p}}_0 \,, 
\end{align*}
where $\tilde{K}^{+}$ is the Morse-Penrose inverse of $\tilde{K}$. Hence, the mean first passage time to the final state $P_X$ ($n=N)$, $\tau_{E \to P_X}$, is given by
\begin{equation}
   t_{\rm{nostall}}= \tau_{E \to P_X}  = \tau_{MFPT} = -\sum_{i\neq N}[ \tilde{K}^{+} \vec{\tilde{p}}_0 ]_i\, .
\end{equation}

\item \textit{Definition 2.} In the limit of slow product formation from $EX$ to $P_X$, we can approximate the equation above by just the final state occupancy at steady-state, $t_{\rm{nostall}}\approx 1/p_{N-1}^R$.  When catalysis is slow, the inverse of the first passage time to the product state ($R$ or $W$) is equal to the flux to the same state \cite{hill_free_2012}.
 
\end{enumerate}

These definitions consider solely the rate of forming the right product. Some prior work includes instead the time to make the wrong product in the calculation of total time to form a product. These two times are numerically similar for any reasonable mutation rate, including the relatively high mutation rates considered in this paper. Our results would remain essentially unchanged if we were to include the time for the wrong product in our definition of $t_{\rm{nostall}}$.

\subsubsection{Rate of entropy production}

The rate of entropy production is computed as defined in \cite{schnakenberg_network_1976}
\begin{equation}
    \sigma = \frac{1}{2}\sum_{i,j}(k_{ij}p_j-k_{ji}p_i) \log \frac{k_{ij}p_j}{k_{ji}p_i}\, ,
\end{equation}
with $k_{ij}$ the rate constants as defined above, and $p$ the steady-state vector.

\subsection{Extending proofreading models to include stalling}
\label{sec:stallingmodels}

We extend the proofreading model introduced above for an entire polymer. This extension is necessary to introduce the experimentally observed effect of stalling upon mismatch incorporation in the current proofreading model. 

We emphasize again that it is easy to construct alternative proofreading models including stalling effects that predict the intuitive speed-accuracy relationship; such theoretical work has been carried out many times \cite{hopfield_kinetic_1974,ninio_kinetic_1975,murugan_speed_2012,bennett_dissipation-error_1979,lan_energyspeedaccuracy_2012,murugan_discriminatory_2014,ouldridge_robustness_2014,francois_phenotypic_2013,francois_case_2016,estrada_information_2016,cui_identifying_2018,galstyan_allostery_2019,poulton_nonequilibrium_2019,pineros_kinetic_2020, mallory_we_2020, midha_synergy_2023}. Our goal here is to demonstrate that simple models can also predict the counterintuitive speed-accuracy relationship observed empirically when hundreds of variants are compared in experiments (Fig 3). See Sec. \ref{sec:altstall} for details experimental evidence for stalling and alternative more complex models of combining stalling and proofreading; here, we focus on the simplest model that predicts the counterintuitive ``faster is more accurate'' result seen empirically in Fig 3.

DNA polymerases are known to stall after incorporating a mismatch \cite{perrino_differential_1989,mendelman_base_1990,huang_extension_1992,johnson_structures_2004,ichida_high_2005,baranovskiy_structural_2022}. We model this effect as a slowing down of the catalysis at the next base: the incorporation of a mismatch affects only the catalysis of the following nucleotide. This one-step memory effect can be modeled in multiple ways, either as a slow-down of the rates of the reaction following the mismatch, or as adding an extra time contribution for each mismatch incorporated. We implement the latter in our model. As depicted in Fig. \ref{fig:SI_proofreadingth}a(iii), incorporation of a wrong base causes the system to enter a stalled state $s$ with a typical holding time $\tau_{\rm{stall}}$ before catalysis of the next base is completed. Given an error rate per base pair $\mu$, the probability of observing $m$ mismatched bases in the product is given by the binomial distribution, $P(m) = {L\choose m}\mu^m(1-\mu)^{L-m}$.

The time to form a product with $m$ mismatches is
\be
T(m) = \sum_{i = 1}^L \tau_{E \to P_{X_i}} + \sum_{j=1}^{m} \tau_h^j
\ee
\noindent where $\tau_h^j$ is the stochastic holding time after mismatch $j$, and $\tau_{E \to P_{X_i}}$ is the time to add base $i$: i.e., $\tau_{E \to P_R}$ if the base is correct, and $\tau_{E \to P_W}$ is the base is incorrect.

\indent In the limit of low error rates, $\mu L \ll 1$, at most one mistake per product will be observed, and the average time to copy a strand is
\bea \label{eq:trep_main}
\langle T \rangle &\equiv& \sum_{m=0}^L P(m) T(m) \nn \\
&\approx& (1-\mu L) \, T(m = 0) + \mu L \, T(m = 1) \nn \\
&=& L \tau_{E \to P_R} + \mu L \tau_{\text{stall}}\\
&\equiv& T_{\rm{rep}} = L t_{\rm{nostall}}(\mu) + L \mu  \tau_{\rm{stall}} \nn 
\eea
\noindent where we have further assumed that $\tau_{E \to P_R} = \tau_{E \to P_W}$, and replaced each holding time $\tau_h^j$ with the characteristic stalling time $\tau_{\text{stall}}$. Eq. \ref{eq:trep_main} is the definition of the time to replicate, $T_{\rm{rep}}$ used in the main paper, where $t_{\rm{nostall}}$ is defined as $\tau_{E \to P_R}$ .

With this assumption, stalling affects the branched network of Fig. \ref{fig:SI_proofreadingth}a(ii) by adding an additional time $\tau_{stall}$ to the total time to replicate each time the wrong branch is taken, as depicted in Fig. \ref{fig:SI_proofreadingth}a(iii). It is crucial to note that this prescription does not affect proofreading, since it introduces a time cost after the proofreading step is performed, and it only applies to the base directly following the mismatch incorporation, not propagating further in the network.

The model presented here is the simplest model of stalling consistent with experimental evidence. This evidence and alternative models, e.g. slow down of specific rates in the next base or extent of proofreading being affected by stalling time, are discussed in detail in Sec. \ref{sec:altstall} below. 

\subsubsection{}

\begin{figure*}[h!!!]
    \centering
    \includegraphics[scale=1]{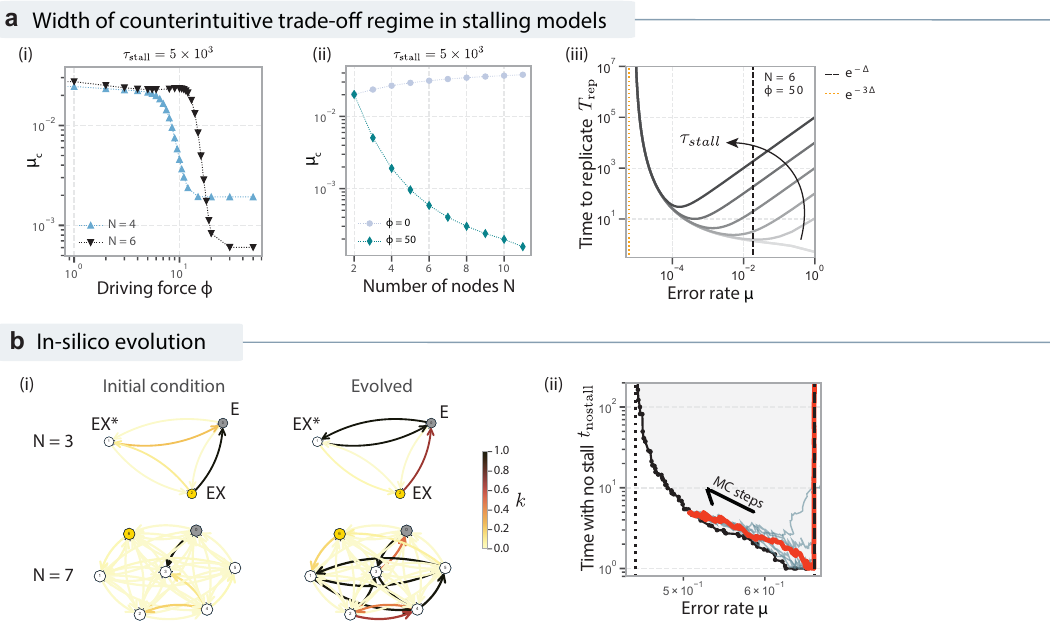}
    \caption{\textbf{Extending simple proofreading models with the stalling effect.} (\textbf{a}) The crossover error rate $\mu_c$ is shown as function of the driving force $\phi$ (i) and the number of nodes $N$. (iii) The time to replicate Eq. \ref{eq:trep_main} is shown as function of error rate for increasing values of the stalling time $\tau_{\rm{stall}}$. (\textbf{b}) The Markov chain networks of the in-silico evolution are shown in (i) for $N=3$ and $N=7$: the in and out rates for each nodes are colored according to the kinetic rate observed in the initial condition (left) and in the evolved state (right). Slow kinetic rates in yellow, while fast kinetic rates in black arrows~---~see color map. (ii) shows the trade-off plot of the time with no stalling $t_{\rm{nostall}}$ and error rate for $N=3$. The Pareto front is highlighted in black, and ten different trajectories~---~from an in-silico evolution with parameters as reported in Tab. \ref{tab:insilicoevo}~---~are plotted, with one highlighted in red. }
    \label{fig:SI_proofreadingcs}
\end{figure*}

\subsection{Specific model of proofreading}
\label{sec:specificnets}

We can apply these formulas to specific proofreading networks considered in Figs. 2a, 6a of the main paper. For networks of the structure shown in Fig \ref{fig:SI_proofreadingth}b(i), the rates for a network of $N$ nodes are defined as follows. We introduce both forward and reverse rates between neighbouring nodes ($k_f$ and $k_r$) and forward and reverse rates from the origin to any other node ($\gamma$ and $\gamma_r$). We assume $k_r=a k_f$ and apply Kolmogorov's loop criterion to properly define the kinetic constants at detailed balance
\bea
k_f^n \gamma^{n\to 0} &=& \gamma_r^{0\to n}\left( \prod_{m=1}^n a_m \right) k_f^n\\
\gamma_r^{n\to 0} &=& \gamma^{0\to n} a^n
\eea
supposing $a_m = a \, \, \forall\, m$.  

The discriminatory reactions between $X=R,W$ are defined so that at equilibrium the discrimination is $\mu_{eq} = e^{-\Delta}$, making the discriminatory reactions ($\gamma_r^{W}$ and $k_r^W$) faster
\bea
\gamma_r^{W, \,n\to 0} &=& \gamma_r^{R, \, n\to 0}e^{\frac{n}{N-1}\Delta}\\
k_r^W &=&  k_r^R e^{\frac{\Delta}{N-1}}\, .
\eea
The definition above sets a maximal proofreading of $\mu^* = e^{-\frac{N}{2}\Delta}$, since the $N-1$ loops that go back to the origin bear a factor of $e^{\frac{n}{N-1}\Delta}$: $\sum\limits_{n=1}^{N-1}\frac{n}{N-1}\Delta=\frac{N}{2}\Delta$\,.

As sketched in Fig. \ref{fig:SI_proofreadingth}b(i), a driving force $\phi$ is coupled to the reactions $\gamma^{0\to n}$, as $e^{-\phi}\gamma^{0\to n}$, to control the non-equilibrium driving and hence the amount of proofreading: an increase of the driving force $\phi$ slows the associated reaction, breaking detailed balance. In Fig. 2a (main paper) the driving force takes the values $\phi = 0, 13,15,17,19,50$, showing that a higher driving force leads to higher discrimination, i.e. lower error rates.

\subsubsection{$N=3$ networks}

For such a network of size $N=3$, we can explicitly write the resulting analytic expressions for speed, as final state occupancy, and error rate both defined in Sec. \ref{sec:proof_model}. These two quantities can be written as function of the kinetic rates ($k, \gamma$), scaling factor $a$, discrimination factor $\Delta$ and driving force $\phi$

\begin{align}
    t_{\text{nostall}} &= \frac{1}{p^R_{N-1}}= \frac{\gamma\left(k+ka(3+a)+\gamma +a\gamma\right)+e^{\phi}a^2(k^2(1+a+a^2)+2k(1+a)\gamma +\gamma^2)}{e^{\phi}k^2a^2 +\gamma(k+2ka+\gamma)}  \label{eq:tnostall3} \\
    \mu &= \frac{e^{-\Delta}\left[k^2a^2+\gamma^2+k(\gamma+2\alpha\gamma)\right]\left[ e^{\Delta/2}\gamma(ka+\gamma)+k(\gamma+a(e^{\phi} k a+\gamma) \right]}{\left[ e^{\phi}k^2a^2+\gamma(k+2ka+\gamma) \right]\left[ e^{\Delta/2}\gamma(ka+\gamma)+k(\gamma+a(ka+\gamma)) \right]}\, . \label{eq:mu3}
\end{align}

\subsubsection{Characterization of the trade-off}

Fig. 2a in the main paper shows that including stalling in the model introduces a crossover error rate, $\mu_c$, which separates two regions of opposite trade-offs: an intuitive one for $\mu<\mu_c$ (``faster is less accurate'') and a counterintuitive one for $\mu>\mu_c$ (``faster is more accurate''). In this paragraph, we characterize further how the trade-off curve depends on the parameters of the model $\phi$, $\tau_{stall}$ and $N$. Fig. \ref{fig:SI_proofreadingth}a(i) shows that the crossover error rate $\mu_c$ decreases monotonically as function of the driving force $\phi$, where networks of size $N=4$ and $N=6$ are used to numerically determine the value of $\mu_c$ as function of $\phi$. Fig. \ref{fig:SI_proofreadingth}a(ii) shows that $\mu_c$ decreases as function of the size of the networks, $N$, at fixed and non-zero driving force ($\phi=50$), in contrast with $\phi=0$. The effect of varying $\tau_{stall}$ on the trade-off is shown in Fig. \ref{fig:SI_proofreadingth}a(iii) for a network of size $N=6$ with driving force $\phi=50$: increasing the stalling time increases the size of the region with the ``faster is more accurate'' trade-off. 

The value of $\mu_c$ can be determined analytically by solving the following optimization, using Eq. \ref{eq:trep_main}
\begin{align}
    \frac{\partial T_{\rm{rep}}}{\partial \mu}&=0 \nn \\
    \frac{\partial t_{\rm{nostall}}}{\partial \mu} + \tau_{stall} &= 0 \label{eq:muctstall} \, .
\end{align}

Eqs. \ref{eq:tnostall3}-\ref{eq:mu3} for $N=3$ can be used to find the value of $\mu_c$ by computing the ratio of derivatives with respect to a parameter, e.g. $a$. To make the optimization analytically tractable, we focus on the limit $\gamma \gg 1$,  which pushes the network to maximal discrimination, and choose $k=1$. In this limit, Eqs. \ref{eq:tnostall3}-\ref{eq:mu3} become
\begin{align}
    t_{\text{nostall}} &\approx \frac{ (1+a+a^2e^{\phi})\gamma^2 }{ \gamma^2+a^2e^{\phi} }\\
    \mu &\approx \frac{ e^{-3/2 \Delta}(\gamma^2e^{\Delta/2}+a^2e^{\phi}) } { \gamma^2 +a^2e^{\phi} }\, .
\end{align}
Solving Eq. \ref{eq:muctstall} for $a$ gives two roots, one of which is negative, hence unphysical. Keeping the root with positive sign, and substituting it in Eq. \ref{eq:mu3}, gives us an analytical form for $\mu_c$ in the limit $\gamma \gg 1$ and $k=1$ that is monotonically decreasing as function of $\phi$
\begin{equation}
    \mu_c \approx e^{-3/2\Delta}\frac{ e^{\Delta/2}\gamma^2 + f(\phi, \tau_{\rm{stall}},\gamma, \Delta)^2 }{ \gamma^2+f(\phi, \tau_{\rm{stall}},\gamma, \Delta)^2 }
\end{equation}
with
\begin{align}
    f(\phi, \tau_{\rm{stall}},\gamma, \Delta)=e^{-3\Delta-\phi}&e^{\phi}\left[-\tau_{\rm{stall}}+e^{\Delta/2}\left(\tau_{\rm{stall}}-e^{\Delta}(-1+\gamma^2) \right)\right]  \nn \\
    &+  \sqrt{e^{\phi} \left\{e^{3\Delta}\gamma^2+e^{\phi}\left[ 
 \tau_{\rm{stall}} +e^{\Delta/2}\left( -\tau_{\rm{stall}} +e^{\Delta}(-1+\gamma^2)\right)\right]\right\}} \, . \nn
\end{align}

\subsection{In-silico evolution of kinetic proofreading}
\label{sec:insilicoevo}

We set up a Markov network of the type in Eq. \ref{eq:markov} with $N$ states, focusing on $N=7$ and $N=3$ for simulations, as discussed below. A network of sizes $N=3$ and $N=7$ is depicted in Fig. \ref{fig:SI_proofreadingcs}b(i), before and after evolution for faster replication times. Comparing the rates of the $N=3$ initial condition and evolved network is consistent with the understanding of kinetic proofreading discussed in the original works \cite{hopfield_kinetic_1974,ninio_kinetic_1975}: (i) the wrong nucleotide needs to preferentially enter state $EX$ through $EX^*$ and not $E$, (ii) the dominant rate outgoing $EX$ needs to be the one back to the initial state $E$ and (iii) the rate out of $EX^*$ needs to be smaller or equal than the rate coming in. Overall, these conditions tell that the reaction $EX^*\to EX$ needs to be driven strong enough to reach maximal discrimination. For $N=7$, the evolution scheme is finding more complex prescriptions to achieve proofreading, as seen in Fig. \ref{fig:SI_proofreadingcs}b(i).

\subsubsection{Fitness function}
A simple definition of the time to copy a strand $T(k)$ does not involve its length, but just considers the average effect of stalling upon incorporation of mistakes by adding a term proportional to the error rate $\mu$ (see Eq. \ref{eq:trep_main})
\bea
\label{eq:time_stall}
T(k) &=& \tau_{MFPT}^{R} + \tau_{\text{stall}}\mu\\
&=& t_{\rm{nostall}} + \tau_{\text{stall}}\mu
\eea
where $\tau_{FP}^R$ is the mean first-passage time from state $E$ to state $P_R$ defining $t_{\rm{nostall}}$ (see Eq. \ref{eq:mfpt}) and $\tau_{\text{stall}}$ is the stalling time upon insertion of a wrong base. In the main figures we used Eq. \ref{eq:time_stall} to define fitness, $F(k)=1/T(k)$.

\subsubsection{Parameterization}
The kinetic rates depend on the energy of state $i$, $V_i$, and the energy barrier between state $i$ and $j$, $B_{ij}$, thus the model is defined by parameterizing the energy landscape
\bea 
 k_{ji} &=& e^{V_i - B_{ji}-\phi_{ij}}\\
 k_{ij} &=& e^{V_j - B_{ij}}
\eea
with $B_{ij} = B_{ji}$. To allow for the coupling of reactions to external energy sources, an external driving potential between states $i$ and $j$, $\phi_{ij}$, is introduced, with $\phi_{ij}=\phi_{ji}$. Two sets of parameters are needed to account for both the insertion of the right and the wrong nucleotide. The parameters for the wrong nucleotide $\{V^W, B^W\, \phi^W\}$ are defined as
\bea
V_i^W = V_i^R + \epsilon^V_i \\
B_{ij}^W = B_{ij}^R + \epsilon^B_{ij} \\
\phi_{ij}^W = \phi_{ij}^R
\eea
An arbitrary split point $s = \text{int}(N/2)$ defines which of the reactions are \textit{discriminatory}
\begin{align}
\epsilon^V_i = \epsilon^B_{ij} = 0 \quad &\text{for}& \quad i,j < s\\
\epsilon^V_i = \epsilon^B_{ij} = \Delta>0 \quad &\text{for}& \quad i,j \geq s\\
\epsilon^B_{ij} = 0 \quad &\text{for}& \quad i<s<j\, .
\end{align}
The prescriptions above effectively slow down the discriminatory reactions by a factor $e^{\Delta}$, in a way that the error rate at equilibrium is $\mu=e^{-\Delta}$ and the maximal discrimination reached by the network is $\mu=e^{-(N-s)\Delta}$, where $N-s$ is the number of discriminatory loops with factor $e^{\Delta}$ that go back to the origin. 

\subsubsection{Evolution}

The network is initialised at equilibrium $\phi_{ij}=0 \,\, \forall i,j$. The in-silico evolution follows a Monte Carlo algorithm with a Metropolis update rule, the Monte Carlo steps being our evolutionary time. The energy function to be minimised is the time to copy a strand defined as in Eq. \ref{eq:time_stall}. As stated above, the main figures are produced with Eq. \ref{eq:time_stall} to define fitness.

During the in-silico evolution, each of three parameters are allowed to vary in a defined grid $\{V^R,B^R,\phi\} \in [l, h+d] $, allowing for mutations to drive some reactions out of equilibrium, $\phi_{ij}>0$. For the simulations presented in this work, we choose $l=-7$, $h=5$ with a spacing of $d=0.2$, setting the fineness of the grid. A mutation will act on the parameters by adding a perturbation
\bea \{V^R_m,B^R_m,\phi_m\}&=&\{V^R+\delta_1,B^R+\delta_2,\phi+\delta_3\}\\
&\in& [l, h+d]
\eea
with $\delta_1$, $\delta_2$, $\delta_3$ each sampled uniformly at random in the interval $[-d,d]$. The mutated parameters are enforced to be in the same grid $[l, h+d]$.

The parameters of the simulation of the $N=3$ and $N=7$ networks are reported in Table \ref{tab:insilicoevo}; all the evolution experiments were run for $10^4$ Monte Carlo steps. 
\begin{center}
\captionof{table}{Parameters of the in-silico evolution.}
\begin{tabular}[h]{llll}
    {Size} & {Temperature $T_e$} & Stalling time $\tau_{\rm{stall}}$& $\Delta$ \\ \midrule
    $N=3$  & $10^{-3}$ & $10^3$& 0.4 \\
      $N=7$  & $10^{-2}$ & $10^3$& 0.4
     \\ 
\end{tabular}
\label{tab:insilicoevo}
\end{center}

\subsubsection{Trajectories exploring the intuitive trade-off plot}

Fig. \ref{fig:SI_proofreadingcs}b(ii) shows the trade-off plot of the time without stalling $t_{\rm{nostall}}$ and error rate $\mu$ computed with a network of size $N=3$, the intuitive Pareto front is shown in solid black: lower error is achieved by spending more time at inserting each base. The trajectories of the in-silico evolution (one example highlighted in red) follow the intuitive trade-off for $t_{\rm{nostall}}$ and error rate $\mu$, while still showing the counterintuitive trade-off for the time to replicate, as shown in the main figures.

\section{Experimental evidence for stalling and alternative models}
\label{sec:altstall}

Stalling effects have been empirically observed in a wide range of systems in which information encoded in a polymer template is copied. Stalling does not appear to be specific to a particular step of the central dogma~---~since it is observed for DNA, RNA and protein synthesis~---~and does not appear to be specific to an enzyme~---~since non-enzymatic replication of RNA templates, relevant for origin of life scenarios, also demonstrates stalling.  

Here we review this broad experimental evidence for stalling. We then outline alternative models~---~to the highly simplified model in Sec. \ref{sec:proof_model}~---~of combining proofreading and stalling that also predict the counterintuitive ``faster is more accurate'' relationship seen in data (Fig. 3). We also contrast these models to ones that predict the intuitive ``faster is less accurate'' trade-off.

\subsection{Methods to characterize stalling}

Experimental techniques used to quantify stalling broadly fall into two classes: (a) kinetic measurements of extension past mismatches, (b) inference from sequencing of transcripts of different length.

\begin{itemize}

    \item \textit{Technique (a)}: \textit{Direct kinetic measurements:} This approach measures the rate of extension past a mismatch. The mismatches are induced in varying ways: (i) by studying extension of a primer with a mismatch at the 3' end \cite{joyce_reactions_1992,johnson_structures_2004,petruska_comparison_1988,huang_extension_1992,esteban_fidelity_1993,perrino_differential_1989,thomas_transcriptional_1998,mendelman_base_1990,weiss_interaction_1992,beard_influence_2004,baranovskiy_structural_2022,sydow_structural_2009}, (ii) by providing only incorrect nucleotides \cite{shaevitz_backtracking_2003}.  The kinetics of such extension past mismatches is measured by: (i) single molecule experiments, (ii) bulk assays measuring yield such as PAGE gels. 

\subitem (i) Single molecule experiments allow to control the incorporation of a mistake and the subsequent extension \textit{in vitro} \cite{johnson_structures_2004}. The protein of interest is isolated by purification and then either used directly in solution or crystallized to probe its structure. One typical experiment consists of a template/primer (often radio-labeled) complex that is extended by the enzyme in controlled conditions: the insertion of a mismatch in the primer is induced by providing only the corresponding wrong nucleotide, and its extension by adding different concentrations of the next right nucleotide and metal activators (Mn$^{2+}$ or Mg$^{2+}$) and incubating for different times depending on the mismatch.

\subitem (ii) In solution, the extension rate is estimated by measuring the intensities of the bands from polyacrylamide gel electrophoresis (PAGE), densitometry and autoradiography in controlled conditions as discussed above. Two bands appear: the first one $I_1$ corresponds to the fraction of primer extended by one nucleotide in time $t$; the second $I_2$ the fraction extended by two nucleotides in the same time $t$. The velocity to go from state $1$ to $2$ is $v=v_{01}I_2/I_1$ with $v_{01}=(I_1+I_2)/t$ the velocity of extension by one nucleotide \cite{randall_nucleotide_1987,mendelman_base_1990}. The stalling factor is defined as the ratio of extension rates from a mismatched versus properly matched terminus and is possibly different for each kind of mismatch introduced \cite{esteban_fidelity_1993}. It is found that the rate of mismatch extension is considerably reduced with respect to extension of a matched construct, giving rise to large stalling factors; details about DNA and RNA polymerases, non-enzymatic replication of RNA and ribozyme are discussed in the following. 

    \item \textit{Technique (b)}: \textit{Sequencing:} An indirect, but possibly higher throughput way of inferring stalling is to sequence the population of transcripts during templated replication at a given moment in time. If this time is chosen correctly, many transcripts will not be of full length: the fraction of incomplete transcripts that end in a mismatch relative the number of full transcripts and transcripts that do not end in mismatches can be used to infer stalling factors. Two recent datasets of this type include works by Szostak and Joyce's groups \cite{tjhung_rna_2020,duzdevich_deep_2020,Duzdevich2021-co}.

\end{itemize}

\subsection{Observations of stalling}

\subsubsection{RNA and DNA polymerases}

The characterization of mismatch incorporation and extension in DNA and RNA polymerases is pivotal to the understanding of the consequences of errors and the emergence of fidelity and has been widely addressed in different DNA and RNA polymerases.

Several works across different DNA polymerases characterized the rate of elongation of a mismatch reporting a wider range of values for the stalling factor, from 10 to $10^6$ \cite{joyce_reactions_1992,johnson_structures_2004,petruska_comparison_1988,huang_extension_1992,esteban_fidelity_1993,perrino_differential_1989,thomas_transcriptional_1998,mendelman_base_1990,weiss_interaction_1992,beard_influence_2004,baranovskiy_structural_2022,sydow_structural_2009}. The seminal work by Salas focused on family B $\phi29$ polymerase, discovered by her group and Luis Blanco's \cite{blanco_relating_1996,de_vega_primerterminus_1996,salas_my_2012,esteban_fidelity_1993,Kamtekar2004-li}~---~more about this polymerase is discussed in Section \ref{sec:phi29-changliu}. The $\phi29$ polymerase can use two different molecules as primers to catalyze nucleotide incorporation: the initiation of replication is protein-primed and the hydroxyl group comes from the terminal protein attached to $\phi29$, but during the following DNA polymerization a DNA primer serves as donor of free 3'-OH groups. The two reactions display very different stalling factors: $10^4-10^6$ for DNA polymerization and $2-6$ for protein-primed initiation \cite{esteban_fidelity_1993}. This result, combined with the low insertion discrimination of protein-primed initiation and inability of the 3'-5' exonuclease activity to cleave the mismatch, highlights the peculiarity of protein-primed initiation and the likelihood of inaccurate products of this reaction. It also confirms the large stalling factors of DNA polymerization found in other families. 

RNA polymerases have also been shown to have low efficiency of extending a mismatch, with a mismatch-specific stalling factor in the range $10^2-10^3$ \cite{sydow_structural_2009,thomas_transcriptional_1998}. In RNA polymerase (pol) II the mismatch stabilizes fraying of the RNA 3' end causing the pausing of the polymerase, which likely prepares the ground for backtracking during proofreading \cite{shaevitz_backtracking_2003}.

Structural studies are necessary to probe the mechanistic causes of stalling by looking at the effects of mismatch incorporation. As the stalling factors vary for every mismatch pair, so their effect on the structure of the primer-enzyme complex. Catalytic activity of BF polymerase (family A) crystals allowed to investigate the structure of all mismatch pairs and four mechanisms that lead to mismatch-induced stalling of the polymerase have been found \cite{johnson_structures_2004}. The purified protein is crystallized and polymerization undergoes in the crystal. Mismatch extension is controlled in a similar way as in the solution experiment described above, the structure is then measured via X-ray diffraction at different steps of the extension. The structural perturbation of the mismatch is transmitted up to six base pairs from the primer terminus back to the active site of the polymerase, suggesting a short-term memory effect of errors in this DNA polymerase. Similarly, the structural effects of incorporating one mismatch (T-U) in RNA polymerase pol II show that the mismatch displays a wobble base pair that triggers misalignment of the nucleophilic RNA 3' end (e.g. 3' hydroxyl group) with the catalytic site and NTP, leading to a non-optimal geometry for catalysis.

\subsubsection{Ribosomes}
Ribosomal stalling has diverse causes and has been extensively studied given its role in triggering key processes like mRNA decay, nascent protein degradation and ribosome recycling \cite{chandrasekaran_mechanism_2019,Joazeiro2019-wq}. The diverse causes include the presence of rare codons that slow down the translation due to limited availability of corresponding tRNAs, presence of stable mRNA secondary structures, and certain peptide sequences, such as polyproline stretches, that are not easily accommodated in the ribosome exit tunnel \cite{lassak_stall_2016}.  These stalling events can trigger numerous downstream events like ribosome-associated quality control and stress response pathways \cite{Joazeiro2019-wq, snieckute_ribosome_2022, yan_ribosome_2021}. These broad classes of ribosomal stalling processes could potentially have implications for speed and fidelity. 

The most direct analog of the models considered in this paper is stalling due to misincorporated amino acids, i.e. incorporation of an amino acid that does not match the codon on the mRNA. Note that such errors are due to incorrect tRNA whose anticodon does not match the mRNA codon; but the tRNAs are themselves charged with the correct amino acid with respect to their anticodon. These errors can disrupt the progression of the ribosome in multiple ways~---~much like with DNA and RNA elongation, a non-optimal fit in the ribosome peptidyl transferase center, can cause delays in peptide bond formation. In addition, incorrect amino acids can result in the formation of abnormal peptide sequences that generate problematic secondary structures within the ribosome exit tunnel. Any of these structural aberrations can impede the ribosome movement along the mRNA, resulting in stalling \cite{doma_endonucleolytic_2006, joazeiro_ribosomal_2017}.

\subsubsection{RNA ribozyme}
Several works \cite{johnston_rna-catalyzed_2001,attwater_-ice_2013,tjhung_rna_2020} in the origin of life context investigated the ability of class I RNA polymerase ribozymes to synthesize RNA strands approaching sizes of their same length. Such ability is a prerequisite for RNA self-replication, the basis of a primordial RNA world. The observation of shorter strands not elongated further in the given incubation time brought the focus on understanding the causes of polymerization interruption, mismatch extension being a possible culprit. The products of the polymerization are analyzed via PAGE and Sanger or Illumina sequencing. The results from older class I RNA polymerase ribozymes from \cite{johnston_rna-catalyzed_2001,attwater_-ice_2013} show the evidence of stalling when mismatches are incorporated. Most error types in \cite{attwater_-ice_2013} can be further extended, however, a couple of the rarer errors, such as G$\to$U transversions, seem to effectively terminate ribozyme-catalyzed extension, and indeed no such errors are found to occur within sequences. Moreover, even if the substitution rate of the last two bases is 7.1\

\subsubsection{Non-enzymatic replication of RNA}
Stalling upon mismatch incorporation has been observed also in non-enzymatic replication of RNA in the context of the origin of life. The 3' hydroxyl group of the mismatched nucleotide is not aligned for proper extension to occur at the next step. This causes a slowdown by more than 2 order of magnitudes relative to a matched extension, i.e. a stalling factor in the range of $10-300$, as shown by Chen et al. by measuring the rate of primer extension when each nucleotide is provided for incorporation at the 3' end \cite{rajamani_effect_2010}. Mismatch incorporation is also found to cause another error to be introduced at the following site 54-75\

\subsection{Alternate models of stalling} 

Our model of stalling in Sec. \ref{sec:stallingmodels} is the simplest model of combining stalling with proofreading models that is consistent with the experimental evidence presented in Fig. 3 and discussed in Sec. \ref{sec:phi29-changliu}. However, there are several variants of these models, some of which also lead to the same prediction as here (``faster is more accurate''), while others predict the other conclusion (``faster is less accurate'').  

Here, we present a brief outline of some of these alternative possibilities for completeness; detailed investigations of these models and experimental support for their specific mechanism of proofreading and stalling is left to future work. Across diverse models, the relationship between speed and accuracy boils down to two key factors:
\begin{enumerate}
    \item Time cost of correcting errors (e.g., proofreading costs a time $t_{nostall}$).
    \item Time cost of making errors (e.g., time spent stalled). 
\end{enumerate}

In general, if the time cost of correcting errors is lower than time cost of making errors, we find a ``faster is more accurate’’ relationship. If not, a ``slower is more accurate’’ relationship holds. The models listed below (and others) provide different definitions for these two quantities as a function of the error rate and thus potentially distinct conclusions on the trade-off between time and error.

\subsubsection{1. Base excision after stalling}

In the simple stalling model used in this paper, we effectively assume that the polymerase cannot exploit the stall time after a misincorporation to backtrack and excise the last base. Such a restriction would be accurate, e.g., for enzymes that are unable to excise the (incorrect) base at site $i$ after having translocated to site $i+1$ and are stalled while trying to catalyze a new (correct) base at site $i+1$. However, backtracking during a stall is thought to occur in several cases in RNA polymerases \cite{shaevitz_backtracking_2003} and in DNA polymerase stalling is linked to a higher likelihood of transition to the exonuclease domain \cite{ibarra_proofreading_2009}.

Allowing for excision during stalling strengthens our results by expanding the counterintuitive ``faster is more accurate'' regime to lower error rates than predicted in our simple model. 

Consider a polymerase, that after incorporating a nucleotide at site $i$, can either pass the strand to the exonuclease domain or translocate to site $i+1$ and process the next nucleotide. In a simple scenario, we can model the rate of attempted exonuclease use by a rate $k_{exo}$, while the latter is modeled by a kinetic constant $k_{fwd}(X)$ that depends on the current nucleotide $X$. The forward constant for the wrong nucleotide W is much slower than for the right nucleotide R, $k_{fwd}(W) \ll k_{fwd}(R)$ due to stalling. To simplify notation, we define $k_{stall} \equiv k_{fwd}(W)$. Note that in this simple model, we are not distinguishing between stalling of translocation to site $i+1$ vs stalling of catalysis at site $i+1$.

The probability of using the exonuclease on a misincorporated nucleotide $W$ is $$p_{exo} = \frac{k_{exo}}{k_{exo} + k_{stall}}.$$

where $k_{stall}$ is the analog of $ 1/\tau_{stall}$ in the model detailed earlier. If we define $\mu_{pre-exo}$ to be the rate at which the polymerase domain by itself incorporates incorrect nucleotides~---~without accounting for error reduction due to the exonuclease~---~the actual error rate is, 
$$ \mu = \mu_{pre-exo} (1-p_{exo}) \approx \mu_{pre-exo} \frac{k_{stall}}{k_{exo}}$$ where we assume that $k_{stall} \ll k_{exo}$.

We can compute the time to copy a strand of length $L$ as the sum of two times: (a) time spent on misincorporations removed by the exonuclease; there are $L \mu_{pre-exo} p_{exo}$ such misincorporations and each of these misincorporations takes a time $1/k_{exo}$ to be removed. 
(b) time spent on misincorporations that are not removed by the exonuclease; there are $L \mu_{pre-exo} (1 - p_{exo})$ such misincorporations and it takes $1/k_{stall}$ time to get past each of these misincorporations, which are not proofread. 

Thus, the total time is given by $$ T_{rep} = L \mu_{pre-exo} \frac{1}{k_{exo}} \left( 2 - \frac{k_{stall}}{k_{exo}} \right) $$ where we assume that $k_{stall} \ll k_{exo}$. 

These equations predict a counterintuitve ``faster is more accurate’’ relationship as $k_{exo}$ is changed among variants. See the following section, \textit{Mutability of stalling}, for mutations that might change $k_{stall} = 1/\tau_{stall}$.

However, compared to the simpler model presented earlier, this one predicts the counterintuitive relationship down to lower error rates. To see this, note that the time wasted due to errors is no longer proportional to the actual error rate $\mu$, measured e.g. by sequencing, but is instead proportional to $\mu_{pre-exo}$, the frequency at which the polymerase domain misincorporates nucleotides. This frequency can be significantly higher than the real error rate, i.e. $\mu_{pre-exo} \gg \mu$. As a consequence, this model predicts a counterintuitive trade-off to significantly lower error rates $\mu$ than in the simpler model analyzed earlier and thus is more relevant to the data in Fig. 3 where the error rate $\mu \sim 10^{-9}$ can be much smaller than the length of the genome copied. 

\subsubsection{2. Mutability of stalling}

In the theoretical analyses so far, we assumed that mutations cannot reduce the stalling effect itself, i.e. we assume that mutations do not change $\tau_{stall}$. 

In principle, mutations could reduce the stall time $\tau_{stall}$, i.e. produce polymerases that avoid stalling by having a more permissive active site as seen in trans-lesion repair polymerases (family Y) \cite{ling_crystal_2001,goodman_translesion_2013}. If such mutations were permitted and stalling itself could be strongly alleviated, we expect our results to no longer hold and a ``faster is less accurate'' result to hold instead. 

However, the data on pGKL1 mutagenesis shown in Fig. 3 can be seen as evidence that stalling is not easily alleviated through mutations, at least in the sequence neighborhood of functional highly processive polymerases. Polymerases such as trans-lesion repair polymerases with lower stalling appear to be strongly compromised in terms of both processivity and accuracy and are used only for filling in short stretches of DNA. Finally, the fact that stalling is observed in non-protein catalyzed, i.e. ribozyme-based, replication and also in templated replication without any enzymes (protein or RNA) at all suggests that stalling is a relatively robust property intrinsic to templated replication. Thus, while changing the extent of stalling is possible, the ``faster is more accurate'' tradeoff described here is likely relevant at least in the vicinity of functional processive polymerases in sequence space.

Finally, we note that the existence of mutations that reduce stalling can still be compatible with the observed Pareto front in Fig. 3 as long as such mutations have other effects, i.e. pleiotropic, that prevent them from occupying the Pareto front. More quantitatively, in the \textit{Base excision after stalling} mode above, $p_{exo} \sim k_{exo} \tau_{stall}$. If mutations that reduce stalling also have other deleterious impacts, e.g. the kind of reduced processivity seen in trans-lesion family Y polymerases, then these variants would be away from the Pareto front. On the other hand, mutations that change exonuclease activity $k_{exo}$ would populate the Pareto front and thus such a model would still be consistent with the counterintuitive ``faster is more accurate'' in Fig. 3.

\subsubsection{3. Correcting errors takes time} 

For completeness, we discuss an alternative model for stalling where the delay is the time taken for the exonuclease to cleave a misincorporated nucleotide, i.e. if correcting errors takes time. After passing the elongating strand to the exonuclease domain, governed by a rate $k_{exo}$, cleaving the last nucleotide could incur a time cost $\tau_{stall}$. 

While superficially similar to the model discussed in this paper, this model predicts the \emph{intuitive} trade-off between speed and accuracy, i.e., ``faster is less accurate'', in tension with the pGKL1 DNA-polymerase data presented in the paper. To see this, note that the stalling time cost is now proportional to the frequency of using the exonuclease domain. Mutations that reduce the frequency of using the exonuclease will lead to faster replication and also higher error rate, i.e. ``faster is less accurate''.

\subsubsection{4. Incorporation of wrong bases takes time} 

An alternative model of stalling is that incorporating a wrong base in the elongating polymer is slow. Even though the extensive experimental data described earlier shows that stalling is a ``prior base is incorrect'' effect instead, we discuss this model here for completeness.

Consider two distinct models of proofreading. In the exonuclease-based proofreading found in DNA polymerases, incorrect bases are first catalytically incorporated into the elongating strand, i.e. the phosphodiester bond is formed, before the strand is potentially handed to the exonuclease for cleaving. This model would predict an intuitive ``faster is less accurate'' trade-off since the  total stalling time is not reduced by reducing (sequenced) error rates through proofreading activity. 

Now consider an alternative model where incorrect bases are proofread without first incorporating them into the elongating strand. In this hypothetical model, ``incorporating wrong bases take time'' would indeed predict our counterintuitive ``faster is more accurate'' trade-off. However, known proofreading polymerases do not operate this way. 

\subsection{Evolution of stalling}

Is stalling an inevitable biophysical constraint or is stalling an evolved property? The evidence detailed above suggests that stalling is inevitable in templated replication at least to some extent, since non-enzymatic replication also displays stalling \cite{rajamani_effect_2010}. 

However, since enzymes can modulate stalling to some extent, we can ask which are the conditions that would select for stalling. One straightforward answer is that stalling evolved to mechanistically enable kinetic proofreading. While this is a reasonable possibility, note that, in principle, proofreading does not require stalling since several non-processive systems, such as tRNA synthetase and T-cell receptors, can proofread even though the stalling framework does not apply to them, since each ligand is processed separately. 

Here, we discuss an alternative possibility of selection for stalling for reasons of `evolvability'. In this second-order selection \cite{good_impact_2015} for stalling, stalling is not assumed to be mechanistically necessary for kinetic proofreading. Instead, stalling is selected because lineages with stalling are able to evolve error correction without a valley crossing and thus realize greater future fitness benefits unlike lineages without stalling. We describe this scenario with a thought experiment in the following.

Model: We consider two lineages, A and B. A and B are identical in all respects, except that lineage A's replication chemistry involves no stalling while lineage B's chemistry involves stalling of some fixed extent.    

Both lineages are assumed to have a genome $g$ of length  $1+ L_c + L_{nc}$; for simplicity, we assume that each base of this genome can be either $0$ or $1$. The first bit of the genome dictates the error rate; $0$ results in a high error rate $\mu_H$ while $1$ results in a low error rate $\mu_L$ due to an error correction mechanism à la kinetic proofreading. 

We assume a simple fitness function that decays with the Hamming distance of the genome $g$ from $000\ldots 00$. We assume that $f(g) = f_c(g_c) + f_{nc}(g_{nc})$ where $g_c$ and $g_{nc}$ are the $L_c$ and $L_{nc}$ bits of the genome. Further, $f_c(g_c) = 1$ if $|g_c|=0$ and $f_c(g_c) = -\infty$ otherwise, and $f_{nc}(g_{nc}) = 1/(1 + \kappa |g_nc|)$. Here, $|g|$ is defined to be the number of $1$s in the genome.

Thus, the first $L_c$ bits of the genome, $g_c$, model a deleterious load since any mutation away from $00\ldots 0$ is lethal. The last $L_{nc}$ bits have a more forgiving fitness function and represent a non-coding region. High error rate $\mu_H$ strains are unable to exploit this region for fitness gains if $\mu_H L_{nc}$ is larger than $\kappa$; but low error rate $\mu_L$ strains can increase their fitness by evolving $g_{nc}$ to be closer to $000\ldots 0$. 

In such a model, consider adaptive paths towards increased long term fitness that start with a wild-type strain with high error rate $\mu_H$. This strain then evolves an error correcting mechanism, reducing the error rate to $\mu_L$. Subsequently, the non-coding region $g_{nc}$ is able to evolve towards the highly fit sequence $00\ldots 0$ and thus gain additional fitness benefits. Note that these benefits cannot be realized without an error correction mechanism since $\mu_H L_{nc}$ is too high. While the above path is available to both lineages, the fitness landscape along this path is qualitatively different. 

In lineage A, the evolution of error correction comes at a fitness cost $f_{speed} - f_{del} > 0$ where $f_{speed}$ is the fitness cost due to reduced speed in an error-correcting variant with lower error rate $\mu_L$ and $f_{del} = (\mu_H - \mu_L) L_c$ is the fitness gain due to the reduction in deleterious load. Subsequent mutations in $g_{nc}$ increase the fitness of this lineage. 

In contrast, in lineage B, the evolution of error correction comes at a fitness gain $f_{speed} + f_{del} > 0$ relative to the wild-type, where $f_{speed}$ is the fitness gain due to increased speed in an error-correcting variant with lower error rate $\mu_L$ and $f_{del} = (\mu_H - \mu_L) L_c$ is the fitness gain due to the reduction in deleterious load. Note that all genotypes in lineage B have lower fitness than the corresponding genotypes in lineage A because of stalling. 

However, the path to evolving an error correction mechanism and subsequent higher fitness for lineage B is easier, i.e. no valley crossing is needed. Thus, lineage B can outcompete lineage A in the long term, provided B is not outcompeted at early times, e.g. if protected by spatial structure or large population sizes. In this sense, selection for evolvability, i.e. long term fitness benefits, can select for stalling.

\section{Speed-accuracy measurements for hundreds of variants of a DNA polymerase} 
\label{sec:phi29-changliu}

Family B polymerases are involved in both DNA replication and repair, are found in eukaryotes and prokaryotes and display $3'-5'$ exonuclease proofreading. The most used polymerases in DNA amplification by PCR are from family A and B, and the most accurate ones, e.g. Q5 HF DNA polymerase (proprietary), are part of family B. Given the interest in the origin of accuracy, members of this particular family have been extensively studied. 

\begin{figure*}[h!!!]
    \centering
    \includegraphics[scale=1]{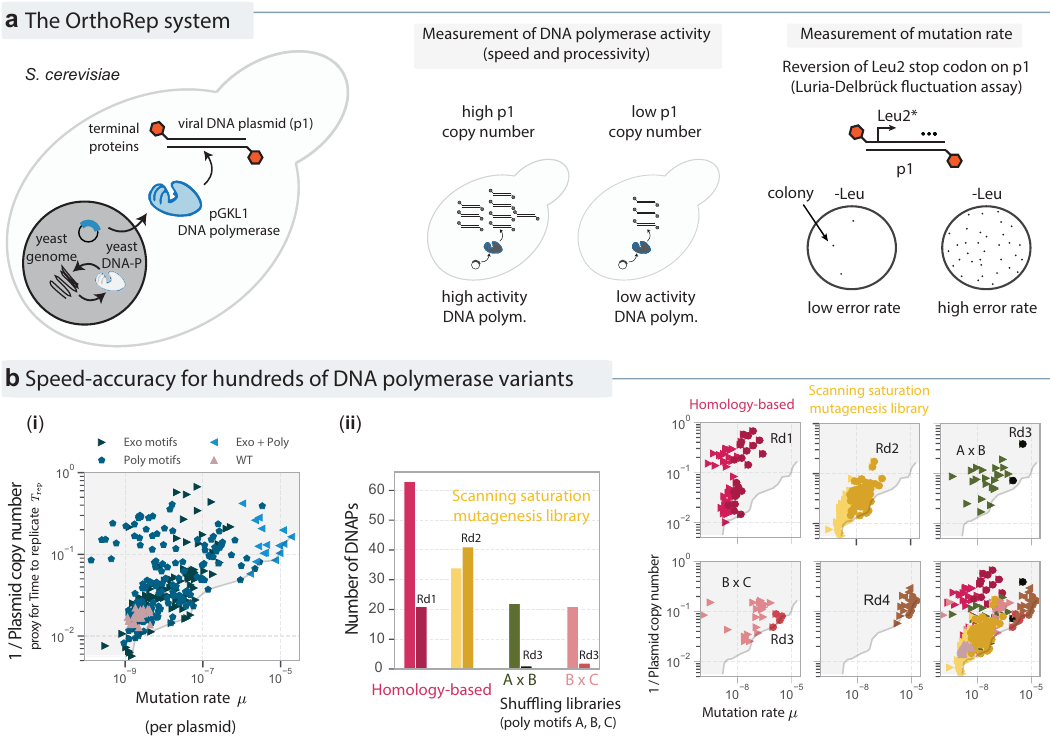}
    \caption{\textbf{The largest dataset to date on speed and accuracy of DNA polymerase variants is provided by the OrthoRep platform.} (\textbf{a}) Sketch of the OrthoRep system \cite{ravikumar_orthogonal_2014,ravikumar_scalable_2018,rix_scalable_2020} in \textit{S. cerevisiae}, where a  cytosolic, multicopy, linear DNA plasmid (`p1') of viral origin is shown. Plasmid p1 can only be copied by a dedicated DNA polymerase (pGKL1) (blue) and not by the yeast machinery due to presence of terminal proteins on p1. The orthogonality of the replication system of p1 and the yeast genome allows for measurement of activity and accuracy of the pGKL1 polymerase. Activity is measured by determining the copy number of p1 at steady state.  Accuracy is measured through a Luria-Delbruck fluctuation assay, based on the reversion of a stop codon in Leu2 gene placed on p1 in a Leu-auxotroph strain. (\textbf{b}) (i) Data plotted from Table S2 of \cite{ravikumar_scalable_2018} with all the DNA polymerases for which p1 copy number and mutation rates have been characterized. Mutations in  different domains is shown using color and shape reported in the legend. (ii) Subdivision of the same data as (i), but according to different libraries constructed in \cite{ravikumar_scalable_2018}: based on mutations in known homologs such as $\phi$29, scanning saturation mutagenesis, shuffling, with the definition of the different rounds used for the discovery of highly-error-prone DNA polymerase variants, with acceptable activity. }
    \label{fig:SI_chang}
\end{figure*}

One example is the $\phi29$ DNA polymerase from bacteriophage $\phi29$, which is characterized by a first step of protein-primed replication followed by DNA polymerization, feature shared only by family B members of viral origin. Salas and collaborators extensively characterized $\phi29$: from its protein-primed initiation \cite{Kamtekar2004-li} to structure-function studies \cite{blanco_relating_1996,longas_functional_2006} building a map from mutations in the sequence, to structural perturbations, to changes in $\phi29$ function, e.g. speed, accuracy and strand-displacement activity~---~ the ability to open the double-stranded DNA in order to extend a primer at a replication fork. The latter is a crucial feature that makes $\phi29$ stands out with respect to other polymerases in the family: strand displacement, combined with the high processivity of $\phi29$ (the intrinsic high processivity is probably due to a region of the polymerase holding the enzyme in place during elongation) allows for complete replication of its genome without any helicase or accessory processivity factors.  Indeed, most replicative polymerases rely on accessory proteins to achieve higher processivity, as the Sso7d DNA binding domain fused into Q5 HF DNA polymerase. 

\subsubsection{Higher speed, higher accuracy for a library of a a $\phi29$-like polymerase}

The most extensive DNA polymerase library to the date of writing has been created for a homolog of $\phi29$, the polymerase that replicates a linear, high-copy, cytoplasmic DNA plasmid, pGKL1 (p1 for brevity), which is originally part of the toxin-antitoxin system of \textit{Kluyveromyces lactis} \cite{stark_killer_1986}. pGKL1 is effectively a vertically transmitted double stranded DNA virus that lives in the cytoplasm of \textit{Kluyveromyces lactis}. The unique mechanism of p1 replication initiation via protein-priming and the spatial separation of p1 and the associated polymerase (TP-DNAP1) from nuclear DNA made this toxin-antitoxin system the basis for OrthoRep \cite{ravikumar_orthogonal_2014,ravikumar_scalable_2018,rix_scalable_2020}, a platform for directed evolution in \textit{S. cerevisiae}. The orthogonal replication of p1 allows for speeding up the evolution of a gene of interest encoded on p1 by modulating the mutation rate of the polymerase without influencing genomic replication in the nucleus. 

The interest in finding error-prone variants of the wild-type TP-DNAP1 for this purpose led to 13625 clones from a scanning saturation mutagenesis library combined to an homology-based library and selected combinatorial libraries, resulting in 213 unique variants for which p1 replication activity and substitution mutation rate have been measured (Fig. \ref{fig:SI_chang}b(i) and Fig. 3 in the main text, data from Table S2 in \cite{ravikumar_scalable_2018}). 

First, let us review how the variants in \cite{ravikumar_scalable_2018} were obtained, see Fig. \ref{fig:SI_chang}b. The Round 1 (Rd1) variants from homology studies mainly gave low activity (slow) polymerases, due to the peculiarities of protein-priming DNA polymerases, like the TP-DNAP1 present in OrthoRep. Hence, the authors \cite{ravikumar_scalable_2018} performed a comprehensive study with the scanning saturation mutagenesis library, identifying 41 unique Rd2 mutants that retained high activity~---~ see Fig. \ref{fig:SI_chang}b(ii). The authors \cite{ravikumar_scalable_2018} also noted that only 9/41 of the Rd 2 polymerases contain mutations at position identified from the homology studies, suggesting that position affecting fidelity can be outside the most-conserved regions. Rd1 and Rd2 variants were then used to generate combinatorial libraries, focusing only on inter-motif combinations between polymerase motifs A, B, C and Exo I, II and III. The shuffling libraries $A\times B$ and $B\times C$ produced 3 unique Rd3 variants, then crossed with all the Exo mutations present in the dataset, leading to 4 unique Rd4 variants, 2/4 having high error rate and high enough activity.

The proxy for the speed (accounting for both the elongation rate and the processivity of the polymerase) of the DNA polymerase is the copy number of p1, as sketched in Fig. \ref{fig:SI_chang}a, which measures how many copies of the plasmid are present in a cell. Intuitively, faster (or potentially more active) polymerases create more copies of p1 before the yeast cell division and thus maintain a large copy number of p1 at steady state. The assay uses a calibration curve relating p1 copy number, determined via quantitative PCR (qPCR), with fluorescence of p1-encoded mKate2: p1 copy number can be then accessed indirectly by higher-throughput fluorescence measurements via flow cytometry.

Mutation rate is measured by the rate of reversion of LEU2 (Q180*) encoded on p1~---~see Fig. \ref{fig:SI_chang}a: LEU2 (Q180*) contains a C$\to$T mutation at base 538 in the Leu2 gene introducing a stop codon. Reversion to the functional gene is detected on selective media lacking leucine and a Luria-Delbr\"uck fluctuation assay \cite{luria_mutations_1943} can be performed to estimate the mutation rate.

We repurposed the data from Table S2 in \cite{ravikumar_scalable_2018} as a window into the proofreading process, and the relationship between speed and accuracy in DNA polymerases. The variants of the wild-type polymerase suggest a counterintuitive trade-off for which the ones with largest copy number are also the most accurate ones. 

\section{Prior work on speed and accuracy trade-offs} 
\label{sec:priorspeedaccuracy}

We begin by pointing out a potential hazard in inferring trade-offs from limited observations, rather than comprehensive mutagenesis. We then review prior theoretical and experimental work on speed and accuracy trade-offs in a diverse range of systems.

\subsection{Inferring trade-offs from comprehensive vs isolated perturbations}

The primary distinction between the counterintuitive result presented here and the intuitive ``faster is less accurate’’ trade-off discussed in some prior studies could stem from the number of variants looked at. Here, we point out a potential hazard in inferring a trade-off from limited data that might partly explain how our ``faster is more accurate’’ results~---~based on $\sim O(100)$ variants of a DNA polymerase~---~could be compatible with prior ``faster is less accurate’’ reports~---~typically based on $\sim O(2)$ variants or perturbations in each system. 

\begin{figure*}[h!!!]
    \centering
    \includegraphics[scale=0.9]{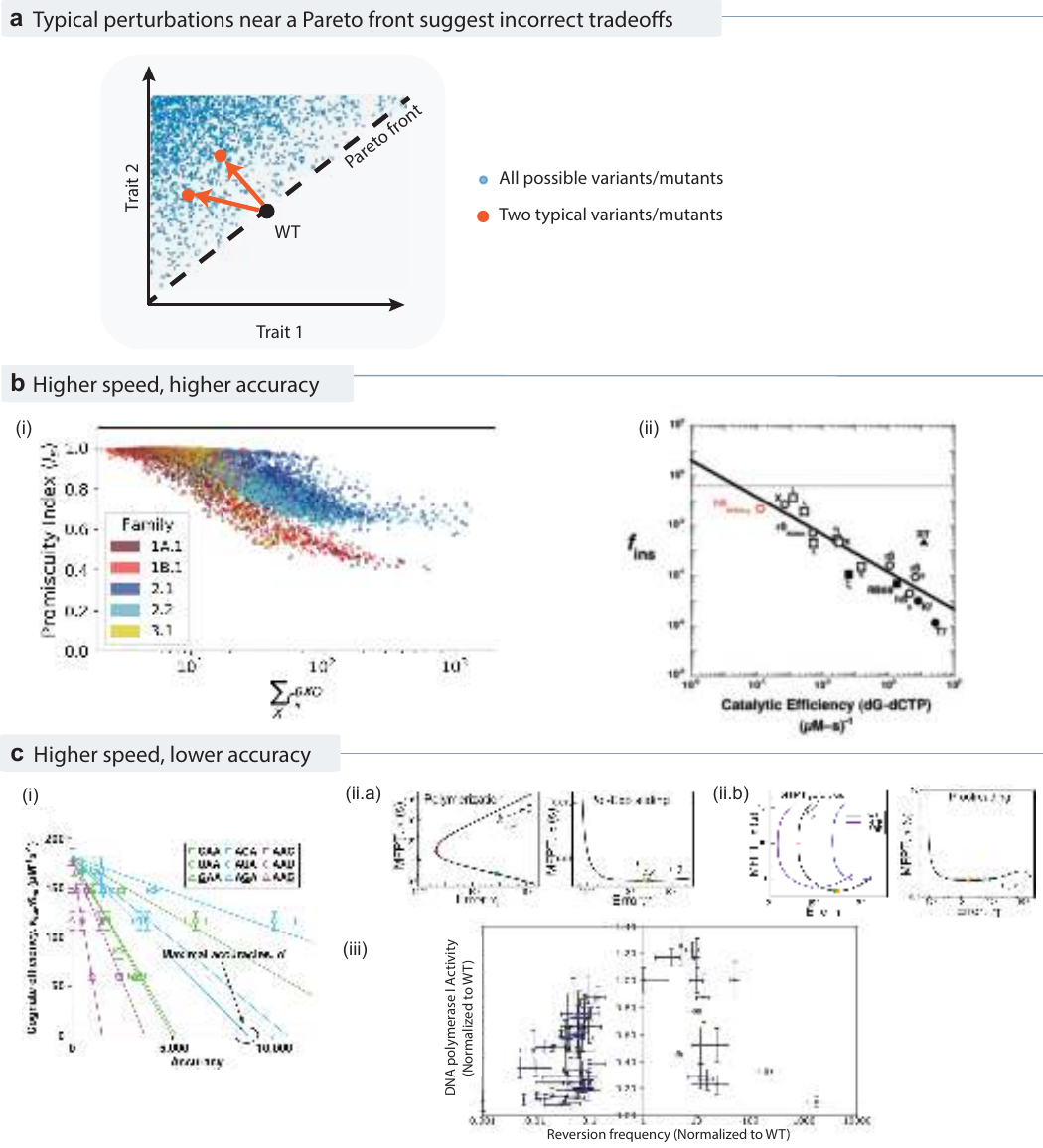}
    \caption{\textbf{Limited data can lead to misleading trade-offs and other existing data on speed-accuracy trade-offs.} (\textbf{a})  A potential hazard in inferring trade-offs from a limited set of perturbations or variants. The sketch shows a scatter plot of two traits over all variants (blue) in a mutational neighborhood of a wild-type (WT) molecule or organism. Together, these variants show a Pareto front, defining a trade-off between the two traits. Generically, the density of variants is expected to be lower in the vicinity of the Pareto front (since each trait is at its maximum with the other trait held fixed). If the wild-type (WT) is near the Pareto front, red points show the perturbation of the WT by two typical mutations, given the distribution of variants in the mutational neighborhood. Red arrows suggest an incorrect trade-off converse to the real trade-off defined by the Pareto front. On the other hand, making all (or many) possible mutations (blue points) will reveal rare perturbations that move the WT along the Pareto front and thus reveal the true trade-off.
    (\textbf{b}) Higher speed, higher accuracy: (i) Trade-off data for ribozymes obtained by comprehensive mutagenesis libraries. Adapted from \cite{janzen_emergent_2022}. (ii) Trade-off data for exonuclease-deficient DNA polymerases across five different families and a mutant of DNA polymerase $\beta$ (in red). Adapted from \cite{beard_efficiency_2002}. (\textbf{c}) Higher speed, lower accuracy: (i) Trade-off data for ribosomes obtained by perturbing the concentration of Mg$^{2+}$. Adapted from \cite{Ehrenberg1980-qz}. (ii) Trade-off data for T7 DNA polymerase (a) and E. coli ribosome (b) obtained by varying one parameter in the model. Adapted from \cite{banerjee_elucidating_2017}. (iii) Trade-off data for mutants of exonuclease-deficient DNA polymerase \textit{pol A} obtained by error-prone PCR. Adapted from \cite{loh_highly_2007}.}
    \label{fig:SI_reftradeoff}
\end{figure*}

Fig.~\ref{fig:SI_reftradeoff}a shows a hypothetical trade-off between two traits: the blue points represent all possible variants in a mutational neighborhood, e.g., obtained by making all possible mutations, or other relevant perturbations. The black line shows the Pareto front, i.e. points that are not entirely dominated in both traits by another point.  Consider the case of a wild-type (WT) that is located at, or near, the Pareto front. If this WT is perturbed by one or two select mutations~---~or environmental conditions such as the concentration of Mg$^{2+}$ ions~---~and we examine the impact on trait 1 and 2 (e.g. speed and accuracy), we will typically obtain variants away from the Pareto front. This bias arises because, generically, there are relatively few ways to perturb a system and stay on the Pareto front, but many ways to break something and be away from the Pareto front. As a consequence, making a generic perturbation will lead to inferring the exact opposite trade-off of the actual trade-off between the two traits. Instead, one could perform a comprehensive examination of all relevant perturbations and populate a full scatter plot. In this case, we can ensure that rare perturbations that move the system along the Pareto front are part of the study and thus reveal the real trade-off.

Finally, the problem illustrated in Fig.~\ref{fig:SI_reftradeoff}a raises the question of what all possible perturbations entails. The answer depends on why one is interested in establishing the existence, or not, of a trade-off. Our interest here is to determine how the biophysics of polymerases constraints or enables evolution, e.g. could proofreading evolve from a need for speed? Does the evolution of proofreading come at a cost in speed or a benefit in speed?  These questions are impacted by the rare mutations that populate the Pareto front. For example, even if most mutations have an impact of the type shown in red in Fig.~\ref{fig:SI_reftradeoff}a, adaptive evolution can select the rare mutations along the Pareto front if the WT shown is pressured to evolve to lower values of trait 1 and thus evolve to lower values of trait 2. Hence, experiments of the type shown in Fig. 3~---~or other comprehensive deep mutational scans~---~that examine evolutionarily accessible mutations, e.g. all mutations within a mutational neighborhood, are needed to establish evolutionarily-relevant trade-offs. 

\subsection{Prior reports of higher speed, higher accuracy}

To our knowledge, only a few other works in the literature discuss the possibility of a counterintuitive trade-off, with faster enzymes also being more accurate. 

One work that posits the possibility of the evolution of accuracy due to a need for speed is a recent paper \cite{pressman_mapping_2019} from Irene Chen's group on speed and specificity of aminoacylation activity of a ribozyme.  In \cite{pressman_mapping_2019},  a library of $10^4-10^5$ ribozyme variants were selected for displaying aminoacylation activity, through an exhaustive search of a central 21 nucleotides region. Three distinct, evolutionarily unrelated catalytic motifs have been identified from this exhaustive search. In a later work \cite{janzen_emergent_2022}, Chen et al. characterized ribozyme catalytic activity and specificity of all single- and double-mutants related to five wild-type ribozymes, each from a different familiy, against six alternative substrates. Each family contains one motif of the three identified previously: motif 1 in families 1A.1 and 1B.1, motif 2 in 2.1 and 2.2, motif 3 in 3.1. The goal is to address the evolutionary potential of self-aminoacylating ribozymes to adopt new amino acid substrates. The results show that across all families specificity increases as overall activity increases, see Fig. \ref{fig:SI_reftradeoff}a. Activity is measured indirectly by fitting a kinetic model to the reacted fraction obtained by sequencing ribozymes reacted with substrate at multiple concentrations. Specificity is measured in two ways, giving similar results: (i) the ability of a sequence to react with multiple substrates at similar activity, (ii) the preference to aromatic amino acids, since some families seem to distinguish aromatic vs non-aromatic side chains by displaying preferential activity with aromatic amino acids. The correlation found is proposed to lead to a scenario where selection for greater activity would also lead to error minimization, as a by-product. Note that these ribozymes are not thought to be processive; hence they do not immediately fit into the modeling framework of this paper.

Beard et al. \cite{beard_efficiency_2002, beard_structural_2003, beard_structure_2014} investigated the relationship between the misinsertion efficiency of a particular mismatch (e.g. dG-dTTP) $(k_{cat}/K_M)^W/(k_{cat}/K_M)^R$, a proxy for mutation rate, and the catalytic efficiency for correct nucleotide insertion $(k_{cat}/K_M)^R$ (e.g. dG-dCTP) in exonuclease-deficient polymerases across five families (A, B, RT, X and Y) and a mutant of DNA polymerase $\beta$, see Fig. \ref{fig:SI_reftradeoff}b(ii). A correlation suggesting ``faster is more accurate' holds when the catalytic efficiency of inserting the correct nucleotide is shown as function of the misinsertion efficiency, e.g. guanosine and cytosine (dG-dCTP).  No strong correlation holds if the catalytic efficiency for mismatch insertion is instead considered. This shows that accuracy correlates with the efficiency of insertion of the right nucleotide, but not of the wrong one. These results suggest that the exonuclease-deficient polymerases that are faster at incorporating the correct nucleotide are also more accurate, while being equally efficient at incorporating the incorrect nucleotide. By doing so, possibly consistent with our framework, they avoid the stalling caused by mismatch incorporation. Note that these studies define speed as elongation rate, i.e. at how fast a correct nucleotide is inserted, without focusing on the processivity of the polymerases, which can be low for exonuclease-deficient polymerases like the ones characterized, e.g. family Y.

On the theoretical side, in a review about speed-accuracy trade-offs in enzymes, Tawfik \cite{tawfik_accuracy-rate_2014} shows that Michealis-Menten kinetics when in the transition-state discrimination leads to the possibility of a counterintuitive trade-off where faster enzymes are more accurate. This is contrasted to a ground-discrimination scenario where the intuitive trade-off holds.

\subsection{Prior reports of higher speed, lower accuracy} 

We report some references where the intuitive correlation between speed and accuracy has been found, where faster enzymes are more error prone. 

Ehrenberg et al. \cite{Ehrenberg1980-qz} measured in vitro the cognate efficiency in terms of Michealis-Menten parameters ($k_{cat}/K_M$) as function of accuracy ($(k_{cat}/K_M)^R/(k_{cat}/K_M)^W$) in a ternary complex consisting of aa-tRNA, elongation factor Tu (EF-Tu) and GTP as a model system for genetic code translation, and in particular of the peptide elongation cycle of the ribosome. The accuracy can be varied by modulating the levels of Mg$^{2+}$ ions, hence the variation explored is only along this one parameter. The accuracy is then measured against the cognate ligand (AAA) for ten values of Mg$^{2+}$ and sparsely in Mg$^{2+}$ space against non-cognate ligands (9 possible combinations), for a total of 30 points. The resulting measurements  each fall on a line that follows a speed-accuracy correlation, see Fig. \ref{fig:SI_reftradeoff}c(i): the faster the translation for a given mismatch, the less accurate. 

Kolomeisky, Igoshin et al. have been interested in how proofreading enzymes balance the opposite objectives of speed and accuracy, probing whether perturbations of kinetic parameters can improve both traits. They built kinetic proofreading models using experimentally measured kinetic rates, allowing to define accuracy and speed from the models. Relevant examples of this line of work are:
\begin{itemize}
\item In \cite{banerjee_elucidating_2017}, they use one-loop kinetic-proofreading networks to control the rate constants for each step of the right and wrong pathways and use kinetic constant data as starting points to explore how changes in the kinetic parameters affect speed and accuracy according to the assumed model. The speed is the defined from the model as the mean first passage time to cross the barrier between reactants and products. The experimental measurements of kinetic parameters are in family A T7 DNA polymerase (1 wild-type, Fig. \ref{fig:SI_reftradeoff}c(ii.a)) and protein synthesis by \textit{E. coli} ribosome (1 wild-type and 2 mutants, Fig. \ref{fig:SI_reftradeoff}c(ii.b)). They find two branches in the speed-accuracy plots: one is the trade-off branch, with higher accuracy corresponding to lower speed, the other instead originates from some kinetic parameters being in a sub-optimal region where a perturbations improves both traits, or improves accuracy with almost no effect on speed. Such sub-optimal regions manifest e.g. if the rate of handling the nucleotide to the cleaving exonuclease domain is higher than the polymerization rate of the correct base, since futile proofreading cycles diminish both speed and accuracy. 

\item In another work Kolomeisky, Igoshin et al. \cite{mallory_theoretical_2021} used the same kinetic proofreading model to investigate speed-accuracy-cost trade-offs in the genomic replication of one coronavirus with proofreading activity (SARS-CoV RdRp complex), for which the kinetic parameters have been measured. They define speed from the model as the stationary flux to create the right product, or in other words, the rate of production of functional copies, i.e. without any deleterious mutations. When plotted against error rate for different values of the rate of switching from the polymerase to the exonuclease domain, the speed has a non-monotonic behavior: it increases as error rate increases (trade-off branch), but then decreases when the rate of mutation is high enough that deleterious mutations become more likely (non-trade-off branch).

\item Finally, in \cite{midha_synergy_2023}, they considered a model of proofreading including stalling effects, backtracking and dependence on accessory proteins for RNA polymerases. They used experimental data from RNA polymerase Pol II to initialize the model and explore how changing its parameters affects the accuracy-speed trade-off. The parameters are varied one at a time, concluding that some perturbations lead to the intuitive trade-off, while others showed no trade-off. Extensive mutagenesis of such RNA polymerases, rather than only varying model parameters from an experimentally determined starting point, is needed to reveal the nature of the underlying Pareto front relevant to answer the evolutionary question of whether higher accuracy can evolve from selection for higher speed. The finding of multiple trade-off branches is indeed indicative that this more extensive exploration is needed.

\end{itemize}

Loeb et al. \cite{loh_highly_2007} investigate how accuracy can be increased without resorting to an increase of the $3'-5'$ exonuclease proofreading intrinsic to family A polymerases that generally leads to lower activity. They build a library of $10^4$ mutants via error-prone PCR of exclusively the polymerase domain of the \textit{pol A} gene (expressing polymerase pol I), starting from a sequence where the exonuclease domain is inactivated by a single substitution. We define this as exo-deficient wild type (exo$^*$-WT). The mean amino acid substitution measured in the library is 4.2 per gene. The mutants were assayed for function in \textit{E. coli} and 408 clones from the surviving cells, with a decreased mean of 2.8 amino acid mutations, were screened for polymerase mutation rate and activity. Mutation rate is estimated by reversion of a stop codon, while activity by in-vitro measurements of incorporation of radiolabeled deoxythymidine monophosphate (dTMP) into DNA. The mutation rate of pol I variants spans a large range, $10^{-3}-10^3$-fold the exo$^*$-WT. The mutations affecting accuracy are reported to be spread across the polymerase domain. A mutation in the M-helix of pol I is present in all higher accuracy variants with additional substitutions elsewhere that are shown to increase polymerase base selectivity with an additive effect. When activity and mutation rate are plotted, no clear overall trade-off front is seen, see Fig. \ref{fig:SI_reftradeoff}c(iii): the more accurate variants (12\

Marx et al. \cite{summerer_enhanced_2005} built a library of 1316 variants of the $3'-5'$ exonuclease-deficient Klenov fragment (KF$^-$) of E. coli DNA polymerase by randomizing only its motif C (three consensus residues). The accuracy of extension is measured by the ratios of fluorescence of primer extensions from matched versus mismatched substrates, while the efficiency by radiometry and kinetics measurements. The three most accurate variants were found to extend mismatches with significantly less efficiency than the wild type KF$^-$, while keeping the same efficiency in extending the matched primer termini, suggesting the slower-more accurate trend.

In summary, as pointed out in the introduction to this section, a search over all parameters is necessary to define the Pareto front. The studies reported here, and other not discussed, e.g. \cite{pigolotti_protocols_2016,yu_trade-offs_2020,mohr_individual_2023}, report only a few variants, or consider only the impact of perturbing one or few environmental parameters. e.g. the concentration of Mg$^{2+}$. The emergence of two branches of the trade-off is also indicative that a more exhaustive exploration, like the one in Fig. 3, is needed to identify the Pareto front. Also, in these works, the measurements of speed do not take into account processivity, the rate of the DNA polymerase detaching from the strand, in contrast with the assay discussed in Sec. \ref{sec:phi29-changliu} \cite{ravikumar_scalable_2018}.

\chapter{Dynamic instability in Self-assembly}
\label{ch:sa}

\begin{figure}
    \centering
    \includegraphics[width=\textwidth]{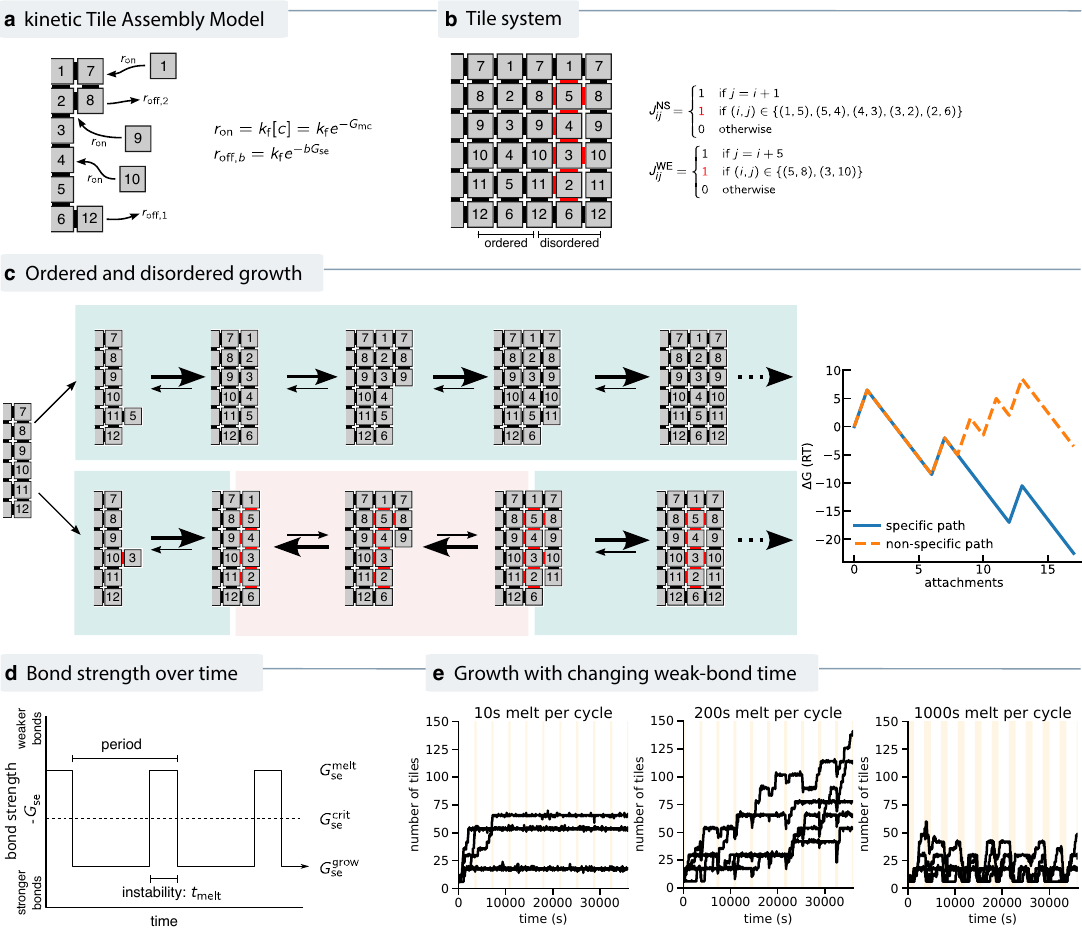}
    \caption{\textbf{Kinetic model of self-assembly and dynamic instability} \figp{a} In the kinetic Tile Assembly Model (kTAM), tiles attach to locations based on the tile concentration $[c]$ (or equivalently, free energy $\gmc$), and detach based on the total strength of interactions $b$ times a (sign-reversed, unitless) unit free energy $\gse$. \figp{b} A ribbon tile system with non-specific interactions exhibiting stalling and errors.  Specific interactions (black) result in fully-connected alternating rows of tiles 1--6 and 7--12.  But non-specific interactions (red) allow a disordered row of tiles 1--6, from which growth cannot continue without incorporating partially-connected tiles. \figp{c} Example pathways for ordered (top) and disordered (bottom) growth.  Both initially grow by primarily favorable, forward-biased steps (green background), but while the ordered pathway can continue favorably, several strongly reverse-biased steps are required for the disordered pathway to continue. \figp{d} Dynamic instability modeled as periodic pulses of lowered bond strength (equivalently, higher  temperature).  Bond strength cycles over a fixed period between some time $\tmelt$ at a strength $\gsemelt$ above the critical bond strength $\gsecrit$ between ribbon growth and melting, and $\gsegrow$ below it. \figp{e} Size of five simulated assemblies over time, with $\tcycle = 3600\;\text{s}$, $\gmc=16$, $\gsegrow=9.5$, $\gsemelt=8.4$.  With a short $\tmelt$, ribbon growth stalls through disordered interactions.  With a longer $\tmelt$, melting cycles detach the growth front, allowing stalled ribbons to grow when bonds become stronger.  With yet longer $\tmelt$, the too much of the ribbon melts each cycle, and thus ribbons do not grow.}
    \label{fig:ext-sa-example}
\end{figure}

\section{Tile self-assembly}

To simulate molecular self-assembly, we used the kinetic Tile Assembly Model (kTAM)~\cite{winfree_simulations_1998}.  From an abstract perspective, in the limit of low monomer concentrations and slow growth, this model can be shown to approximate the behavior of systems in the abstract Tile Assembly Model (aTAM)~\cite{winfree_simulations_1998}, which in turn has been used to investigate the computational ability of self-assembly with many monomer types, showing it to be capable of Turing-universal computation~\cite{Rothemund2000-ov}.  From a physical perspective, in implementations with DNA tiles, the kTAM has been representative of the behavior of experimental systems~\cite{evans_physical_2017}.  The model is based on the assumption that growth occurs through individual monomer attachments and detachments to a lattice (Figure \ref{fig:ext-sa-example}a), with the attachment rate for any monomer tile at any one empty site adjacent to a lattice based only on the concentration of the monomer tile~---~not on the strength of bonds it can form, and the detachment rate for a tile in a structure based on the sum of the strength of the bonds it has with other tiles in the structure, that is,

\begin{align}
    r_\text{on} &= [c] = \kf e^{-\gmc} & r_{\text{off},b} &= k_f e^{-b\gse}\;,
\end{align}

where $\gmc \equiv \log{[c]u_0^{-1}}$ for some standard concentration $u_0$ (usually 1 M) and monomer concentration $[c]$, $\gse$ is the a unitless, sign-reversed (positive is stronger) free energy of a single, normal bond, and $b$ is the total bond strength of a particular tile attached to an assembly\footnote{In standard terms, the change in free energy for a single bond would be $\Delta G = -RT\gse$.  In DNA implementations, experiments~\cite{jiang_understanding_2017} and molecular dynamics simulations~\cite{fonseca_multiscale_2018} have found that the free energy change of a tile attaching by multiple bonds is not a simple sum of changes for each bond---generally, attachment by two or more bonds incurs a penalty compared to a single bond---and additional corrections are required.  However, these corrections can often be handled as a rescaling of $\kf$, $\gse$, and $\gmc$~\cite{Evans2024-se}, and so are ignored here.}.  While the kTAM typically considers bonds in terms of matching `glues' on different tile types, we will consider general interaction matrices for North-South and West-East interactions $\Ins$ and $\Iwe$, such that, for a tile $i$ with adjoining tiles $n$, $e$, $s$, and $w$ in corresponding directions, $b = \Ins_{is}+\Ins_{ni}+\Iwe_{ie}+\Iwe_{wi}$, where any empty site has tile type $0$ and $\Ins_{x0}=\Ins_{0x}=\Iwe_{x0}=\Iwe_{0x}=0$ for any tile type $x$, and interaction strengths are typically either $0$ or $1$. 

In DNA implementations, where bonds are implemented as hybridization of short complementary single-stranded regions of the tile, $\gse$ is temperature dependent, with bonds weakening, and thus $\gse$ becoming smaller, as temperature increases.  Thus, changing temperature is often seen as roughly equivalent to changing $\gse$ in the kTAM~\cite{Evans2024-se}, and lowering/raising temperature is often referred to interchangeably with raising/lowering $\gse$.  Systems are most often designed to grow when tile attachments by two or more bonds are favorable, and attachments by one bond are not, this corresponds with $\gmc = 2\gse - \epsilon$ for some positive $\epsilon$ smaller than $\gse$.

In many tile systems, growth is `seeded', either by a special tile type with stronger bonds to start off growth, or a small pre-assembled structure; in experimental implementations, this is often constructed with DNA origami \cite{Barish2009-ne}.  In conditions with minimal nucleation, and with a significantly higher concentration of each monomer type than the seed structure, this allows growth to be modeled as taking place on a fixed number of assemblies, with the effects on attachment rates from the depletion of free monomer concentrations being insignificant.  Thus, growth can be approximated by assuming that monomer concentrations remain constant, particularly in conditions where growth is sufficiently forward-biased, that is, with $r_\text{on} \gg r_{\text{off},2}$, that the linear decrease in $r_\text{on}$ as concentrations deplete does not significantly change the bias of the system; in conditions where growth is initially only slightly forward-biased, the assumption of constant monomer concentrations may be valid only in the `initial moments' of assembly.

\section{Stalling }

We designed a tile system to intentionally exhibit strong stalling behavior, shown in Figure \ref{fig:ext-sa-example}b.  The system uses 12 tile types, designed to attach in alternating rows of tiles 1--6 and 7--12, in order from north to south.  In `ordered' growth, every tile interacts with strength one with its neighbor in each direction; a full row of six tiles includes six bonds to its west, and five bonds between tiles in the row.  The resulting change in free energy for a row, and critical free energy between growth and melting, is thus

\begin{align}2
    \Delta G_\text{row} &= 6 \gmc - 11 \gse & \gsecrit &= \frac{6}{11} \gmc
\end{align}

With $\gmc = 16$, this means that ordered growth requires $\gse \geq \gsecrit = 8.73$: above this value (`lower temperature' or `stronger bonds') it will grow, and below this value, it will melt.

Outside of this ordered growth pathway, we add additional, `non-specific' interactions to allow a `disordered' growth pathway that will later result in stalling; in principle interactions could be added to allow for many disordered growth pathways and configurations, but for simplicity, we only design one disordered configuration.  In this configuration, rather than a 1--6 row, a row in order (1, 5, 4, 3, 2, 6) grows, through additional non-specific interactions that are added between the tiles in the row, and to the west edges of tiles 2--5; to avoid non-ribbon growth pathways, tiles 1 and 6 remain in the same positions, as they have no interactions to the north and south, respectively.  

With no other interactions, while the growth of a disordered row is itself equally favorable to a disordered row, the ordered row will allow continued favorable growth of a subsequent row, while the disordered row will not.  After an ordered row, tiles 7--12 can attach and form a row where each tile interacts with both a tile in the previous row, and its neighbors in the new row.  After a disordered row, however, tiles able to bind individually to tiles in the disordered row would not form any bonds with each other, and tiles able to bind to their neighbors would result in only two of the tiles, on each edge of the row and thus widely separated, interacting with the previous row.  In both cases, the resulting attachments would be unfavorable and unstable enough that growth of a full row or further rows would be very unlikely.

In conditions where growth is sufficiently forward-biased, the combination of the favorable pathway to a full disordered row (thus making complete detachment of the row unlikely), and the unfavorable pathways to continued growth after a disordered row, result in a disordered row at the growth front being a local energy minimum with substantial kinetic barriers to both growth and melting, causing growth to stall.

In order to allow a possible, but difficult, path to continued growth, we add two additional west-east interactions, between tiles 5 and 8, and tiles 3 and 10.  These interactions, determined through trial and error but explored further in the next section, reduce the barrier to further growth after a disordered row, allowing a small, but measurable, error rate, or rate of disordered growth, in assemblies.

\section{Dynamic instability}

Dynamic instability is a non-equilibrium phenomenon observed most famously in microtubules \cite{horio_visualization_1986,mitchison_dynamic_1984,mitchison_microtubule_1984}. In many conditions, microtubules show behaviors that can be reasonably described as incremental self-assembly, but with occasional rapid disassembly events. Many detailed microscopic models of dynamic instability have been developed over the years with varying degrees of fidelity to the underlying molecular mechanism. For example, one popular model is the GTP cap model \cite{carlier_kinetic_1981} in which the microtubule exhibits one of three phases~---~growth, catastrophes and rescues.  In the growth phase,  tubulin dimers are added to a growing microtubule through detailed balance-preserving self assembly dynamics much like that of the kTAM model above. However, unlike in the kTAM model, each dimer is GTP bound and can eventually be hydrolyzed into the GDP bound state. The catastrophe phase starts when sufficient numbers of dimers are hydrolyzed to the GDP form close to the growing tip; this results in a dramatic disassembly of large parts of the microtubule. Rescue events~---~due to binding of new GTP-bound dimers to the leading tip~---~then limit this disassembly to being only a partial disassembly of the microtubule. Numerous other models that incorporate other known molecular mechanisms have been developed but qualitatively, these three phases constitute dynamic instability.

Here, we focus on a simple model that captures these three key features of dynamic instability and does not attempt to model molecular details.  For the system described in the previous section, growth depends on having bonds be of strength $\gse > \gsecrit$.  If $\gse$ is changed for a period of time, the dynamics of the system will change: for example, if $\gse$ is changed to be significantly smaller than $\gsecrit$, meaning that bonds are too weak to support growth, existing ribbons will rapidly melt.  Thus, we can model a periodic change between a growth and melting phase as a periodic change in $\gse$, which could be physically interpreted as either a change in temperature (with bonds weakening as the system is heated) or a uniform change to all tiles in system.  

For our simulations, we used a periodic switching of $\gse$, with period $\tcycle$, that started melting phase of a low bond strength $\gsemelt$, for a time $\tmelt$, and then spent the remaining $\tcycle-\tmelt$ of each cycle in a growth phase at a high bond strength $\gsegrow$ (Figure \ref{fig:ext-sa-example}d).  Thus at the beginning of each new cycle, whatever ribbon had grown would have a heavy bias toward melting for the duration of the melting phase.  We used a fixed $\tcycle = 3600\;\text{s}$, $\gmc=16$, $\gsegrow=9.5$, $\gsemelt=8.4$, and explored the effects of varying $\tmelt$. The strength of dynamic instability is defined by $\lambda = t_\text{melt}$.  Evolution of $\lambda$ was performed by a hill-climbing algorithm, starting from $\lambda = 0$, and accepting changes to $\lambda$ if the assembly rate, on 768 simulations, increased.

The stalling ribbon system, in growth-biased conditions, will tend to stall when a disordered row attaches.  With a periodic melting phase, however, there is a pathway for the disordered row at the growth front to melt away, allowing growth to continue in the next growth phase.  Overall growth rate, with dynamic instability, is thus dependent on having melting phases of sufficient duration and frequency, and with sufficiently weak bond strength, to ensure that disordered regions at the growth front are likely to melt, while not being so melting-biased or of such duration that the overall melting outpaces the growth of the ribbon in the growth phase.  An example of three choices of $\tmelt$, with too little instability (resulting in continued stalling), instability sufficient to increase overall growth rate, and too much instability (resulting in no overall growth), is shown in Figure \ref{fig:ext-sa-example}e.

\section{Stalling strength and parameter evolution}

The tile system in the previous section exhibits stalling when growing through non-specific pathways in conditions favorable for growth, and when periodic decreases in bond strength are added as a form of dynamic instability, the partial melting during those low-bond-strength periods results, for some choices of dynamic instability parameters, in stalled, disordered regions at the growth front melting and allowing another opportunity for ordered growth in the next high-bond-strength period.  In order for this behavior to result in an increase in assembly speed and accompanying decrease in disordered growth, and for selection for assembly speed to also decrease disordered growth, however, the kinetic barriers involved in the stalling process must satisfy certain constraints:

\begin{itemize}
    \item The barrier to further growth after disordered attachments must be large enough that a disordered state will sufficiently stall further growth of the system: otherwise, optimizing for assembly speed will not select for ordered growth.
    \item The barrier to further growth after disordered attachments, however, must also be small enough that a disordered state will have some slow but significant rate at which growth will continue, incorporating the disordered section; otherwise, disordered growth will never be incorporated into an assembly, and while dynamic instability will increase overall growth rate by removing stalled states, there will be no error rate to reduce.
    \item The barrier to a disordered region detaching and being replaced by an ordered region at the melting temperature, without any introduced dynamic instability, must be large enough that the stalled state remains stalled; otherwise, series of disordered attachments and detachments will occur repeatedly at the growth front until ordered attachments allows growth to continue.
\end{itemize}

The third constraint, on detachment of disordered regions, is simply satisfied in our system by disordered growth pathways having a sufficient amount of favorable attachment steps before any unfavorable steps are necessary, and growing with sufficiently forward-biased conditions: a complete 1, 5, 4, 3, 2, 6 row can grow with the same number of bonds as an ordered 1, 2, 3, 4, 5, 6 row, and it is only in the subsequent row that disordered growth requires attachments by fewer bonds than ordered growth.  However, the first two constraints may depend on the specifics of the barriers to further growth after disordered attachments, in both the number and arrangement of bonds made.

\begin{figure}
    \centering
    \includegraphics[width=\textwidth]{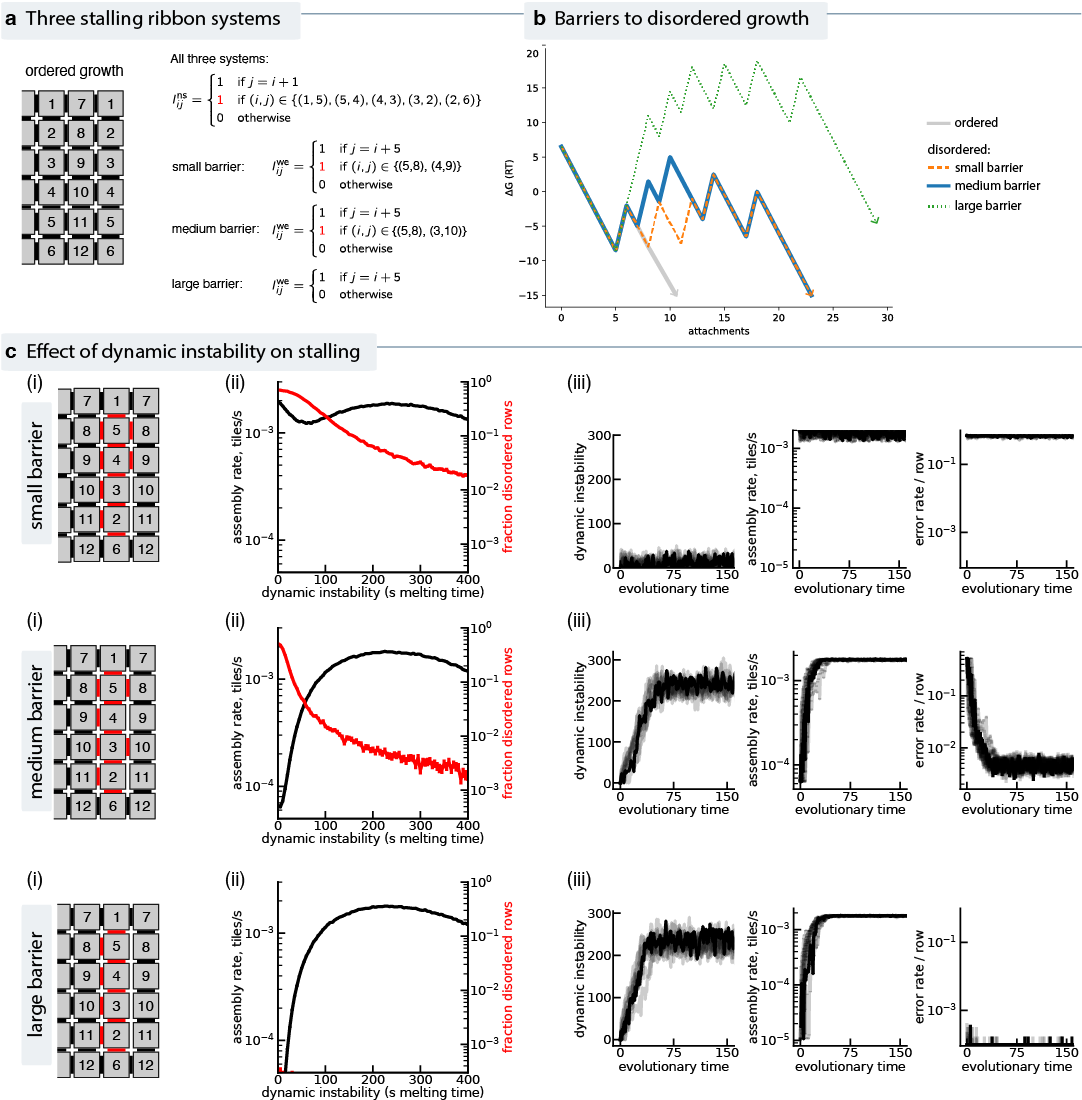}
    \caption{\textbf{Tile sets with too large or too small a stalling barrier do not evolve non-equilibrium order from speed.}\figp{a} Each self-assembling system has the same set of specific interactions, and grows as in Fig. \ref{fig:ext-sa-example}, but differ in their non-specific interactions: four bonds, with three adjacent, in the \textbf{small} barrier system, four bonds spread across the row for the \textbf{medium} barrier system, and two bonds in the \textbf{large} barrier system.  \figp{b} The different non-specific interactions result in different free energy barriers to disordered growth.  \figp{c} (i) Structures and bonds for continued growth after a disordered row in each system.  (ii), as dynamic instability increases (i.e., melting time $t_{melt}$), the small and medium barriers have less disordered growth, and the medium and large barriers have significantly faster growth. (iii), both the medium and large barriers can have melting time $t_{melt}$ evolved to optimize speed, but only in the medium barrier system does this also decrease disordered growth.}
    \label{fig:ext-sa-goldilocks}
\end{figure}

In Figure \ref{fig:ext-sa-goldilocks}, we investigate three similar systems, with small changes to the non-specific bonds between the disordered row and continued growth.  The `medium barrier' system is the same as the system in Figure \ref{fig:ext-sa-example}.  The `small barrier' system allows the same number of bonds between the two rows as the medium barrier system, but arranges them slightly differently.  The `large barrier' system instead allows only two, widely separated bonds between the disordered row and continued growth.  Examples of the resulting energy landscapes, for potential least-unfavorable pathways to a disordered row and continued growth, are shown in Figure \ref{fig:ext-sa-goldilocks}b.  All three systems form 11 bonds for the initial, disordered row.  For the subsequent row, the change in bond arrangement between the small and medium barrier systems---intuitively, the three adjacent bonds between 1 to 7, 5 to 8, and 4 to 9 helping to stabilize the attachment of that part of the subsequent row---result in a smaller barrier for the small barrier system, while the lack of bonds in the large barrier system results in pathways with a much larger barrier to attachment of the subsequent row.

For a growing assembly, the determination of error rate is complicated by the growth front, which at any moment may have disordered attachments that will not be incorporated into the assembly.  For a given target size in tiles $N_\text{target}$, only errors in the first $\frac{1}{6}N_\text{target} - 6$ rows were considered, roughly corresponding to ignoring the last six rows.  

Evolution was performed by a simple hill-climbing algorithm, modifying $\tmelt$.  Starting from $\tmelt = 0$, $\tcycle = 3600\;\text{s}$, $\gsegrow=9.5$ and $\gsemelt=8.4$, 768 assembly simulations were run for a particular set of parameters.  Each simulation was run on a $10\times 128$ canvas until reaching a target size of 480 tiles, corresponding to approximately 40 repeats or 80 rows, or reaching a $10^8\;\text{s}$ time cut-off.  For each new evolutionary step, $\tmelt$ was modified by adding a random time from a normal distribution with a standard deviation of $15\;\text{s}$.  The change was accepted if it increased the mean of the inverse of the times taken to reach the target size.  This hill-climbing process was repeated for 200 steps; the entire evolutionary process was run 10 times.

\subsection{Relationship to prior work on error correction in algorithmic self-assembly}

The self-assembly systems presented in Chapter \ref{ch:sa} all share the property that disordered growth through non-specific attachments results in configurations where there is a kinetic barrier to further growth, and growth thus `stalls', while ordered growth results in configurations where growth can continue unimpeded.  In principle, as our model of tile assembly allows for tile detachments, there is also some rate at which the disordered attachments at the growth front of a stalled assembly will detach, allowing different tiles to attach, potentially in an ordered configuration that will allow continued growth.  While this pathway is unlikely when far from equilibrium, in near-equilibrium conditions, this process of stalling configurations detaching and allowing non-stalling configurations to attach could serve as a form of error correction. 

\begin{figure}
    \centering
    \includegraphics{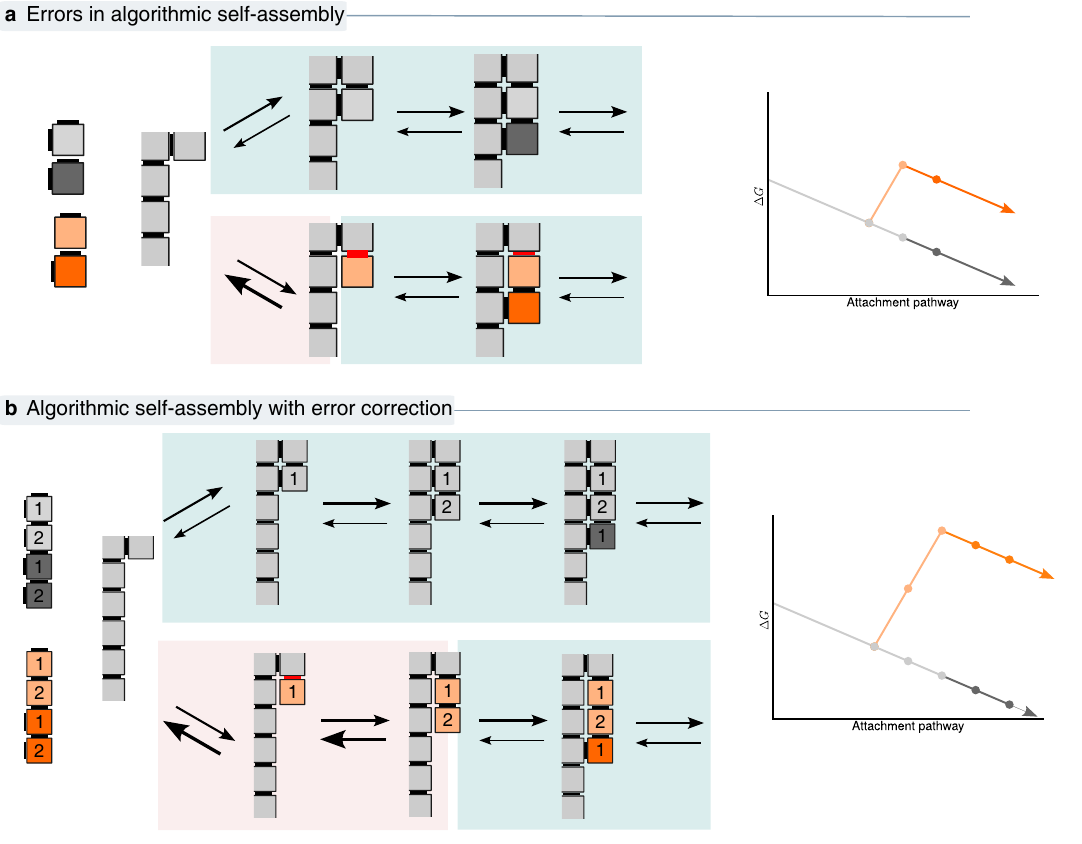}
    \caption{\textbf{Review of prior work on algorithmic error-correction.} \figp{a} In prior work on `algorithmic tile assembly' \cite{Winfree1998-vj}, there are locations on the growth front of a correct assembly where there are both tiles that can attach by two bonds (gray), and tiles that can attach by only one of the two available bonds (orange).  While a tile attaching by one bond will detach much faster than it attaches, further attachments by two or more bonds may be able to stabilize the initial attachment and allow growth to continue.  \figp{b} In the prior work on error-correcting tile systems \cite{{winfree_proofreading_2004,chen_error_2005,evans_optimizing_2018}}, interactions are such that a tile attaching by one bond will require one, or $k$, further attachments by one bond before growth can continue by two-bond attachments. Error-correcting constructions can be interpreted as stalling continued growth after disordered attachments, at a point where pathways to detachment of the disordered tiles remain likely.}
    \label{fig:ext-sa-proofreading}
\end{figure}

In tile self-assembly, and particularly in algorithmic self-assembly \cite{Winfree1998-kc}, this form of error correction has been studied under the name of `proofreading' (but note that this concept is more closely related to stalling in this paper rather than kinetic proofreading as used in this paper).  `Proofreading' tile systems, which we will describe as `algorithmic error-correcting tile systems' (to avoid confusion with kinetic proofreading), either transform existing systems or design interactions such that incorrect attachments to correct assemblies will necessarily have these kinetic properties~\cite{winfree_proofreading_2004,chen_error_2005,evans_optimizing_2018}.  A simplified form of transformation is shown in Figure \ref{fig:ext-sa-proofreading}.  In a typical tile system design, a location at the growth front of a correct assembly will have one tile that can attach `correctly', by two bonds, but potentially many tiles, here shown in orange, that can attach `incorrectly' by one bond.  Having a higher detachment rate than attachment rate, this incorrect pathway has no barrier to proceeding back to the initial state, but at the same time, there may be tiles allowing growth to continue by favorable, two-bond attachments, incorporating the incorrect tile, thus `trapping' the incorrect attachment in place, and, because the subsequent tile attachments are based on the interactions of the incorrect tile, potentially changing subsequent growth entirely.

Error-correcting tile systems reduce incorrect growth by ensuring that after an incorrect attachment, further unfavorable attachments will be required for continued growth: in the simplest case, this means ensuring that, for any correct structure, a incorrect tile attaching by one bond where the correct tile can attach by two will not result in adjacent sites where any tile in the system can attach by two bonds: one to the existing structure, and one to the incorrect tile. For many systems, if considering only sites where a correct attachment is possible, this property can be achieved through a scale-up transformation, replacing each tile in a non-error-correcting system with a $2 \times 2$ square of tiles that attach and detach individually, but have unique internal interactions~\cite{winfree_proofreading_2004}: in this case, where, close to the melting temperature, an error rate of a non-error-correcting system would be expected to scale as $e^{-\gse}$, the error-correcting system will scale as $e^{-2\gse}$.  When considering all sites, including sites where no correct tile can attach, or no tile can attach by two bonds and facet nucleation is possible, more complex transformations are required~\cite{chen_error_2005}.

Algorithmic error-correcting tile systems can be interpreted as reducing error rates by `stalling' growth after incorrect attachments, allowing continued growth after correct attachments without stalling, and having sufficiently reversible attachments—often involving incorrect attachments that are purely unfavorable—such that the incorrect attachments resulting in stalling will be more likely to detach than allow further growth past the stalled state.  In contrast, the stalling systems of the chapter, on the other hand, have significant barriers to both attachment and detachment when a disordered row attaches, with the values of $\gmc$ and $\gsegrow$ used: once a disordered row is attached, growth is likely to remain stalled until the periodic dynamic instability of $\gse$ lowering to $\gsemelt$ makes detachment favorable, as in Figure \ref{fig:ext-sa-slow}b.

\chapter{Non-equilibrium homeorhesis through resets}

Homeorhesis is a generalization of homeostasis to dynamical trajectories \cite{waddington_canalization_1942}; while homeostasis refers to a system being able to maintain a fixed state, homeorhesis refers to the ability to maintain a stereotyped trajectory between two states. Homeorhesis has been chiefly discussed in the context of Waddington's framework for developmental biology \cite{waddington_canalization_1942,waddington_strategy_1957}; e.g., Waddington's famous landscape depicts how an energy landscape can effectively reduce variance in (developmental) trajectories from a fertilized egg to a fully developed organism. 

Here, we explore an abstracted error-correcting mechanism~---~stochastic resets \cite{evans_stochastic_2020}~---~that effectively reduces variance in trajectories through non-equilibrium dynamics rather than with a Waddington-like energy landscape. We study an abstract model of replication in which  a cell starts at a \textit{Birth} state and must reach a final \textit{Division} state before dividing. These states are connected by a large number of distinct trajectories through the high-dimensional state space of a cell, e.g. these trajectories could represent different ways of assembling macromolecular structures \cite{phillips_molecular_2020}, copying genetic information with and without errors, achieving mitotic segregation, accomplishing metabolic tasks or other tasks at larger scales in the life of a cell.  Mathematically, we assume that each path $x(t)$ from \emph{Birth} to \emph{Division} is realized with a probability distribution $P[x(t)]$.  The completion time $T_{\rm{rep}}$ of these trajectories, which is also the replication time, is consequently stochastic, and we denote it's distribution as $P(T_{\rm{rep}})$.

Here, we show that, under specific conditions on $P(T_{\rm{rep}})$, the average replication time is \emph{decreased} by a Maxwell demon that randomly resets \cite{evans_stochastic_2020} the system back to the \emph{Birth} state on occasion. Critically, we argue that such a reset mechanism also decreases the entropy of possible paths $x(t)$ used to reach \emph{Division}. Consequently, non-equilibrium reset mechanisms speed up replication but also ensure that the \emph{Birth}- \emph{Division} journey is completed in a relatively stereotyped way, i.e. non-equilibrium homeorhesis. The reduced variance in trajectories can be seen an abstraction of the specific error correction mechanisms like kinetic proofreading and dynamic instability.

\section{Fitness as long-term exponential growth rate}

Consider a self-replicative system, an organism or a cell, with a distribution of replication times $P(T_{\rm{rep}})$. We define the fitness $f$ of such an organism to be the long-term exponential growth rate, $$\langle N(t) \rangle \sim e^{f t}$$ where $N(t)$ is the number of offspring after time $t$. 

The fitness of a self-replicative system is not simply given by $1/\langle T_{\rm{rep}} \rangle$ where $\langle T_{\rm{rep}} \rangle$ is the average replication time from $P(T_{\rm{rep}})$. Such definition would be an underestimate. Slow replication (large $T_{\rm{rep}}$) events do not affect fitness as much as this expression implies, since sister cells can continue dividing. Instead, we can derive an approximate mean fitness through a recursion relationship, in contrast with other resetting models already discussed in the literature. Let $N(t)$ be the population size of a lineage seeded by one cell after time $t$ with generation times for each division drawn independently from $P(T_{\rm{rep}})$. The population size, $\langle N(t) \rangle$, averaged over many such realizations, satisfies the recursion relationship
\begin{equation}
    \langle N(t) \rangle = 2 \int_0^\infty dT P(T) \langle N(t-T) \rangle 
\end{equation}
where the integral is over the time taken for the first division of one cell into two. Setting
\begin{equation}
\langle N(t) \rangle \sim e^{f t}\, ,
\end{equation}

where fitness $f$ is the long-term exponential growth rate, we find
\begin{equation}
    \tilde{P}(-f) = \frac{1}{2}\, ,
    \label{eq:fitness}
\end{equation}
where $\tilde{P}(w) = \int_0^\infty e^{w T} P(T) dT$ is the moment generating function for $P(T_{\rm{rep}})$. Thus, we can find fitness $f$ by solving the above equation $2 \tilde{P}(-f) = 1$ \cite{jafarpour_bridging_2018}.

\section{Resets speed up replication and decrease entropy of paths}
Here, we derive conditions on general $P(T_{\rm{rep}})$ under which ordering spontaneously emerges in a self-replicative system. We compute the fitness in the presence of resetting mechanisms at time $T_r$, $f(T_r)$, defining the distribution of times $P^{neq}(T_{\rm{rep}})$ in the presence of resets and computing its moment generating function. For simplicity of notation, $P(T_{\rm{rep}})$ is the equilibrium distribution $P^{eq}(T_{\rm{rep}})$. We then find an expression for the reset time $T_r$ that optimizes $f(T_r)$.

\subsubsection{Demon resets at time $T_r$: definitions}
Let us consider a Maxwell demon that randomly resets at time $T_r$. We can define the probability distribution of replication times cutting all times larger than $T_r$ as
\begin{eqnarray}
    P^r(T_{\rm{rep}}) &= & \frac{ P(T_{\rm{rep}})}{a(T_r)} \quad \mbox{ if } 0\leq T_{\rm{rep}} <  T_{r}  ,\, 0 \mbox{ otherwise} \nonumber\\
    a(T_r) & = & \int_0^{T_r} P(T_{\rm{rep}})dT_{\rm{rep}}\, .\nonumber
\end{eqnarray}

The moment generating function reads
\begin{eqnarray}
    \tilde{P}^{r}(w) &=& \int_0^\infty e^{w T_{\rm{rep}}} P^{r}(T_{\rm{rep}}) dT_{\rm{rep}} \nonumber \\
    &=& \frac{1}{a} \int_0^{T_r} e^{w T_{\rm{rep}}} P(T_{\rm{rep}}) dT_{\rm{rep}} 
\end{eqnarray}

Note that $a(T_r)$, $a$ for brevity, is the probability of not getting lost, i.e. picking a trajectory of length shorter than $T_r$ and serves as the normalization factor for $P^r(T_{\rm{rep}})$.

\begin{figure*}[h!!!]
    \centering
    \includegraphics[scale=1]{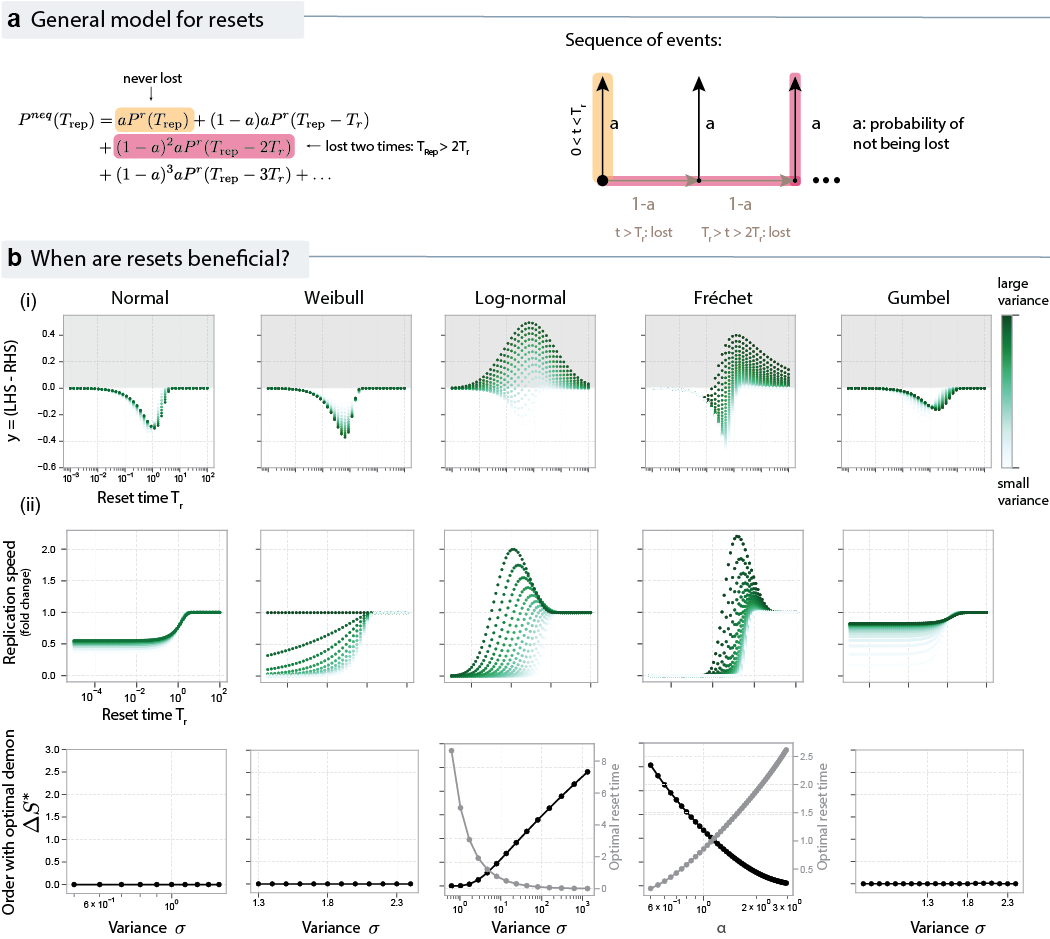}
    \caption{\textbf{Resets are beneficial only for wide distributions of replication times.} (\textbf{a}) The relationship between the sequence of events and terms in Eq. \ref{eq:ptneq} is sketched to show the intuition behind Eq. \ref{eq:ptneq}. (\textbf{b}) (i) The residual $y=(\rm{LHS - RHS})$ of the inequality in Eq. \ref{eq:beneficialresets} is plotted as function of reset time $T_r$ and for different values of variance~---~see colorbar: when $y>0$, resets are beneficial, i.e. increase replication speed. We find that $y>0$ for some choices of $T_r$ for the Log-normal and Fréchet distribution. (ii) The fold-change in replication speed (Eq.~\ref{eqn:generalresetfitness}) is shown for different choices of variance parameters and for different probability distributions of replication times: Normal, Weibull, Log-normal, Fr\'echet and Gumbel. The order $\Delta S^*$ with the optimal demon is computed as in  Eq.~\ref{eq:order_times} imposing Eq.~\ref{eqn:optimalfT}, which sets the constraint for an optimal demon $(f^*,T_r^*)$. In the case where resets provide a fitness benefit, for Log-normal and Fr\'echet, the optimal reset time is shown in the plots (grey, right y-axis). We fix the mean of all distributions to unity, but for the plots with the Fr\'echet distribution in (ii).} 
    \label{fig:SI_resets}
\end{figure*}

\subsubsection{Probability distribution of times with resets}
We can calculate how the distribution of times $P(T_{\rm{rep}})$ is modified when a demon resets at time $T_r$, giving $P^{neq}(T_{\rm{rep}})$
\begin{eqnarray}
 P^{neq}(T_{\rm{rep}})  &=&  a P^r(T_{\rm{rep}}) + (1-a) a P^r(T_{\rm{rep}} - T_r) \nonumber \\
 &+&(1-a)^2 a P^r(T_{\rm{rep}} - 2T_r)\nonumber \\
 &+&  (1-a)^3 a P^r(T_{\rm{rep}} - 3T_r) + \ldots
 \label{eq:ptneq}
\end{eqnarray}
where $P^r(t) \neq 0 $ only for $0\leq t<T_r$; thus each term is non-zero in a distinct interval of $T_{\rm{rep}}$: being lost only once happens with probability $(1-a)a$, as seen in the factor multiplying $P^r(T_{\rm{rep}} - T_r)$ which is non-zero only in the interval $T_{\rm{rep}}\in [T_r,2T_r)$, possible sequence of events are sketched in Fig. \ref{fig:SI_resets}(a).

The moment generating function for $P^{neq}(T_{\rm{rep}})$ is $\tilde{P}^{neq}(w) = \int e^{w T_{\rm{rep}}} P^{neq}(T_{\rm{rep}}) dT_{\rm{rep}}$ as follows,
\begin{eqnarray}
 \tilde{P}^{neq}(w)  &=&  a \tilde{P}^r(w)  + (1-a) a e^{w T_r}  \tilde{P}^r(w)\nonumber \\
 &+&   (1-a)^2 a e^{2 wT_r} \tilde{P}^r(w) + (1-a)^3 a e^{3 w T_r} +  \ldots \nonumber \\
 & =& a \tilde{P}^r(w) [1 + (1-a)e^{wT_r} + (1-a)^2 e^{2wT_r} + \ldots] \nonumber \\
 &=& \frac{a \tilde{P}^r(w)}{1 - (1-a) e^{w T_r}}\nonumber
\end{eqnarray}
where $\tilde{P}^r(w) = \int_0^\infty e^{w T_{\rm{rep}}} P^r(T_{\rm{rep}}) dT_{\rm{rep}} = \frac{1}{a(T_r)} \int_0^{T_r} e^{w T_{\rm{rep}}} P(T_{\rm{rep}}) dT_{\rm{rep}}$.
Hence, this simplifies to
\begin{equation}
    \tilde{P}^{neq}(w) = \frac{a_w}{1 - (1-a e^{w T_r}}\label{eq:mgf}
\end{equation}
with $a_w(T_r) = \int_0^{T_r} e^{w T_{\rm{rep}}} P(T_{\rm{rep}}) dT_{\rm{rep}} $. In the formula above we drop the dependence on $T_r$ for both $a$ and $a_w$ to ease the notation.

\subsection{Resets decrease the entropy of paths}

Paths are defined as sequences of states, $\boldsymbol{x}\in\{ABDEFG, ACYTGS, \dots \}$, while trajectories are the combination of a path with its respective replication time drawn from $P(T_{\rm{rep}})$: $\{(\boldsymbol{x_1}, t_1), (\boldsymbol{x_2}, t_2), \dots \}$, defining a joint distribution $P[x(t),t]$. 

The order imposed by the demon is measured by the reduction in the entropy of paths $x \in \{ABDEFG, ACYTGS, \dots \}$ successfully taken from start to finish without being reset. Denote by $P^{eq}[x(t)]$ the distribution of paths taken in the absence of the demon, and by $P^{neq}[x(t)]$ the distribution in the presence of a demon that resets at a time $T_{\rm{r}}$. We seek to compute the Kullback-Leibler (KL) divergence between these two distributions:

\begin{equation} 
\Delta S \equiv D_{\rm{KL}} \left( P^{neq}[x(t)] \vert\vert P^{eq}[x(t)] \right).
\end{equation}

To obtain $P^{neq}[x(t)]$, consider the joint distributions of paths and completion times in the absence of resets: $P^{eq}(x,t)$. The demon acts to kill any pair $(x,t)$ with $t > T_{r}$; consequently, we can write the joint distribution in the presence of resets, $P^{neq}(x,t)$, as\footnote{For the calculation of the entropy of paths, we assume that the non-equilibrium distribution $P^{neq}[x(t)]$ is composed of paths going from the initial to the final state after resetting has happened, i.e. not taking into account the loops going back to the initial state. This assumption is supported by real systems, as self-assembly, where the order created by resets depends on the structure that is assembled after the previous one has been destabilized, not on the details of the disassembling process. Hence, the non-equilibrium distribution can be written as the distribution after resetting at time $T_r$, $P^{neq}[x(t)]=P^r[x(t)]$. }

$$ P^{neq}(x,t) = \frac{1}{C} \, P^{eq}[x(t)] \, P^{eq}(t \vert x) \, \left[ 1 - \Theta(t-T_{r}) \right] $$

\noindent where $\Theta$ is the Heaviside function, $C$ is a constant to ensure normalisation, and we have split $P^{eq}(x,t)$ into the marginal $P^{eq}[x(t)]$ and the conditional $P^{eq}(t \vert x)$. Note that the time $t$ refers only to the time between the last reset and completion; it does not include the time lost due to futile cycling.

The unknown constant of proportionality $C$ is found by normalisation, $\sum_x \int_0^{\infty} dt \, P^{neq}(x,t) = 1$:

$$
   C = \sum_x \int_0^{T_{\rm{r}}} dt \, P^{eq}(x,t) = \underbrace{\int_0^{T_{\rm{r}}} dt \, P^{eq}(t)}_{a(T_{\rm{r}})}
$$

\noindent which gives us $P^{neq}[x(t)]$ by marginalising out $t$:

\begin{equation}
P^{neq}[x(t)] = \frac{1}{a(T_{\rm{r}})} P^{eq}[x(t)] \, \underbrace{\int_0^{T_{\rm{r}}} dt \, P^{eq}(t \vert x)}_{z(T_r \vert x)}
\end{equation}

\noindent where we have introduced the conditional probability of not getting lost, $z(T_r \vert x)$. Then,

\begin{eqnarray}
\label{eq:order_times}
    \Delta S &=& \sum_x P^{neq}[x(t)] \ln \frac{P^{neq}[x(t)]}{P^{eq}[x(t)]} \nonumber \\
    &=& \sum_x P^{neq}[x(t)] \ln \frac{z(T_r | x)}{a(T_r)} \nonumber \\
    &=& -\ln a(T_r) + \sum_x P^{neq}[x(t)] \, \ln z(T_r | x) \nonumber \\
    &\leq& -\ln a(T_r)
\end{eqnarray}
 
In the last step, note that $0< z <1$, and so $\ln z < 0$. We conclude that $\Delta S$ is at most $-\ln a$, and the bound is saturated when $z(T_r | x) = 1 \, \, \forall \, x$ s.t. $P^{neq}[x(t)] > 0$. This latter condition is satisfied when, for instance, each path $x$ is completed in a deterministic time (i.e. $P_0(t |x)$ is a delta-function for all $x$).

\subsection{Resets speed up replication}

\subsubsection{Fitness of a resetting demon}
To find the fitness $f(T_r)$ in the presence of such resets at time $T_r$, we use Eqs. \ref{eq:fitness} and \ref{eq:mgf} to find,
\begin{equation}
   2 a(T_r) \tilde{P}^r(-f(T_r))= 1 - (1-a(T_r)) e^{- f(T_r) T_r}
\end{equation}
which can be simplified to given an equation for fitness $f(T)$ in the presence of a demon that resets at time $T$:
\begin{equation}
   {2 \int_0^{T_r} e^{- f(T_r) T_{\rm{rep}}} P(T_{\rm{rep}}) dT_{\rm{rep}} = 1 - (1-a(T_r)) e^{- f(T_r) T_r}}\, .\label{eqn:generalresetfitness}
\end{equation}

\subsubsection{Optimal fitness}
To find the reset time $T_r$ that confers optimal fitness $f^*$, we differentiate the equation above with respect to $T_r$ and set $df/dT_r=0$. We get,
\begin{eqnarray}
   2  e^{- f(T_r) T_r} P(T_r) &=&  \partial_{T_r}a(T_r) e^{- f(T_r) T_r}    + (1-a(T_r)) e^{- f(T_r) T_r} f(T_r) \nonumber 
\end{eqnarray}

but since $\partial_{T_r} a(T_r)  = P(T_r)$, we get
\begin{equation}
   { f^*(T_r^*)  = \frac{P(T_r^*)}{1-a(T_r^*)}}
   \label{eqn:optimalfT}
\end{equation}
for the optimal fitness $f^*$ and the corresponding optimal reset time $T_r*$.

Solving Equation \ref{eqn:generalresetfitness} and \ref{eqn:optimalfT} together for the two unknowns $f,T_r$ will give us the optimal $f^*,T_r^*$. Note that the equations involve the CDF $a(T_r)$ and a partial transform $\int_0^{T_r} e^{-f T_{\rm{rep}}} P(T_{\rm{rep}})$ of $P(T_{\rm{rep}})$.

\subsubsection{When do resets speed up replication?}

Here we derive a condition on when resets are beneficial to fitness when we consider the approximation of fitness as mean time. The mean time after including resets at time $T_r$ is
 \begin{equation}
     \langle T_{\rm{rep}} \rangle_{T_r} = \int_0^{T_r} T_{\rm{rep}} P(T_{\rm{rep}}) dT_{\rm{rep}}  + p_{lost}(T_r) ( T_r + \langle T_{\rm{rep}} \rangle_{T_r})
 \end{equation}
 where $p_{lost}(T_r) = \int^\infty_{T_r} P(T_{\rm{rep}}) dT_{\rm{rep}}$. 
 Solving, we get~---~using $t$ instead of $T_{\rm{rep}}$ inside the following integrals for ease of notation
  \begin{equation}
     \langle T_{\rm{rep}} \rangle_{T_r} = \frac{1}{1 - p_{lost}(T_r)} \left(\int_0^{T_r} t P(t) dt  + p_{lost}(T_r)  T_r \right) 
 \end{equation}

A demon that resets at time $T_r$ is beneficial only if $ \langle T_{\rm{rep}} \rangle_{T_r} < \langle T_{\rm{rep}} \rangle_{\infty} \equiv E(T_{\rm{rep}}) $. That leads to
\begin{eqnarray}
       \frac{1}{1 - p_{lost}(T_r)} \left(\int_0^{T_r} t P(t) dt  + p_{lost}(T_r)  T_r \right)  \nonumber < \int_0^\infty t P(t) dt
 \end{eqnarray}

 \begin{eqnarray}
       \int_0^{T_r} t P(t) dt  &+&  p_{lost}(T_r)  T_r   < \int_0^\infty t P(t) dt\nonumber \\
       &-& p_{lost}(T_r) \int_0^\infty t P(t) dt
 \end{eqnarray}

 \begin{eqnarray}
         p_{lost}(T_r)  T_r   < \int_{T_r}^\infty t P(t) dt - p_{lost}(T_r) \int_0^\infty t P(t) dt \nonumber
 \end{eqnarray}
 
 which simplifies to
\begin{eqnarray}
    \int_{T_r}^{\infty}t P(t) dt &>& P_{lost}(T_r)\left( T_r + \int_0^{\infty}t P(t)dt\right)\\ \label{eq:beneficialresets}
    E(T_{\rm{rep}} | \text{ lost}) &>&  T_r + \int_0^{\infty}t P(t)dt\\
    &=&  T_r + E(T_{\rm{rep}})\\
    &=&  T_r + \left[ (1-P_{lost})E(T_{\rm{rep}}|\text{ not lost})+P_{lost}E(T_{\rm{rep}}|\text{ lost})\right]\\
    E(T_{\rm{rep}} | \text{ lost}) &>& \frac{T_r}{1-P_{lost}} + E(T_{\rm{rep}}|\text{ not lost})\, .
\end{eqnarray}
The numerical implementation of this condition shows good agreement with when the resets are beneficial to fitness, yet being approximate, see Fig. \ref{fig:SI_resets}(b)-i where the condition is computed for different probability distributions $P(T_{\rm{rep}})$.

\subsubsection{Numerical criterion}
The fitness can be computed numerically via Eq. \ref{eqn:generalresetfitness} as function of the reset time $T_r$ to determine whether a resetting demon is beneficial to the replication speed, once a distribution of times $P(T_{\rm{rep}})$ is assumed. The fold change in fitness $f(T_r)/f_0$ resulting from different distributions $P(T_{\rm{rep}})$ (Gaussian, Weibull, Fréchet and Gumbel) is shown in Fig. \ref{fig:SI_resets}(b)-ii.

\section{Minimal model of templated replication: two $\delta$ functions}

We work out the case of
\begin{equation}
P(T_{\rm{rep}}) = (1-\mu) \delta(T_{\rm{rep}} - \tau_f) + \mu \delta(T_{\rm{rep}} - \tau_s)
\end{equation}
with $\tau_s = k \tau_f$ and $k>1$ so that $\tau_f<\tau_s$ and $\tau_f,\tau_s$ represent fast and slow replication times, respectively. The optimal reset time is $T_r \approx \tau_f$. We find that from the above condition Eq. \ref{eq:beneficialresets} for evolution of resetting translates to
\begin{eqnarray}
    \frac{\tau_s}{\tau_f} > 1 + \frac{1}{1-\mu}\, , \nonumber \\
\end{eqnarray}
and the order (Eq. \ref{eq:order_times})
\begin{equation}
    \Delta S = D_{KL}(P^{neq} \vert \vert P^{eq}) = - \log (1-\mu)\, .
\end{equation}
Spontaneous order and dissipation diverge as $\mu \to 1$, which requires strong stalling effects $\frac{\tau_s}{\tau_f} \to \infty$. 

We find the moment generating function to be $\tilde{P}(w) = (1-\mu) e^{w \tau_f} + \mu e^{w k \tau_f}$. Let us assume resets at time $T_r$ with $\tau_f < T_r < k \tau_f$. The distribution of generation times with resets is then given by 
\begin{eqnarray}
 P^{neq}(T_{\rm{rep}})  &=&  (1-\mu) \delta(T_{\rm{rep}} - \tau_f) + \mu (1-\mu) \delta(T_{\rm{rep}} - (T_r + \tau_f)) \nonumber \\
 &+&\mu^2 (1-\mu) \delta(T_{\rm{rep}} - (2T_r + \tau_f)) \nonumber \\
 &+&  \mu^3 (1-\mu) \delta(T_{\rm{rep}} - (3T_r + \tau_f)) + \ldots
\end{eqnarray}
where subsequent terms represent $n$ reset events, due to $n$ attempted long trajectories with probability $\mu^n$, followed by attempting the fast trajectory with probability $1-\mu$. 
The corresponding moment generating function is,
\begin{eqnarray}
 \tilde{P}^{neq}(w) = \frac{(1-\mu) e^{w \tau_f}}{1 - \mu e^{w T_r}}
\end{eqnarray}

\textbf{The templated replication case}
The above example of two $\delta$ functions can be rewritten as a model of templated replication: consider the case of templated replication of a sequence of length $L$ with mutation rate $\mu$. The relevant trajectories are:
\begin{enumerate}
    \item $RR\ldots R$ (correct sequence, fast completion time $\tau_f$) with $\tau_f \sim L \tau_0$, $p_f= (1 - \mu)^L \approx 1 - \mu L$,
    \item $RR\ldots R W R \ldots R$ (sequence with errors, slow completion time $\tau_s$) with $\tau_s \sim (L-1)\tau_0 + \tau_{stall}$, $p_s = \mu L$, summed over the class of all trajectories with one mistake $W$ in them. 
\end{enumerate}

With these definitions, the equation for fitness benefit of resets Eq. \ref{eq:beneficialresets} gives
\begin{eqnarray}
\label{eq:delta1}
\int_{T_r}^\infty T_{\rm{rep}} P(T_{\rm{rep}}) dT_{\rm{rep}} &=& \mu L (L \tau_0 + \tau_{stall}) \nonumber \\
\int_{T
_r}^\infty P(T_{\rm{rep}}) dT_{\rm{rep}} & = &\mu L \nonumber \\
\int_0^\infty T_{\rm{rep}} P(T_{\rm{rep}}) dT_{\rm{rep}} &=& \mu L (L \tau_0 + \tau_{stall}) + (1-\mu L) (L \tau_0)\nonumber \nonumber \\
& = & L \tau_0 + \mu L \tau_{stall} + \mathcal{O}(L^2)
\end{eqnarray}
and
 \begin{eqnarray}
 \label{eq:delta3}
          \mu L (L \tau_0 + \tau_{stall})  &>& \mu L ( L \tau_0 + \mu L \tau_{stall} + T_r) \nonumber \\
            \tau_{stall} ( 1-  \mu L) &>&  T_r
 \end{eqnarray}
hence, using again $T_r\approx\tau_f$
 \begin{eqnarray}
 \label{eq:delta2}
    \tau_{stall} &>& \frac{L \tau_0}{1 - \mu L} \nonumber \\
    \Delta S = D_{KL}(P^{neq} \vert \vert P^{eq}) &=& - \log (1 - \mu L) \approx \mu L\, .
\end{eqnarray}

\section{Results for other distributions $P^{eq}(T_{\rm{rep}})$}
In the following we define different distribution of replication times used in Fig. \ref{fig:SI_resets}(b) where for simplicity of notation $P(\tau)$ is the distribution of replication times $P^{eq}(T_{\rm{rep}})$.
\subsubsection{Gaussian}
We first test these ideas in the simple case of Gaussian distributed replication times, $P(\tau) = \mathcal{N}(\mu,\sigma)$. The moment generating function is $\tilde{P}(w) = e^{\mu w + \sigma^2 w^2/2}$. 

Eq. \ref{eqn:generalresetfitness} reduces to
\begin{eqnarray}
2 \int_0^{T_r} e^{- f(T_r) \tau} P(\tau) d\tau &=&   
e^{\frac{1}{2} f \left(f \sigma ^2-2 \mu \right)} \nonumber \\
& & \left(\mbox{erf}\left(\frac{\mu -f \sigma ^2}{\sqrt{2} \sigma }\right)+\mbox{erf}\left(\frac{f \sigma ^2-\mu +T_r}{\sqrt{2} \sigma }\right)\right) \nonumber \\
a(T_r) & = &\frac{1}{2} \text{erfc}\left(\frac{\mu -T_r}{\sqrt{2} \sigma }\right)\, .
\end{eqnarray}

In our analysis, we consider the normal distribution with centered mean $\mu=1$, and values of variance $\sigma$ in the interval $[0.8,1.5]$.

\subsubsection{Log-normal}

We consider the log-normal distribution with mean fixed to 1 ($\exp\left(\mu +\frac{\sigma^2}{2}\right)=1$)

\begin{equation}
P(\tau) = \frac{1}{\tau\sigma\sqrt{2\pi}  }    \exp\left(-\frac{(\log \tau + \frac{\sigma^2}{2})^2}{2\sigma^2}\right)\, .
\end{equation}
In our analysis, we explore values of $\sigma$ in the interval $[0.6, 2]$.

\subsubsection{Weibull}

The Weibull distribution with mean fixed to 1 ($\lambda \Gamma(1+1/k)=1$) reads

\begin{equation}
P(\tau) = k\Gamma(1+1/k) \left(\tau \Gamma(1+1/k)\right)^{k-1}\exp\left[ -(\tau \Gamma(1+1/k))^k \right]
\end{equation}
where $\Gamma(x)$ is the gamma function, and we vary $k$ in the interval $[1.3, 2.5]$.

\subsubsection{Fr\'echet}

In our analysis, we fix the value of $\mu=0$ and $\beta=0.5$ and consider $\alpha$ in the interval $[0.5,2]$. Note that the mean diverges for $\alpha \leq 1$ and the variance diverges for $\alpha \leq 2$.

\begin{equation}
P(\tau) = \frac{\alpha}{\beta  }   \exp\left[-\left(\frac{\tau-\mu}{\beta }\right)^{-\alpha } \right]\left(\frac{\tau-\mu}{\beta }\right)^{-(\alpha +1)}
\end{equation}

In order to solve the integrals in Eq. \ref{eq:beneficialresets}, we fixed the mean to 1, and vary parameters in a regime such that the mean does not diverge, $\alpha\in[1.3,3]$. For the rest of the analysis, the mean is allowed to vary, and also diverge since $\alpha \leq 1$ as stated above.

\subsubsection{Gumbel}

The Gumbel distribution with mean fixed to 1 ($\mu + \beta\gamma = 1$) reads

\begin{equation}
    P(\tau) = \frac{1}{\beta} \exp \left[{-\frac{\tau-(1-\beta\gamma)}{\beta}} + \exp\left(-\frac{\tau-(1-\beta\gamma)}{\beta}\right) \right]\, ,
\end{equation}
where $\gamma$ is the Euler-Mascheroni constant, and we vary $\beta$ in the range $[0.5,2.5]$.

\section{Relation to reset literature}

Resets have been extensively explored in prior literature~---~see \cite{evans_stochastic_2020} for a review~---~in fields ranging from e.g. computer algorithms \cite{luby_optimal_1993,villen-altamirano_restart_nodate} to biology \cite{roldan_stochastic_2016,morris_anillin_2020} and ecology \cite{pal_search_2020}. First, studies within statistical physics \cite{evans_diffusion_2011,evans_stochastic_2020} involved diffusion processes in continuous spaces. Technical generalization including going beyond Poisson-distributed reset times \cite{eule_non-equilibrium_2016,pal_diffusion_2016,nagar_diffusion_2016}. Non-equilibrium statistical physics approaches have also considered the dissipation inherent to the detailed balance-breaking resetting procedure \cite{fuchs_stochastic_2016,busiello_entropy_2020,eliazar_entropy_2023}. 

Our results on resets builds on these works and stands out in a couple of ways. 

\begin{enumerate}

  \item While the prior literature focused on the impact of resets on the speed up in a search, our claim is that not only resets contribute to a reduction in the distribution of replication times \cite{eliazar_entropy_2023}, but also to a reduction in the entropy of the paths taken, see Eq. \ref{eq:order_times} where we derive a bound for the entropy of paths, then used to numerically compute the order. 
  Other works \cite{fuchs_stochastic_2016,busiello_entropy_2020,eliazar_entropy_2023} looked at the thermodynamic cost of the resetting process, but the reduction in entropy discussed in those cases is directly linked to the resetting process itself, e.g. a diffusing particle that is reset at position $x_0$ will have lower entropy in its position after resetting \cite{fuchs_stochastic_2016}. \cite{eliazar_entropy_2023} brought this forward by identifying conditions under which resetting increases or reduces the entropy of the completion time distribution. 
  Our interest here is in the reduction in entropy of paths, which we associate with the apparent evolution of order. 
  
  \item In particular, restart processes have been used to describe backtracking and cleavage in RNA replication \cite{roldan_stochastic_2016}, the focus being on building diffusion processes with resetting in recovery times~---defined as mean first passage times to an absorbing state~---~and accessing recovery time distributions. The connection between resets and DNA replication is brought up as a future research direction in the Discussion section of \cite{rotbart_michaelis-menten_2015}, expanding the use of resets from speeding up the search of a target \cite{viswanathan_optimizing_1999} to being linked to the unbinding rate of the wrongly inserted nucleotide. 
  Our work focuses on the interaction between stalling and kinetic proofreading is  formalize a connection to the theory of resets, and broadly make a link between resetting and error-correcting mechanisms in biology and self-assembly. The link with kinetic proofreading is made explicit in the two $\delta$-function toy model, see Eqs. \ref{eq:delta1}, \ref{eq:delta3} and \ref{eq:delta2} and Fig. 5 in the main paper.

  \item In our work we identify a condition, Eq. \ref{eq:beneficialresets}, for which resetting is beneficial to fitness~---~as defined in Eq. \ref{eq:fitness}. This condition is positing a constraint on the distribution of replication times: the average time cost of slow trajectories must be large and the probability of taking these slow trajectories must be small. This condition has some resonance with the inspection paradox, which is connected to stochastic resetting in \cite{pal_inspection_2022}: resets provides a mechanism to escape the long time intervals of the inspection paradox,  reducing sampling bias, hence the waiting time.

  \item Our work focuses on self-replicating systems, thus resetting acts on replication times, where a lineage fitness is not the mean of the first passage time distribution as considered in prior works, but rather a different biased average defined by Eq. \ref{eq:fitness} that is weighted towards the faster trajectories \cite{jafarpour_bridging_2018}. Even though the effect of resetting is fully determined by the distribution of replication rimes and of resetting times (Eq. \ref{eq:mgf}), the renewal approach developed in \cite{reuveni_optimal_2016,pal_first_2017} can also be used to compute the way the resetting process affects fitness.
  
\end{enumerate}

\chapter{Spontaneous pressure to dissipate}

\section{Equilibrium and non-equilibrium error correction}

Hopfield and Ninio's model of kinetic proofreading \cite{Hopfield1974-wa,ninio_kinetic_1975} are often celebrated as an example of non-equilibrium dynamics being exploited to achieve error correction in enzymatic reactions. However, dissipation is not neither necessary nor sufficient for suppressing errors~---~in fact, in several contexts, e.g. seeded crystal growth, templated polymerization \cite{Bennett1979-kz}~---~being near equilibrium results in fewer errors than being far from equilibrium. 

Here we contrast different families of error correcting mechanisms that achieve the same amount of order through different amounts of dissipation. We find three key results:

(i) selection for speed preferentially selects for more dissipative mechanisms over less dissipative mechanisms, even when they result in the same amount of order. 

(ii) this effect can be understood in terms of a balance between the time savings by reducing errors, e.g. less time spent stalled, and the time cost of error correction; dissipation reduces the latter. 

(iii) the above effects result in a spontaneous pressure to dissipate due to fast self-replication, even without any selection on increased order; we estimate this pressure by introducing a fitness cost of dissipation that prevents the evolution of increased order.

\subsection{Comparing reversible and irreversible error correction}

\begin{figure}[h]
    \centering
    \includegraphics[scale=1]{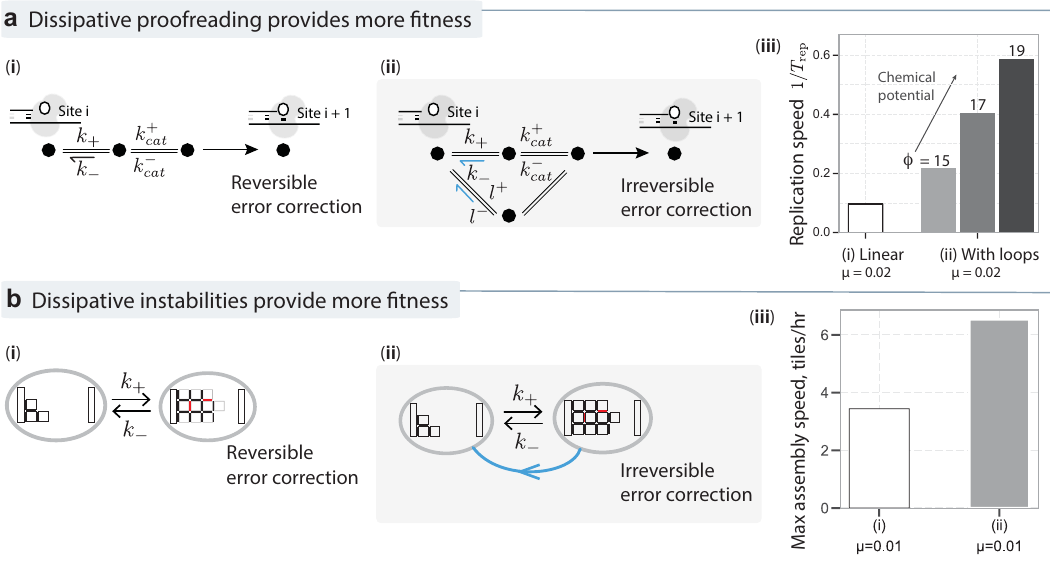}
    \caption{\textbf{Fast replication preferentially selects for more dissipative order-maintaining mechanisms.} (\textbf{a}) We compare polymerases (i) that remove errors by microscopically reversing polymerization pathways and (ii) with exonuclease-based error-correction, i.e. use alternative irreversible pathways. Reversible error correction is achieved in silico with a network of size $N=2$ (linear), while for irreversible error correction we choose a network of $N=6$ (with loops). We fix the stalling time to be $\tau_{\rm{stall}}=5$. Upon selecting for speed, both networks evolve towards lower error rate. However, when compared at the same error rate $\mu = 0.02$, more dissipative mechanisms have higher replication speed. 
     (\textbf{b}) We compare reversible and irreversible error correction in self-assembly: (i) through near-equilibrium self-assembly that exploits the microscopic reversal of assembly pathways, and (ii) through dynamic instability that uses distinct irreversible disassembly pathways. 
    Both mechanisms reduce defects while speeding up assembly; but the assembly speed of dynamic instability-based dissipative mechanisms is higher when compared at the same defect rate $\mu = 0.01$. }
    \label{fig:Dissipation}
\end{figure}

We compare two distinct models of error correction in templated replication: 

\begin{enumerate}
    \item an `irreverisble' model of error correction, i.e. kinetic proofreading in which misincorporated nucleotides are removed by exonuclease activity. Note that exonuclease activity is not simply the microscopical reversal of the forward polymerase activity; as a consequence, kinetic proofreading models always have loops of paths connecting enzyme-substrate states as seen in the looped network as shown in Fig. \ref{fig:Dissipation}(a.ii) used for simulations in this section.
    \item a `reversible' model of error correction in which misincoporated nucleotides must be removed by the microscopic reversal of the polymerization pathway itself. Such a model, shown in Fig. \ref{fig:Dissipation}(a.i), has no loops of paths connecting enzyme-substrate states.  
\end{enumerate}

Note that the `reversible' model is a less dissipative mode of error correction: it is most effective at correcting errors when it is operated near equilibrium, while the irreversible kinetic proofreading model, with loops, is most effective when strongly driven out of equilibrium.

We compare these two models at a value of error rate that is achievable by both of them. Fig.  \ref{fig:Dissipation}(a.iii) shows that the irreversible error correction mechanism, i.e. kinetic proofreading models with loops, replicate templates faster~---~and would thus be evolutionarily prefered~---~than reversible mechanisms that achieve the same error rate. Note that this effect cannot be explained by the usual picture of dissipation being needed to reduce error rates, e.g. because of the second law, since both networks are being compared when they achieve the same error rate. The speed to replicate defined in Fig.  \ref{fig:Dissipation}(a.iii) is computed as in Eq. \ref{eq:time_stall} with $\tau_{\rm{stall}}=5$ and is compared between the equilibrium network ($N=2$) and three values of the driving force in the loop network ($N=6$), $\phi=15,17$ and $\phi=19$, at the same error rate $\mu = 0.02$. Note that when $\tau_{\rm{stall}} \gg 1$, the curves asymptote to $\mu \tau_{\rm{stall}}$ for large $\mu$, which pushes the networks to have the same speed when compared at the same, large value of error rate where the reversible error correction curve ($N=2$) lies, see e.g. Fig. 2(a)iii (main text). Hence, even if increasing the stalling time reduces the differences between the speed, the networks with loops are always faster then linear one, until $\tau_{\rm{stall}}\sim10^5$ where the speeds are all comparable.

\subsection{Preference for dissipative error correction}

To quantitatively understand the preference for irreversible error correction seen above, we consider continuously varying the non-equilibrium driving force in an error correcting mechanism.

Consider a lineage with a looped kinetic proofreading-based error correction mechanism, driven out of equilibrium to an extent $\phi$, that replicates at a rate $F = 1/T_{\rm{rep}}$ with 
\begin{eqnarray}
T_{\rm{rep}} = L t_{\rm{nostall}}(\mu) + L \mu  \tau_{\rm{stall}}
\end{eqnarray}

Here $t_{\rm{nostall}}(\mu)$ is the time taken by a proofreading network, with non-equilibrium drive $\phi$, to achieve a mutation rate $\mu$ without accounting for stalling, as calculated in many prior works \cite{hopfield_kinetic_1974, ninio_kinetic_1975, murugan_discriminatory_2014}. 

\begin{figure}[h]
    \centering
    \includegraphics[scale=1]{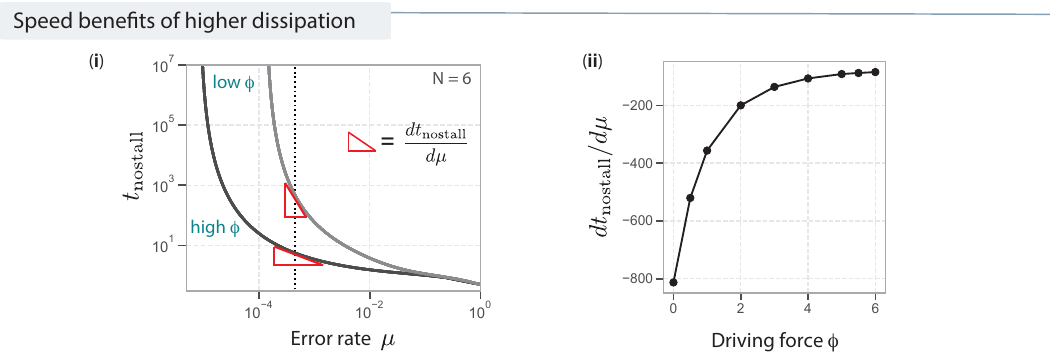}
    \caption{\textbf{Fast replication preferentially selects for more dissipative order-maintaining mechanisms.} (\textbf{c}) The preference for dissipative error correction in proofreading can be understood from how $dt_{\rm{nostall}}/d\mu$ depends on the driving force, $\phi$. (i) Plots of the time to replicate with no stalling $t_{\rm{nostall}}$ as a function of error rate $\mu$ shown for an $N=6$ proofreading network driven out of equilibrium to two different extents $\phi$. Both curves show that reducing error by an amount $\Delta \mu$ costs time $\Delta t_{\rm{nostall}}$, but the time cost is lower for highly driven $\phi$ networks.  (ii) The derivative $dt_{\rm{nostall}}/d\mu$ is plotted as a  function of the driving force $\phi$ for the same network of size $N=6$. Thus more dissipative mechanisms achieve the same error correction at a smaller cost in time. }
    \label{fig:Dissipation2}
\end{figure}

A given change in mutation rate $\delta \mu$ results in a change in fitness,
\begin{equation}
\frac{\Delta F}{\Delta \mu} = -\frac{1}{T_{\rm{rep}^2}} \left( L \frac{\Delta t_{\rm{nostall}}(\mu) }{\Delta \mu} + L \tau_{\rm{stall}} \right) 
\end{equation}

The first term here, $\frac{\Delta t_{\rm{nostall}}(\mu) }{\Delta \mu}$, is the slope of the speed-error trade-off of proofreading models in the absence of stalling, i.e. the slope of the intuitive trade-off explored in prior literature \cite{Hopfield1974-wa,Ninio1975-ky,munsky_specificity_2009,sartori_kinetic_2013,murugan_discriminatory_2014} and also plotted in Fig. \ref{fig:Dissipation2}. The derivative $\frac{\partial t_{\rm{nostall}}(\mu) }{\partial \mu}$ is computed numerically in a proofreading network of the type introduced in Sec. \ref{sec:specificnets} with $N=6$ and $\tau_{\rm{stall}}=5\times 10^3$, keeping the error rate in the range $[0.02,0.0202]$. As seen there, this slope  $\frac{\Delta t_{\rm{nostall}}(\mu) }{\Delta \mu}$ is negative and increases as the non-equilibrium drive $\phi$ in the proofreading network is increased. 

Hence, the fitness increase $\Delta F$ due to a given reduction $\Delta \mu$ is higher if that reduction was achieved by a more dissipative proofreading mechanism, rather than less dissipative mechanism. Intuitively, dissipative mechanisms achieve the same reduction $\Delta \mu$ using a smaller increase in the time cost $\Delta t_{\rm{nostall}}(\mu)$ error correction. 

\subsection{Penalty for dissipation}

The above results suggest that fast self-replication can produce a spontaneous selection pressure $s_\epsilon$ on living organisms to dissipate more energy rather than less; note that this pressure is not just the often-discussed energy cost of accuracy since the above results show a (speed) advantage for more dissipative systems, even when compared at fixed accuracy. We call this effect a `spontaneous pressure' to dissipate since this selection pressure is intrinsic to exponential growth due to self-replication, even without assuming any fitness advantage of accuracy.

\begin{figure}[h]
    \centering
    \includegraphics{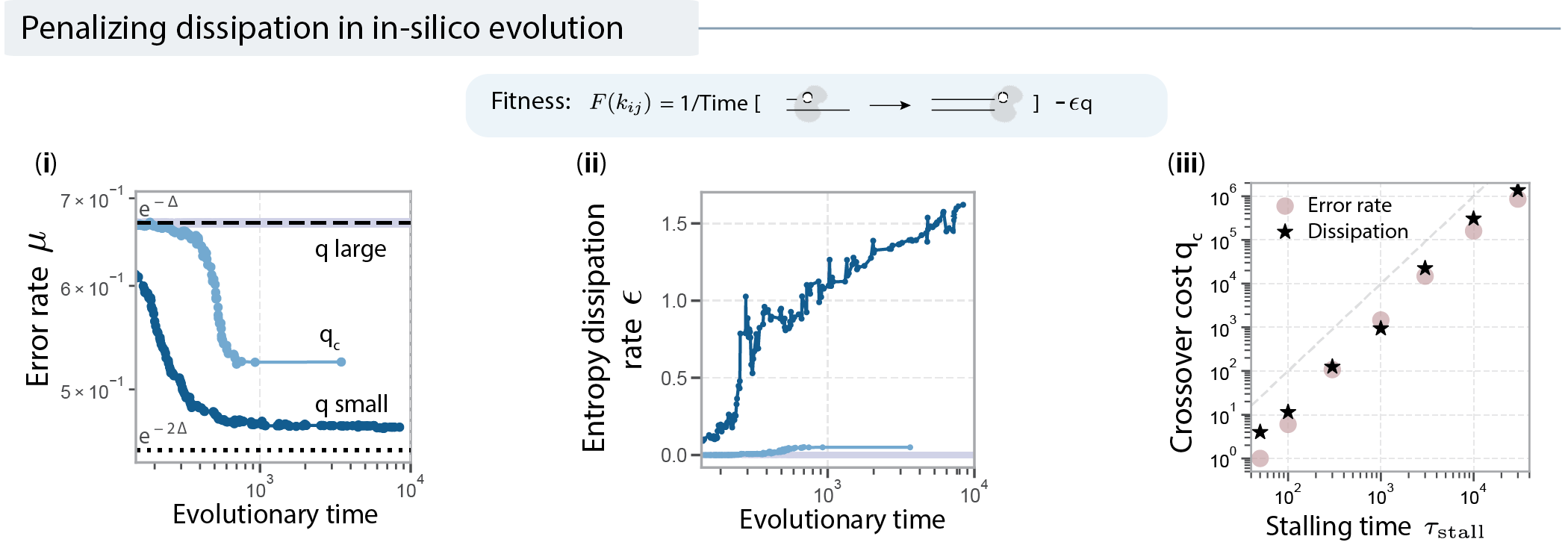} 
    \caption{\textbf{In silico evolution with fitness penalty for dissipation} To study the effect of a fitness penalty for dissipation we introduce a cost proportional to dissipation,  $q\epsilon$, in the fitness function, Eq.~\ref{eq:trep_main}, introducing a pressure between time minimization and dissipation. The in-silico evolution is performed for a proofreading network of size $N=3$ and $\tau_{\rm{stall}}=10^3$ as in Sec.~\ref{sec:insilicoevo}, but with the new fitness function including the fitness cost $q\epsilon$, in addition to minimizing time. The effect of such a fitness cost of dissipation is explored for (i) error rate $\mu$, (ii) entropy dissipation rate $\epsilon$, which are shown as function of the evolutionary time. Beyond some critical cost of dissipation $q_c$, in silico evolution no longer evolves proofreading mechanisms, i.e. low $\mu$ or high $\epsilon$. 
    (iii) The crossover cost $q_c$ is identified numerically as the value of $q$ for which error rate no longer decreases or at which dissipation no longer increases. We measure $q_c$ as function of the stalling time, $\tau_{\rm{stall}}$, and show that it increases with stalling time; the dashed line is $q_c\sim\tau_{\rm{stall}}^2$. This crossover cost $q_c$ can be interpreted as the spontaneous fitness pressure to dissipate due to fast self-replication. }
    \label{fig:SI-DNA-dissipation}
\end{figure}

We can quantify this spontaneous selection pressure to dissipate by introducing a fitness penalty for dissipation as shown in Fig. \ref{fig:SI-DNA-dissipation}. That is, we now assume that the fitness function includes a cost proportional to the dissipation rate $\epsilon$,
\begin{equation}
    F(q)=1/T_{\rm{rep}}-q\epsilon\, .
\end{equation}

We carried out \emph{in silico} evolution of proofreading networks as earlier but with this modified fitness function. That is, we took a random copier network as defined in Sec. \ref{sec:insilicoevo} of size $N=3$ and carried out \emph{in silico} evolution with the parameters defined in Tab. \ref{tab:insilicoevo}. 

The error rate and the entropy dissipation rate can be measured as function of evolutionary time for different values of $q$: if $q$ is small, then there is almost no effect of the dissipation cost; if $q$ is large then the dissipation cost is so large that evolution does not result in lower error rates $\mu$ or in higher dissipation. In between, there exists a crossover $q_c$ for which the error rate starts dropping from the equilibrium value $e^{-\Delta}$ and the entropy dissipation rate starts increasing. 
The value of $q_c$ can be estimated by screening across different values of $q$, monitoring changes in error rate and entropy dissipation rate, which give two independently measured estimates of $q_c$ reported in panel (iii) of Fig. \ref{fig:SI-DNA-dissipation}. Since $q_c$ is the dissipation penalty at which fast self-replication no longer selects for higher accuracy, we can interpret $q_c$ to be a measure of the spontaneous pressure to dissipate. To see this, note that $\frac{dF(q)}{d\epsilon} = \frac{dT_{\rm{rep}^{-1}}}{d\epsilon} - q$; at $q_c$, $\frac{dF(q)}{d\epsilon} = 0$ and hence $q_c \sim \frac{dT_{\rm{rep}^{-1}}}{d\epsilon}$.

Fig. \ref{fig:SI-DNA-dissipation}(iii) shows that $q_c$ and hence the spontaneous pressure to dissipate increases with the stalling time $\tau_{\rm{stall}}$. More broadly, outside of templated replication, we expect this pressure to dissipate to scale with the variance of the distribution of replication times as discussed in the section of resets.

\section{Self-assembly}

\subsection{Reversible and irreversible error correction in self-assembly}

The paper explored dynamic instability as a mechanism of error correction in self-assembly. This mechanism corrects errors by introducing new disassembly pathways that are not simply the microscopic reversible of assembly pathways.

However, errors in self-assembly can also be corrected by an alternative near-equilibrium reversible mechanism. If assembly is carried out close to the melting temperature, errors can be corrected by simply reversing assembly pathways with a high frequency. Unlike dynamic instability, this mechanism is most effective when self-assembly is carried out near equilibrium with minimal non-equilibrium drive and dissipation. As with templated replication earlier, reversible error correction models do not contain loops of pathways, while irreversible dynamic instability models do contain loops since they introduce novel disassembly pathways.

\begin{figure}
    \centering
    \includegraphics{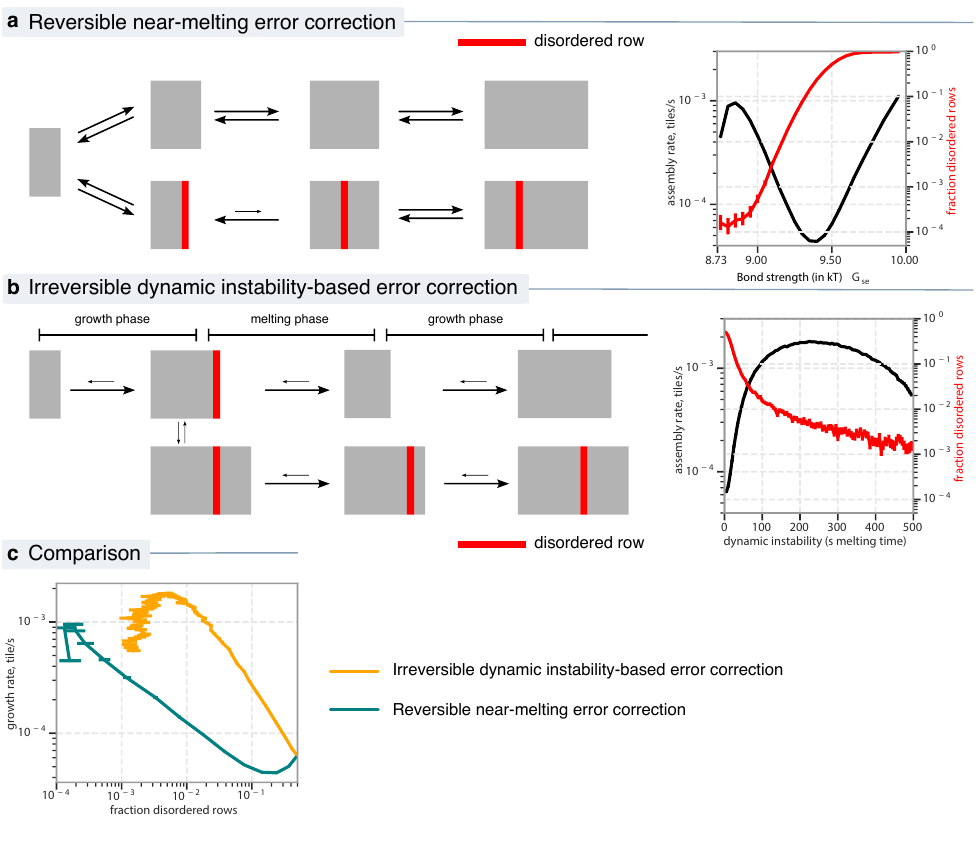}
    \caption{\textbf{Preferential evolution of irreversible error correction over reversible  mechanisms in self-assembly} \figp{a} 
    Reversible error correction mechanism: At near-melting point conditions ($\gse \sim \gsecrit = 8.73$), errors (red columns = disordered rows) are disassembled through fluctuations that microscopically reverse the assembly pathways. Consequently, the error rate (fraction of disordered rows) is lower for $\gse \sim \gsecrit = 8.73$ (red curve on right) than when assembly is strongly forward biased (e.g., $\gse \sim 10$). The assembly rate (black curves) shows complex non-monotonic behavior; assembly rate initially decreases with increasing $\gse$ (i.e., lowered temperature) because disordered rows are created more frequently and stall further growth. However, at even larger $\gse$ (i.e., even lower temperatures), the stalling effect itself is reduced and assembly proceeds past disordered rows with ease.
    \figp{b} Irreversible dynamic instability corrects errors by disassembly pathways distinct from the reversal of assembly pathways; in our model, disassembly and assembly occur in alternating time periods of high and low temperature (i.e., low and high $\gse$). 
Assembly rate is maximal for a specific length of the melting $\gsemelt$ phase while the probability of continued disordered growth decreases with increasing $\gsemelt$ phase time. \figp{c} Reversible error correction results in lower assembly rates than the irreversible dynamic instability-based error correction when compared at the same achieved error rate (defined as the fraction of disordered rows in completed assemblies).} 
    \label{fig:ext-sa-slow}
\end{figure}

To explore this possibility of reversible near-equilibrium error correction and compare it to irreversible dynamic instability, constant bond strength simulations were run using the medium barrier system with $\gmc = 16$, and $\gse$ ranging from 8.75 to 10 in increments of 0.05.  At each point, 3,072 simulations of individual assemblies were run, up to a target size of 480 tiles, with a maximum time cutoff of $10^8$ seconds.  The results are shown in \ref{fig:ext-sa-slow}a.

At high enough constant bond strength (equivalently, low fixed temperature), growth becomes essentially disordered, as kinetic barriers to disordered attachments are easily surmounted: increasing bond strength from $\gse = $ 9.5 to 10, assembly rate increases, but most rows contain errors.  At around $\gse = 9.4$, the stalling effect of disordered growth is strongest, and growth is both slow and error-prone.  Decreasing bond strength from this point, while individual tile attachments become less forward-biased, the rate at which stalled structures will have disordered tiles detach increases, resulting in both an overall in growth rate, and decrease in error rate, as structures become more likely to escape stalled states through detachments.  This trend continues to slightly above the critical $\gsecrit = 8.73$: below that point, while error rates remain near the theoretical minimum estimate of $e^{-\gsecrit} \sim 10^{-4}$, growth speed decreases as the reduction in stalling no longer outpaces the reduction in the un-stalled growth rate of the system.

Figure \ref{fig:ext-sa-slow}c shows the highest growth achievable for a target error rate, choosing the optimal $\gse$ constant-bond-strength growth and optimal $\tmelt$ for growth with dynamic instability, corresponding to the plots in \ref{fig:ext-sa-slow}a and b.  While, in the simulated conditions, lower error rates (below $10^{-3}$ per row) can be achieved with constant-temperature growth than for growth with dynamic instability, for a range of error rates, centered around $10^{-2}$ per row, dynamic instability allows the system to grow faster than it would grow at any constant temperature.  This comparison also only considers dynamic instability with a fixed period (3600 s), $\gsegrow$, and $\gsemelt$, varying only $\tmelt$.  It is likely that optimizing all parameters would result in lower error rates, and faster growth.

\bibliography{ravasio,paperpile,cge}